\documentclass[a4paper,11pt]{article}
\usepackage{jheppub}
\usepackage{lineno}
\usepackage{bm}
\usepackage{graphicx}
\usepackage{multirow}
\usepackage[table,xcdraw]{xcolor}
\usepackage{hhline}
\usepackage[rightcaption]{sidecap}
\usepackage{url}
\usepackage{tablefootnote}
\usepackage{fancyhdr}
\usepackage{blindtext}
\usepackage{scrextend}
\addtokomafont{labelinglabel}{\sffamily}
\usepackage{subfigure}
\usepackage{gensymb}
\usepackage{wrapfig}
\usepackage{float}
\usepackage{slashed}
\usepackage{tikz}
\usetikzlibrary{backgrounds}
\usepackage[compat=1.1.0]{tikz-feynhand}
\usepackage{subcaption}

\title{\boldmath Non-eikonal corrections to dijet production in DIS}

\author[a]{N\'estor Armesto}
\author[a,b]{Fabio Dom\'{\i}nguez}
\author[a]{and Adri\'an Romero}
\affiliation[a]{Instituto Galego de F\'{\i}sica de Altas Enerx\'{\i}as IGFAE,
Universidade de Santiago de Compostela,\\ 15782 Santiago de Compostela, Galicia-Spain}
\affiliation[b]{CPHT, CNRS, \'Ecole Polytechnique,  Institut Polytechnique de Paris, 91120 Palaiseau, France}

\emailAdd{nestor.armesto@usc.es, fabio-alejandro.dominguez-gonzalez@polytechnique.edu, adrian.romero.gonzalez@usc.es}

\abstract{We compute non-eikonal corrections to dijet production in deep inelastic scattering off a nucleus. Such corrections are expected to be quantitatively important at the energies of the future Electron Ion Collider. We focus on those corrections stemming solely from the finite longitudinal size of the nucleus. For both longitudinally and transversely polarized photons, we provide general, all-order expressions in terms of two-dimensional path integrals. To proceed further, we use the harmonic oscillator approximation for the target averages of Wilson lines. We then expand the general expressions order by order beyond the shockwave limit which provides the eikonal results, up to next-to-next-to-eikonal accuracy. We observe that next-to-eikonal corrections to this observable vanish for the mentioned approximation for target averages, as previously found for single gluon production in proton-nucleus collisions. Finally, we calculate the back-to-back of correlation limit of our expressions.}

\begin{document}
\maketitle
\flushbottom

\section{Introduction}
\label{sec:intro}

In the last few decades, the study of strongly interacting matter under extreme conditions has attracted significant attention, leading directly to the development of theoretical methods to describe the vast array of interactions taking place in such environments. In particular, high-energy collision experiments involving nuclei allow us to probe the properties of Quantum Chromodynamics (QCD) at very high temperatures and/or densities, where high-energy particles experience multiple interactions which must be resummed in theoretical calculations.

In such cases, there is a clear separation between projectile and target, where one is a very high-energy particle and the other is composed of soft modes which can be accounted for by a large classical field. The energy of the projectile is much larger than any transverse scale in the process, which leads to an effective separation of the transverse and longitudinal dynamics where the interactions between projectile and target are purely transverse and can be resummed in terms of Wilson lines along the light-cone.

One case of particular interest is the study of the small-$x$ degrees of freedom and how they affect hadron structure and particle production in nuclear collisions. The usual framework employed in those cases is the Color Glass Condensate (CGC)~\cite{Gelis:2010nm,Kovchegov:2012mbw}, which is a weak-coupling, non-perturbative effective field theory of high-energy QCD. For most processes of interest, the leading contribution can be obtained using the eikonal approximation, in which the fast moving target is taken as a background field localized in the longitudinal direction, akin to a ``shockwave" where their longitudinal extent is neglected and its internal dynamics is ignored. In light-cone coordinates\footnote{$(x^+, x^-)$, $x^\pm = (x^0 \pm x^3)/\sqrt{2}$ for a left-moving target.}, the target field then takes the following form
\begin{equation}
	A^\mu_a(x^+,x^-,\mathbf{x}) \approx \delta^{\mu -}\delta(x^+) \alpha_a(\mathbf{x}),
\end{equation}
with $a=1,\dots,N_c^2-1$ the colour index, transverse coordinates written in boldface and function $\alpha_a$ representing the transverse dependence of the gauge field. The projectile then probes the target at a fixed transverse coordinate, which is the same as assuming that the transverse coordinate of the projectile is ``frozen" while traversing the target when considering a different reference frame.

It is important to note that this eikonal approximation consists of neglecting three types of contributions which are suppressed by powers of the energy of the projectile: the effect of the finite extent of the medium, the contribution from other components of the background field which do not couple to the large momentum component, and the time dependence of the background field encompassing its inner dynamics.

High-energy scattering in a medium with a finite length is a phenomenon which has been widely studied in a slightly different way, also in the context of high-energy nuclear collisions. Specifically, in jet quenching studies where the modification and energy loss of jets traversing a quark-gluon plasma (QGP) is considered~\cite{Blaizot:2015lma,Apolinario:2022vzg,Mehtar-Tani:2025rty}. In that case the strict eikonal approximation is not enough to account for the leading effects since the probes are created in the same collision event as the medium and the energy lost due to interactions with such medium is dominated by gluon radiation inside the medium. The use of jet quenching techniques to calculate non-eikonal corrections to CGC calculations was first presented in~\cite{Altinoluk:2014oxa,Altinoluk:2015gia} where a systematic expansion in terms of the length of the medium is performed while keeping a full resummation of multiple scatterings.

Other non-eikonal corrections to CGC calculations have also been computed by either considering other components of the background field or considering the time evolution of such fields. In particular, corrections for scalar, quark, and gluon propagators, also including quark exchanges with the target, have been calculated in~\cite{Altinoluk:2020oyd,Altinoluk:2021lvu,Agostini:2023cvc,Balitsky:2015qba,Balitsky:2016dgz,Chirilli:2018kkw,Chirilli:2021lif,Altinoluk:2023qfr,Jalilian-Marian:2017ttv,Jalilian-Marian:2018iui,Chirilli:2026vij}, with many of these studies being done in the context of spin physics, given that spin exchanges are subleading in energy and thus subeikonal~\cite{Kovchegov:2015pbl,Kovchegov:2016weo,Kovchegov:2016zex,Kovchegov:2017jxc,Kovchegov:2018znm,Cougoulic:2019aja,Cougoulic:2020tbc,Kovchegov:2021iyc,Jalilian-Marian:2019kaf}. Also, in the context of jet quenching studies, there has been extensive work to include other non-eikonal corrections beyond those enhanced by the length of the medium in order to include effects coming from flow and anisotropies in the plasma~\cite{Majumder:2008zg,Sadofyev:2021ohn,Barata:2022krd}.

Existing numerical studies of the impact of non-eikonal corrections on unpolarized observables have mostly focused on relaxing the shockwave approximation in single and double inclusive particle production in proton-proton and proton-nucleus collisions~\cite{Altinoluk:2015xuy,Agostini:2019avp,Agostini:2019hkj,Agostini:2022ctk,Agostini:2022oge}. For the remaining non-eikonal corrections, in~\cite{Agostini:2024xqs} the effects on dijet production in Deep Inelastic Scattering (DIS) of the consideration of a finite length and transverse target fields were analysed. All these studies indicate that the considered non-eikonal effects are significant for centre-of-mass energies $\sqrt{s_{NN}} \lesssim 100$ GeV. While these energies lie below the top ones achievable at the Relativistic Heavy Ion Collider, they are precisely those of interest for the Electron Ion Collider (EIC)~\cite{Accardi:2012qut,AbdulKhalek:2021gbh}. Therefore, the theoretical accuracy required to match the experimental measurements at the EIC requires the consideration of non-eikonal effects on DIS observables.
	
Dijet production in DIS has recently become a model observable for several reasons. First, it can potentially reveal reveal non-linear density effects in QCD~\cite{Dumitru:2018kuw,Mantysaari:2019hkq,Zhao:2021kae,Boussarie:2021ybe,Altinoluk:2021ygv,vanHameren:2021sqc}, i.e., saturation. Second, several types of corrections to the leading order eikonal calculation have been computed, including the cross section at next-to-leading order in the coupling constant~\cite{Caucal:2021ent,Taels:2022tza,Caucal:2022ulg,Caucal:2023nci,Caucal:2023fsf}, and also the relevant next-to-eikonal corrections~\cite{Altinoluk:2022jkk,Agostini:2024xqs,Mukherjee:2026cte} (see~\cite{Altinoluk:2025ivn} for recent calculations of next-to-eikonal corrections to quark and gluon propagators at next-to-leading order in the coupling constant).
Third, in this process it is possible to establish a link between CGC calculations and transverse momentum distributions (TMDs), which encode three-dimensional information about the hadron and nuclear structure, in the so-called "correlation limit"~\cite{Boussarie:2023izj,Dominguez:2011wm,Kotko:2015ura,Dumitru:2015gaa,Dumitru:2018kuw,Marquet:2016cgx,Marquet:2017xwy,Altinoluk:2019fui,Altinoluk:2019wyu,Altinoluk:2024zom}\footnote{To define the correlation limit, we consider two jets with transverse momenta $\mathbf{p},\mathbf{k}$ and light-cone momentum fractions $z_{1,2}$ acquired from the (virtual) photon. We define the momentum imbalance as ${\bf q} = {\bf p} + {\bf k}$ and the relative momentum as ${\bf P} = z_2 {\bf p} - z_1 {\bf k}\simeq ({\bf p} - {\bf k})/2$. The correlation limit corresponds to $|{\bf P}| \gg |{\bf \mathbf{q}}|$, where the approximate equality holds.}, under which a connection between target averages of Wilson lines and target TMDs at momentum fraction $x=0$ is found. Last, it has been used as a process for discussion about how to link small and moderate-to-large $x$ QCD evolution~\cite{Caucal:2024bae,Caucal:2025mth}.
	
In this work we study non-eikonal effects on dijet production in nuclear DIS. We focus on those corrections stemming from the relaxation of the shockwave approximation, thus considering a target with a finite extent. While previous calculations~\cite{Altinoluk:2022jkk,Agostini:2024xqs} considered only the first corrections to the eikonal approximation, i.e., at next-to-eikonal order, we go beyond by using the all-order formalism employed in jet quenching~\cite{Casalderrey-Solana:2007knd,Mehtar-Tani:2013pia,Blaizot:2015lma,Apolinario:2022vzg,Mehtar-Tani:2025rty}, where the quark propagator is written as a path integral that resums all orders of such type of non-eikonal corrections. We provide general results but, in order to proceed further with developing this result order by order in the eikonal expansion and analysing the correlation limit, we use a specific model for the target averages of Wilson lines, the harmonic oscillator (HO) or multiple soft approximation frequently employed in jet quenching~\cite{Zakharov:1998sv}, which is equivalent to the Golec-Biernat--W\"usthoff (GBW) model~\cite{Golec-Biernat:1998zce,GolecBiernat:1999qd} used to describe DIS inclusive and diffractive data, see~\cite{Agostini:2022ctk} for the equivalence of both models in the CGC context.

This manuscript is organised as follows. First, in section~\ref{sec:eik} we revisit the eikonal calculation in order to set our notation and conventions. Then, in section~\ref{sec:long} we compute the amplitudes and their squares, for longitudinal and transverse photons. In section~\ref{sec:shockwave} we perform the shockwave expansion of the results, and in section~\ref{sec:correl} we study the correlation limit. Finally, section~\ref{sec:conclu} contains a summary and our conclusions. Technical details of the calculations are given in the several appendices.

\section{Photon-Nucleus interaction in the eikonal approximation}
\label{sec:eik}
\begin{figure}[ht]
\centering
\subfigure[Splitting before the medium]{
\begin{tikzpicture}
\begin{feynhand}

\vertex (a) at (-2,0);
\vertex (b) at (0,0);
\vertex (c) at (4.5,1.5);
\vertex (d) at (4.5,-1.5);

\propag [photon] (a) to [edge label = $q$] (b);
\propag [fer] (b) to [edge label = $p_{0}$, out=90, in=180] (c);
\propag [fer] (d) to [edge label = $k_{0}$, out=180, in=-90] (b);
\node at (4.6,1.8) {$p, r$};
\node at (4.6,-1.8) {$k, s$};
\node at (2,2.3) {$0$};
\node at (4,2.3) {$L^{+}$};
\node at (3.7, 1.2) {${\bm{x}}$};
\node at (3.7, -1.2) {${\bm{z}}$};
\node at (0.8, 0) {$\left(y^{+},\bm{y}\right)$};

\fill[gray, opacity=0.3] (2,-2) rectangle (4,2);

\end{feynhand}
\end{tikzpicture}
\label{Fig:Diagrama1}}
\subfigure[Splitting after the medium]{
\begin{tikzpicture}
\begin{feynhand}

\vertex (a) at (-2.5,0);
\vertex (b) at (0,0);
\vertex (c) at (2.5,1.5);
\vertex (d) at (2.5,-1.5);

\propag [photon] (a) to [edge label = $q$] (b);
\propag [fer] (b) to [out=90, in=180] (c);
\propag [fer] (d) to [out=180, in=-90] (b);
\node at (2.7,1.8) {$p, r$};
\node at (2.7,-1.8) {$k, s$};
\node at (-2.25,2.3) {$0$};
\node at (-0.25,2.3) {$L^{+}$};
\node at (1.4, 1) {${\bm{x}}$};
\node at (1.4, -1) {${\bm{z}}$};
\node at (0.8, 0) {$\left(y^{+},\bm{y}\right)$};
\node at (0, 0.8) {$p_{0}$};
\node at (0, -0.8) {$k_{0}$};

\fill[gray, opacity=0.3] (-2.25,-2) rectangle (-0.25,2);

\end{feynhand}
\end{tikzpicture}
\label{Fig:Diagrama2}}

\caption{Diagrams illustrating the process $\gamma^{\ast}_{L,T}+A\to q\bar{q}X$. $k,p$ denote momenta, $x,y,z$ positions and $r,s$ helicities. The modulus of transverse vectors is defined as $x\equiv |{\bm{x}}|$.}
\label{Fig:Diagrama}
\end{figure}
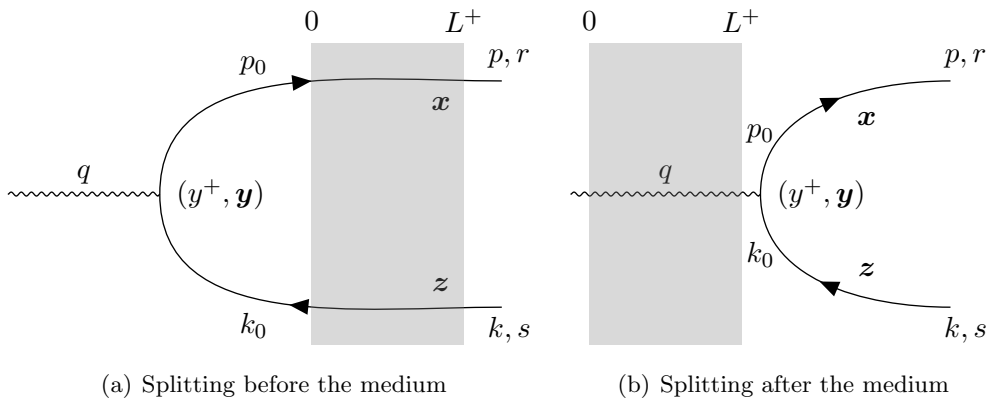

In the dipole picture for the interaction between the virtual photon coming from the electron and the heavy nucleus, the photon splits into a $\displaystyle q\bar{q}$ pair that traverses the target nucleus. In the CGC the target is treated as a set of classical and randomly distributed colour sources that generate a classical field. In the eikonal approximation the longitudinal momentum of the incoming particles, $\displaystyle p^{+}$, $\displaystyle k^{+}$ for the quark and anti-quark respectively, are much larger than any other momenta in the process. In this approximation, the splitting of the photon can only occur either before interacting with the nucleus  or after the virtual photon traverses it, since splittings inside the nucleus are eikonally suppressed; see Fig.~\ref{Fig:Diagrama} where our notations are shown. We also work in the light-cone gauge $A^+=0$.

In this approximation the interaction of the eikonal quark with the medium is given by the all-order resummation of gluon scatterings with the classical field of the nucleus at a fixed transverse position, $\displaystyle A^{\mu}(x^{+},x^{-},\bm{x})=\delta^{\mu-}A^{-}\left(x^{+},\bm{x}\right)$, with the $-$ component of the field being the one enhanced by boosts of the target. This resummation results in a Wilson line\footnote{$e_q$ and $g$ denote the electromagnetic quark charge and the strong coupling constant, respectively.}:
\begin{equation}
U\left(L^{+},0;\bm{x}\right)=U\left(\bm{x}\right)=\mathcal{P}\exp\left[-ig\int_{0}^{L^{+}}dx^{+}A^{-}\left(x^{+},\bm{x}\right)\right],
    \label{eq.1.1}
\end{equation}
where $L^{+}$ is the longitudinal extent of the nucleus in the $+$ direction.

The amplitude for the scattering is then given by
\begin{equation}
\begin{split}
    S=&2\pi\delta\left(p^{+}+k^{+}-q^{+}\right)\int_{-\infty}^{0}dy^{+}\int d^{2}y \ e^{-iq^{-}y^{+}}e^{i\bm{q}\cdot\bm{y}}
    \\
    &\times \int\dfrac{d^{2}p_{0}}{\left(2\pi\right)^{2}}\dfrac{d^{2}k_{0}}{\left(2\pi\right)^{2}}e^{i\frac{p^{2}_{0}}{2p^{+}}y^{+}}e^{-i{\bm{p_{0}}}\cdot\bm{y}}e^{i\frac{k^{2}_{0}}{2k^{+}}y^{+}}e^{-i{\bm{k_{0}}}\cdot\bm{y}}
    \left[\bar{u}_{r}\left(p_{0}\right)\left(-ie_{q}e\slashed{\varepsilon}\left(q\right)\right)v_{s}\left(k_{0}\right)\right]
    \\ &\times  \int d^{2}x \ d^{2}z \ e^{-i\left({\bm{p}}-{\bm{p_{0}}}\right)\cdot\bm{x}}e^{-i\left({\bm{k}}-{\bm{k_{0}}}\right)\cdot\bm{z}}\left[U\left(\bm{x}\right)U^{\dagger}\left(\bm{z}\right)-1\right],
\end{split}
\label{eq.1.2}
\end{equation}
where the first term in the square bracket with the Wilson lines represents the case with the splitting before traversing while the second term corresponds to splitting after traversing the shockwave.

From this amplitude we can obtain the differential cross section for a virtual photon with longitudinal polarization vector $\displaystyle \varepsilon_{L}^{\mu}=\left(\dfrac{q^{+}}{Q},\dfrac{Q}{2q^{+}},\bm{0}\right)$ simply by squaring the amplitude, summing over the polarizations and colours, taking a reference frame with $\displaystyle \bm{q}=\bm{0}$ and integrating the momenta and the vertex position. One gets
\begin{equation}
\begin{split}
    \frac{d\sigma^{\gamma^{\ast}_{L}A\to q\bar{q}X}}{d^{3}p \ d^{3}k}=&\left(2\pi\right)^{2}\delta\left(p^{+}+k^{+}-q^{+}\right)8N_{c}\alpha_{em}e^{2}_{q}\zeta^{2}(1-\zeta)^{2}\frac{Q^{2}}{q^{+}}\\
    &\times \int\frac{d^{2}x}{\left(2\pi\right)^{2}}\frac{d^{2}z}{\left(2\pi\right)^{2}}\frac{d^{2}\bar{x}}{\left(2\pi\right)^{2}}\frac{d^{2}\bar{z}}{\left(2\pi\right)^{2}}e^{-i\bm{p}\cdot\left(\bm{x}-\bm{\bar{x}}\right)}e^{-i\bm{k}\cdot\left(\bm{z}-\bm{\bar{z}}\right)} K_{0}\left(\epsilon\left|\bm{x}-\bm{z}\right|\right)
    \\
    &\times K_{0}\left(\epsilon\left|\bm{\bar{x}}-\bm{\bar{z}}\right|\right)\left[1-S^{\left(2\right)}_{\bm{xz}}\left(L^{+},0\right)-S^{\left(2\right)}_{\bm{\bar{z}\bar{x}}}\left(L^{+},0\right)+S^{\left(4\right)}_{\bm{xz\bar{z}\bar{x}}}\left(L^{+},0\right)\right],
\end{split}
\label{eq.1.3}
\end{equation}
where we have defined the momentum fraction $\displaystyle \zeta\equiv p^{+}/q^{+}=1-k^{+}/q^{+}$, $\displaystyle \epsilon^{2}\equiv Q^{2}\zeta(1-\zeta)$ for massless quarks, and the two- and four-point functions as
\begin{equation}
\begin{split}
&S^{\left(2\right)}_{\bm{xz}}\left(L^{+},0\right)=S^{\left(2\right)}\left(L^{+},0;\bm{x},\bm{z}\right)=\frac{1}{N_{c}}Tr\left\langle U\left(\bm{x}\right)U^{\dagger}\left(\bm{z}\right)\right\rangle_{\left(L^{+},0\right)},\\
&S^{\left(4\right)}_{\bm{xz\bar{z}\bar{x}}}\left(L^{+},0\right)=S^{\left(4\right)}\left(L^{+},0;\bm{x},\bm{z},\bm{\bar{z}},\bm{\bar{x}}\right)=\frac{1}{N_{c}}Tr\left\langle U\left(\bm{x}\right)U^{\dagger}\left(\bm{z}\right)U\left(\bm{\bar{z}}\right)U^{\dagger}\left(\bm{\bar{x}}\right)\right\rangle_{\left(L^{+},0\right)},
\end{split}
    \label{eq.1.4}
\end{equation}
with $\displaystyle \left\langle\cdots\right\rangle$ denoting the CGC average of the colour charges over the nuclear
wave function and $\left(L^{+},0\right)$ indicating the domain of integration for the Wilson lines, given by the extent of the target.

In the same way, we can obtain the cross section for the transverse photon with polarization vector $\displaystyle\varepsilon_{T}^{\lambda}=\left(0,0,\bm{\varepsilon}^{\lambda}\right)$ with $\displaystyle \bm{\varepsilon}^{\lambda}=-\left(\lambda,i\right)/\sqrt{2}$ and $\displaystyle \lambda=\pm1$:
\begin{equation}
    \begin{split}
        \frac{d\sigma^{\gamma^{\ast}_{T}A\to q\bar{q}X}}{d^{3}p \ d^{3}k}=&\left(2\pi\right)^{2}\delta\left(p^{+}+k^{+}-q^{+}\right)2N_{c}\alpha_{em}e^{2}_{q}\zeta(1-\zeta)\left[\zeta^{2}+\left(1-\zeta\right)^{2}\right]\frac{Q^{2}}{q^{+}}\\
        &\times\int\frac{d^{2}x}{\left(2\pi\right)^{2}}\frac{d^{2}z}{\left(2\pi\right)^{2}}\frac{d^{2}\bar{x}}{\left(2\pi\right)^{2}}\frac{d^{2}\bar{z}}{\left(2\pi\right)^{2}}
         e^{-i\bm{p}{}\cdot\left(\bm{x}-\bm{\bar{x}}\right)}e^{-i\bm{k}\cdot\left(\bm{z}-\bm{\bar{z}}\right)}\\
         &\times\dfrac{\left(\bm{x}-\bm{z}\right)\cdot\left(\bm{\bar{x}}-\bm{\bar{z}}\right)}{\left|\bm{x}-\bm{z}\right|\left|\bm{\bar{x}}-\bm{\bar{z}}\right|}K_{1}\left(\epsilon\left|\bm{x}{}-\bm{z}{}\right|\right)K_{1}\left(\epsilon\left|\bm{\bar{x}}-\bm{\bar{z}}\right|\right)
        \\
        &\times \left[1-S^{\left(2\right)}_{\bm{xz}}\left(L^{+},0\right)-S^{\left(2\right)}_{\bm{\bar{z}\bar{x}}}\left(L^{+},0\right)+S^{\left(4\right)}_{\bm{xz\bar{z}\bar{x}}}\left(L^{+},0\right)\right].
    \end{split}
\end{equation}

\section{Photon-nucleus interaction beyond eikonal accuracy}
\label{sec:long}

In this Section we calculate the non-eikonal contributions, stemming from the finite longitudinal extent of the target, to dijet production for both longitudinally and transversely polarised virtual photons. 

\begin{figure}[ht]
\centering
\subfigure[Splitting before the medium]{
\begin{tikzpicture}
\begin{feynhand}

\vertex (a) at (-2,0);
\vertex (b) at (0,0);
\vertex (c) at (4.5,1.5);
\vertex (d) at (4.5,-1.5);

\propag [photon] (a) to [edge label = $q$] (b);
\propag [fer] (b) to [edge label = $p_{0}$, out=90, in=180] (c);
\propag [fer] (d) to [edge label = $k_{0}$, out=180, in=-90] (b);
\node at (4.6,1.8) {$p, r$};
\node at (4.6,-1.8) {$k, s$};
\node at (2,2.3) {$0$};
\node at (4,2.3) {$L^{+}$};
\node at (2.4, 1.2) {${\bm{x_{0}}}$};
\node at (3.7, 1.2) {${\bm{x}}$};
\node at (2.4, -1.2) {${\bm{z_{0}}}$};
\node at (3.7, -1.2) {${\bm{z}}$};
\node at (0.8, 0) {$\left(y^{+},\bm{y}\right)$};

\fill[gray, opacity=0.3] (2,-2) rectangle (4,2);

\end{feynhand}
\end{tikzpicture}
\label{Fig: Diagrama 1}}
\subfigure[Splitting inside the medium]{
\begin{tikzpicture}
\begin{feynhand}

\vertex (a) at (-2.5,0);
\vertex (b) at (0,0);
\vertex (c) at (2.5,1.5);
\vertex (d) at (2.5,-1.5);

\propag [photon] (a) to [edge label = $q$] (b);
\propag [fer] (b) to [out=90, in=180] (c);
\propag [fer] (d) to [out=180, in=-90] (b);
\node at (2.7,1.8) {$p, r$};
\node at (2.7,-1.8) {$k, s$};
\node at (-1,2.3) {$0$};
\node at (1,2.3) {$L^{+}$};
\node at (1.4, 1) {${\bm{x}}$};
\node at (1.4, -1) {${\bm{z}}$};
\node at (0.8, 0) {$\left(y^{+},\bm{y}\right)$};
\node at (0, 0.8) {$p_{0}$};
\node at (0, -0.8) {$k_{0}$};

\fill[gray, opacity=0.3] (-1,-2) rectangle (1,2);

\end{feynhand}
\end{tikzpicture}
\label{Fig: Diagrama2}}

\caption{Diagrams illustrating the next-to-eikonal contributions to the amplitude coming from the relaxation of the shockwave approximation. A third diagram identical to Fig.~\ref{Fig:Diagrama2}, corresponding to splitting after the nucleus, also contributes.}
\label{Fig:Diagramas}
\end{figure}
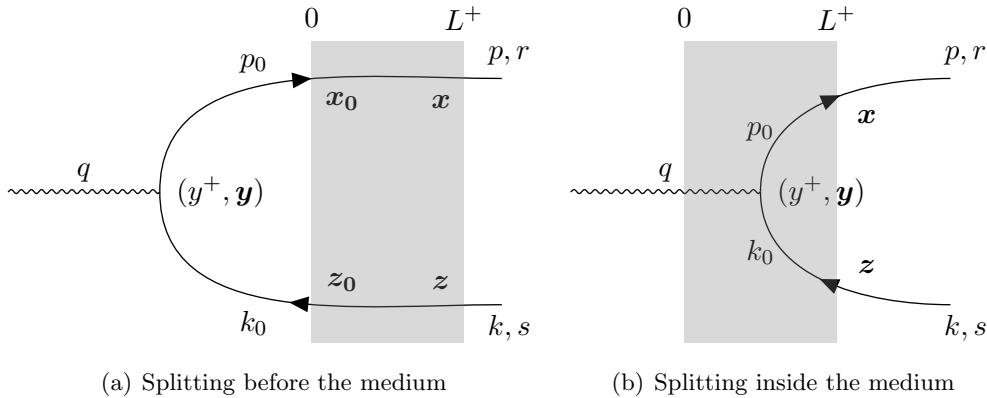

With respect to the eikonal case, there are two differences in the calculation. First, the target cannot be viewed as a localized shockwave in the longitudinal direction, therefore the contribution from emissions inside the nucleus is no longer negligible and must be taken into account. Second, the transverse diffusion of the quark and the antiquark also becomes non negligible and must be included by upgrading the Wilson lines to in-medium propagators given in terms of path integrals~\cite{Zakharov:1998sv}: for a quark propagating from $\displaystyle \left(y^{+},\bm{y}\right)$ to $\displaystyle \left(x^{+},\bm{x}\right)$ with momentum $\displaystyle p^{+}$

\begin{equation}
    G\left(x^{+},\bm{x};y^{+},\bm{y}|p^{+}\right)=\int^{\bm{r}\left(x^{+}\right)=\bm{x}}_{\bm{r}\left(y^{+}\right)=\bm{y}}D\bm{r}\left(s^{+}\right)\exp\left(\frac{ip^{+}}{2}\int^{x^{+}}_{y^{+}}ds^{+}\left[\frac{d\bm{r}}{ds^{+}}\right]^{2}\right)U\left(x^{+},y^{+};\bm{r}\right),
    \label{eq.2.1}
\end{equation}
and, for an antiquark,
\begin{equation}
    \bar{G}\left(x^{+},\bm{x};y^{+},\bm{y}|p^{+}\right)=\int^{\bm{r}\left(x^{+}\right)=\bm{x}}_{\bm{r}\left(y^{+}\right)=\bm{y}}D\bm{r}\left(s^{+}\right)\exp\left(\frac{ip^{+}}{2}\int^{x^{+}}_{y^{+}}ds^{+}\left[\frac{d\bm{r}}{ds^{+}}\right]^{2}\right)U^{\dagger}\left(x^{+},y^{+};\bm{r}\right).
    \label{eq.2.2}
\end{equation}

\subsection{Amplitude}
\label{sec:long_amp}
Let us now compute the relevant amplitude, which we separate in three different terms depending on where the splitting occurs with respect to the target.

\subsubsection{Splitting before entering medium}
The amplitude for the splitting before the medium, Fig.~\ref{Fig: Diagrama 1}, is almost identical to the eikonal case discussed before but substituting the Wilson lines for the new propagators that allow the transverse diffusion when traversing the target:
\begin{equation}
    \begin{split}
        S_{bef}=&\left(2\pi\right)\delta\left(q^{+}-p^{+}-k^{+}\right)e^{i\frac{p^{2}}{2p^{+}}L^{+}}e^{i\frac{k^{2}}{2k^{+}}L^{+}}\int^{0}_{-\infty}dy^{+}d^{2}y \ e^{-iq^{-}y^{+}}e^{i\bm{q}\cdot\bm{y}}\\
        &\times \int\frac{d^{2}p_{0}}{\left(2\pi\right)^{2}}\frac{d^{2}k_{0}}{\left(2\pi\right)^{2}}e^{i\frac{p_{0}^{2}}{2p^{+}}y^{+}}e^{-i{\bm{p_{0}}}\cdot\bm{y}}e^{i\frac{k_{0}^{2}}{2k^{+}}y^{+}}e^{-i{\bm{k_{0}}}\cdot\bm{y}}\left[\bar{u}_{r}\left(p_{0}\right)\left(-ie_{q}e\slashed{\varepsilon}\left(q\right)\right)v_{s}\left(k_{0}\right)\right]\\
        &\times \int d^{2}x_{0} \ d^{2}x \ d^{2}z_{0} \ d^{2}z\\
        &\times e^{-i{\bm{p}}\cdot{\bm{x}}}e^{i{\bm{p_{0}}}\cdot\bm{x_{0}}}e^{-i{\bm{k}}\cdot{\bm{z}}}e^{i{\bm{k_{0}}}\cdot\bm{z_{0}}} \ G\left(L^{+},{\bm{x}}; 0,\bm{x_{0}}|p^{+}\right)\bar{G}\left(L^{+}, \bm{z}; 0, \bm{z_{0}}|k^{+}\right),
    \end{split}
    \label{eq.2.3}
\end{equation}
where the polarization vector $\displaystyle\varepsilon\left(q\right)$ can be either transverse or longitudinal. Integrating over the position of the vertex (with $\displaystyle \bm{q}=\bm{0}\Rightarrow \bm{k_{0}}=-\bm{p_{0}}$) we get
\begin{equation}
    \begin{split}
        S_{bef}=&-\left(2\pi\right)\delta\left(q^{+}-p^{+}-k^{+}\right)2q^{+}e_{q}e \ \zeta\left(1-\zeta\right) \ e^{i\frac{p^{2}}{2p^{+}}L^{+}}e^{i\frac{k^{2}}{2k^{+}}L^{+}}\int\frac{d^{2}p_{0}}{\left(2\pi\right)^{2}}\dfrac{f_{\lambda;L}^{r,s}\left(\bm{p_{0}}\right)}{p^{2}_{0}+\epsilon^{2}}\\
        &\times\int d^{2}x_{0} \ d^{2}x \ d^{2}z_{0} \ d^{2}z \ e^{-i\bm{p}\cdot\bm{x}}e^{-i\bm{k}\cdot\bm{z}}e^{i\bm{p_{0}}\cdot\left(\bm{x_{0}}-\bm{z_{0}}\right)}\\
        &\times G\left(L^{+},\bm{x}; 0,\bm{x_{0}}|p^{+}\right)\bar{G}\left(L^{+}, \bm{z}; 0, \bm{z_{0}}|k^{+}\right),
    \end{split}
    \label{eq.2.5}
\end{equation}
where we have used an adiabatic turn-off to regulate the integration over the longitudinal coordinate, and $\displaystyle f_{\lambda;L}^{r,s}\left(\bm{p_{0}}\right)$ is the result of calculating the corresponding Dirac algebra. It will be given by
\begin{equation}
    f_{\lambda}^{r,s}\left(\bm{p_{0}}\right)=\left.\bar{u}_{r}\left(p_{0}\right)\slashed{\varepsilon}^{\lambda}_{T}\left(q\right)v_{s}\left(p_{1}\right)\right|_{\bm{p_{1}}=-\bm{p_{0}}}=2\dfrac{\delta_{r,-s}}{\sqrt{\zeta\left(1-\zeta\right)}}\left[\zeta\delta_{\lambda,r}-\left(1-\zeta\right)\delta_{\lambda,-r}\right]\left(\bm{\varepsilon}^{\lambda}\cdot\bm{p_{0}}\right),
\end{equation}
when the photon is transverse or
\begin{equation}
f_{L}^{r,s}\left(\bm{p_{0}}\right)=\left.\bar{u}_{r}\left(p_{0}\right)\slashed{\varepsilon}_{L}\left(q\right)v_{s}\left(p_{1}\right)\right|_{\bm{p_{1}}=-\bm{p_{0}}}=\dfrac{-\delta_{r,-s}}{Q\sqrt{\zeta\left(1-\zeta\right)}}\left[p^{2}_{0}-Q^{2}\zeta\left(1-\zeta\right)\right],
\end{equation}
if it is longitudinal.

\subsubsection{Splitting inside medium}
As previously anticipated, there is a new diagram contributing to the amplitude when the eikonal approximation is relaxed, as pictured in Fig.~\ref{Fig: Diagrama2}. This contribution to the amplitude reads
\begin{equation}
    \begin{split}
        S_{in}=&\left(2\pi\right)\delta\left(q^{+}-p^{+}-k^{+}\right)e^{i\frac{p^{2}}{2p^{+}}L^{+}}e^{i\frac{k^{2}}{2k^{+}}L^{+}}\int^{L^{+}}_{0}dy^{+}d^{2}y\,e^{-iq^{-}y^{+}}e^{i\bm{q}\cdot\bm{y}}\\
        &\times \int\frac{d^{2}p_{0}}{\left(2\pi\right)^{2}}\frac{d^{2}k_{0}}{\left(2\pi\right)^{2}} e^{-i{\bm{p_{0}}}\cdot\bm{y}}e^{-i{\bm{k_{0}}}\cdot\bm{y}}\left[\bar{u}_{r}\left(p_{0}\right)\left(-ie_{q}e\slashed{\varepsilon}\left(q\right)\right)v_{s}\left(k_{0}\right)\right]\\
        &\times \int d^{2}x_{0} \ d^{2}x \ d^{2}z_{0} \ d^{2}z e^{-i{\bm{p}}\cdot{\bm{x}}}e^{i{\bm{p_{0}}}\cdot{\bm{x_{0}}}}e^{-i{\bm{k}}\cdot{\bm{z}}}e^{i{\bm{k_{0}}}\cdot{\bm{z_{0}}}} \\
        &\times G\left(L^{+},{\bm{x}}; y^{+},{\bm{x_{0}}}|p^{+}\right)\bar{G}\left(L^{+}, {\bm{z}}; y^{+}, {\bm{z_{0}}}|k^{+}\right).
    \end{split}
    \label{eq.2.6}
\end{equation}
At this point of the calculation we cannot yet perform the integration over the longitudinal coordinate of the vertex as done in the other parts of the amplitude. Integrating over the transverse position of the vertex we get
\begin{equation}
    \begin{split}
        S_{in}=&\left(2\pi\right)\delta\left(q^{+}-p^{+}-k^{+}\right)\left(-ie_{q}e\right)e^{i\frac{p^{2}}{2p^{+}}L^{+}}e^{i\frac{k^{2}}{2k^{+}}L^{+}}\int^{L^{+}}_{0}dy^{+}e^{-iq^{-}y^{+}}\\
        &\times\int\frac{d^{2}p_{0}}{\left(2\pi\right)^{2}}f_{\lambda;L}^{r,s}\left(\bm{p_{0}}\right)\int d^{2}x_{0} \ d^{2}x \ d^{2}z_{0} \ d^{2}z \ e^{-i{\bm{p}}\cdot{\bm{x}}}e^{-i{\bm{k}}\cdot{\bm{z}}}e^{i{\bm{p_{0}}}\cdot\left({\bm{x_{0}}}-{\bm{z_{0}}}\right)}\\
        &\times G\left(L^{+},{\bm{x}}; y^{+},{\bm{x_{0}}}|p^{+}\right)\bar{G}\left(L^{+}, {\bm{z}}; y^{+}, {\bm{z_{0}}}|k^{+}\right).
    \end{split}
    \label{eq.2.7}
\end{equation}

\subsubsection{Splitting after traversing medium}
Finally, if the virtual photon splits after having passed through the target the contribution is simply the same as the vacuum splitting, Fig.~\ref{Fig:Diagrama2}:
\begin{equation}
    \begin{split}
    S_{aft}=&\left(2\pi\right)^{3}\delta\left(q^{+}-p^{+}-k^{+}\right)\delta^{\left(2\right)}\left(\bm{q}-\bm{p}{}-\bm{k}{}\right)\left[\bar{u}_{r}\left(p\right)\left(-ie_{q}e\slashed{\varepsilon}\left(q\right)\right)v_{s}\left(k\right)\right]\\
    &\times \int^{\infty}_{L^{+}}dy^{+}e^{-iq^{-}y^{+}}e^{i\frac{p^{2}}{2p^{+}}y^{+}}e^{i\frac{k^{2}y^{+}}{2k^{+}}}\\
    =&-\left(2\pi\right)^{3}\delta\left(q^{+}-p^{+}-k^{+}\right)\delta^{\left(2\right)}\left(\bm{p}+\bm{k}\right)2q^{+}e_{q}e \ \zeta\left(1-\zeta\right) \\
    &\times e^{i\frac{Q^{2}}{2q^{+}}L^{+}}e^{i\frac{p^{2}}{2p^{+}}L^{+}}e^{i\frac{k^{2}}{2k^{+}}L^{+}}\ \dfrac{f_{\lambda;L}^{r,s}\left(\bm{p}\right)}{p^{2}+\epsilon^{2}}\ .
    \end{split}
    \label{eq.2.8}
\end{equation}

\subsection{Cross section}
\label{sec:long_squ}

The total amplitude of the process will be the sum of the three contributions $\displaystyle S_{Neik}=S_{bef}+S_{in}+S_{aft}$. With the definition $\displaystyle S=\left(2q^{+}\right)2\pi\delta\left(q^{+}-p^{+}-k^{+}\right)i\mathcal{M}$, the differential cross section reads
\begin{equation}
    \frac{d\sigma^{\gamma^{\ast}_{L;T}A\to q\bar{q}X}}{d\mathcal{PS}}=\left(2q^{+}\right)2\pi\delta\left(q^{+}-p^{+}-k^{+}\right)\left\langle\overline{\sum}\left|\mathcal{M}\right|^{2}\right\rangle,
    \label{eq.2.9}
\end{equation}
where $\displaystyle \mathcal{PS}$ is the phase space of the $\displaystyle q\bar{q}$ pair, $\displaystyle \overline{\sum}$ is the sum (average) of the final (initial) helicities and colours, and the average $\left\langle\dots\right\rangle$ is taken over background field configurations.

The medium averages will be taken in the Gaussian approximation~\cite{Baier:1996kr,Baier:1996sk,Casalderrey-Solana:2007knd}, where it is assumed that the correlator of the target fields takes the form
\begin{equation}
    \left\langle A^{-a}\left(x^{+},\bm{x}\right)A^{-b}\left(y^{+},\bm{y}\right)\right\rangle=g^{2}n\left(x^{+}\right)\delta^{ab}\delta\left(x^{+}-y^{+}\right)\gamma\left(\bm{x}-\bm{y}\right),
    \label{eq.2.13}
\end{equation}
where $\displaystyle n\left(x^{+}\right)$ is the longitudinal density of scattering centers in the target and
\begin{equation}
    g^{2}\gamma\left(\bm{r}\right)=\int\frac{d^{2}q}{\left(2\pi\right)^{2}}e^{i\bm{q}\cdot\bm{r}}\frac{d^{2}\sigma_{el}}{d^{2}\bm{q}}\sim g^{2}\int\frac{d^{2}q}{\left(2\pi\right)^{2}}\frac{e^{i\bm{q}\cdot\bm{r}}}{q^{4}}
    \label{eq.2.14}
\end{equation}
is the Fourier transform of the in-medium elastic scattering potential. When calculating averages of products of Wilson lines, such correlators will always appear in the combination
\begin{equation}
\sigma(\bm{r})=g^4[\gamma(\bm{0})-\gamma(\bm{r})]\,,
\end{equation}
often referred to as the dipole cross section. We will work in the harmonic oscillator approximation~\cite{Baier:1996kr,Zakharov:1998sv} where $\displaystyle C_{F}n\left(t\right)\sigma\left(\bm{r}\left(t\right)\right)\simeq\frac{Q_{s}^{2}}{4L^{+}}\bm{r}^{2}\left(t\right)$.

Similarly to the fully eikonal case, we will need to calculate the medium average of traces of two and four Wilson lines, with the difference that in this case the transverse coordinate is allowed to vary throughout the longitudinal extent of the Wilson line. In the Gaussian approximation the two-point function can be expressed in terms of the dipole cross section in the following way
\begin{equation}
    S^{\left(2\right)}_{\bm{xy}}\left(t,t_{0}\right)=\frac{1}{N_{c}}Tr\left\langle U\left(\bm{x}\right)U^{\dagger}\left(\bm{y}\right)\right\rangle_{\left(t,t_{0}\right)}=\exp\left[-C_{F}\int^{t}_{t_{0}}d\xi \ n\left(\xi\right)\sigma\left(\bm{x}\left(\xi\right)-\bm{y}\left(\xi\right)\right)\right].
    \label{eq.2.15}
\end{equation}

Calculating the four-point function is not as straightforward due to the color algebra being more complicated. For a system of two quarks and two antiquarks there are two global singlet states, as opposed to the lone singlet state for the dipole case. The multiple field insertions entering the calculation of the four-point function induce color transitions which prevent us from resumming the multiple interactions in terms of simple exponentials. Nevertheless, these transitions between singlet states are suppressed in the large-$N_c$ limit, allowing us to obtain a closed result in that case~\cite{Blaizot:2012fh,Apolinario:2014csa}. The four-point function then reads
\begin{equation}
    \begin{split}
    S^{\left(4\right)}_{\bm{12\bar{2}\bar{1}}}\left(t,t_{0}\right)&=\frac{1}{N_{c}}Tr\left\langle U\left(\bm{r_{1}}{}\right)U^{\dagger}\left(\bm{r_{2}}{}\right)U\left(\bm{\bar{r}_{2}}{}\right)U^{\dagger}\left(\bm{\bar{r}_{1}}{}\right)\right\rangle_{\left(t,t_{0}\right)}\\
    &=S^{\left(2\right)}_{\bm{1\bar{1}}}\left(t,t_{0}\right)S^{\left(2\right)}_{\bm{2\bar{2}}}\left(t,t_{0}\right)
    \\
    &\quad+\int^{t}_{t_{0}}d\xi \ S^{\left(2\right)}_{\bm{1\bar{1}}}\left(t,\xi\right)S^{\left(2\right)}_{\bm{2\bar{2}}}\left(t,\xi\right)T\left(\xi\right)S^{\left(2\right)}_{\bm{12}}\left(\xi,t_{0}\right)S^{\left(2\right)}_{\bm{\bar{1}\bar{2}}}\left(\xi,t_{0}\right)+\mathcal{O}\left(1/N_{c}\right),
    \end{split}
    \label{eq.2.16}
\end{equation}
where $\displaystyle T\left(\xi\right)=\frac{Q_{s}^{2}}{2L^{+}}\left(\bm{r_{1}}{}-\bm{\bar{r}_{1}}{}\right)\cdot\left(\bm{r_{2}}{}-\bm{\bar{r}_{2}}{}\right)$ is called the transition amplitude and all two-point functions are evaluated within the mentioned approximations:
\begin{equation}
    S^{\left(2\right)}_{\bm{ij}}\left(t,t_{0}\right)=\exp\left[-\frac{Q^{2}_{s}}{4L^{+}}\int^{t}_{t_{0}}d\xi \ \left|\bm{r_{i}}{}\left(\xi\right)-\bm{r_{j}}{}\left(\xi\right)\right|^{2}\right].
    \label{eq.2.17}
\end{equation}
The two terms in~\eqref{eq.2.16} are often referred as the factorizable and non-factorizable terms respectively. In the large-$N_c$ limit the two available singlet states can be pictured as the two possible ways of pairing two quarks and two antiquarks into two dipoles. The factorizable term corresponds to the case where the system remains in the same pairing through the full extent of the longitudinal reegion, while the non-factorizable term corresponds to the case where a transition between the two color states occurs at a given longitudinal position $\xi$. Any further transitions between color states are suppressed by powers of $N_c^2$.

Since the Wilson lines appearing in these averages come from the in-medium propagators appearing in the amplitudes of the previous section and their corresponding complex conjugates, we have to compute path integral over the trajectories of each of the involved partons. Taking into account that all averages have been expressed in terms of two-point functions, it is clear that the only path integrals we will have to calculate explicitly are the following (see~\cite{Blaizot:2012fh,Apolinario:2014csa} for closely related calculations):
\begin{equation}
    \begin{split}
 &\mathcal{F}_{k^{+}}\left(\bm{x_{f}},\bm{\bar{x}_{f}},\bm{x_{i}},\bm{\bar{x}_{i}};t^{+}_{f},t^{+}_{i}\right)\\
 &\hskip 1cm =\int^{\bm{x_{f}}}_{\bm{x_{i}}}D\bm{x}\int^{\bm{\bar{x}_{f}}}_{\bm{\bar{x}_{i}}}D\bm{\bar{x}} \ \exp\left\lbrace\int^{t^{+}_{f}}_{t^{+}_{i}}dt\left[\dfrac{ik^{+}}{2}\left(\dot{x}^{2}-\dot{\bar{x}}^{2}\right)-C_{F}n\sigma\left(\bm{x}-\bm{\bar{x}}\right)\right]\right\rbrace
    \end{split}
    \label{eq:functionf}
\end{equation}
which, with the change of variables given by
\begin{align}
\label{eq:changevar}
    \bm{u}_{x}&=\bm{x}-\bar{\bm{x}}, & \bm{v}_{x}&=\dfrac{1}{2}\left(\bm{x}+\bar{\bm{x}}\right),\nonumber\\
    \bm{u}_{xi}&=\bm{x}_{i}-\bar{\bm{x}}_{i}, & \bm{v}_{xi}&=\dfrac{1}{2}\left(\bm{x}_{i}+\bar{\bm{x}}_{i}\right),\\
    \bm{u}_{xf}&=\bm{x}_{f}-\bar{\bm{x}}_{f}, & \bm{v}_{xf}&=\dfrac{1}{2}\left(\bm{x}_{f}+\bar{\bm{x}}_{f}\right),\nonumber
\end{align}
and using the harmonic oscillator approximation, results in
\begin{align}
    \mathcal{F}_{k^{+}}\left(\bm{u_{f}},\bm{v_{f}},\bm{u_{i}},\bm{v_{i}};\Delta t^{+}\right)=&\left(\dfrac{k^{+}}{2\pi i \Delta t^{+}}\right)^{2}\exp\left\lbrace\dfrac{ik^{+}}{\Delta t^{+}}\Delta\bm{u}\cdot\Delta\bm{v}\right\rbrace\notag\\
    &\times\exp\left\lbrace-\dfrac{Q^{2}_{s}}{12L^{+}}\Delta t^{+}\left(u_{f}^{2}+u_{i}^{2}+\bm{u_{f}}\cdot\bm{u_{i}}\right)\right\rbrace,
    \label{eq:functionfho}
\end{align}
where $\displaystyle \Delta t^{+}=t^{+}_{f}-t^{+}_{i}$, $\displaystyle \Delta\bm{u}=\bm{u_{f}}-\bm{u_{i}}$, $\displaystyle \Delta\bm{v}=\bm{v_{f}}-\bm{v_{i}}$; and
\begin{equation}
    \begin{split}
&\mathcal{K}\left(\bm{x_{f}},\bm{y_{f}},\bm{x_{i}},\bm{y_{i}};t^{+}_{f},t^{+}_{i}\right)\\
&\hskip 1cm =\int^{\bm{x_{f}}}_{\bm{x_{i}}}D\bm{x}\int^{\bm{y_{f}}}_{\bm{y_{i}}}D\bm{y} \ \exp\left\lbrace\int^{t_{f}}_{t_{i}}dt\left[\dfrac{ik^{+}_{0}}{2}\dot{x}^{2}+\dfrac{ik^{+}_{1}}{2}\dot{y}^{2}-C_{F}n\sigma\left(\bm{x}-\bm{y}\right)\right]\right\rbrace
    \end{split}
    \label{eq:functionk}
\end{equation}
which, with the change of variables given by
\begin{align}
    \bm{u}&=\bm{x}-\bm{y}, & \bm{m}&=\zeta\bm{x}+\left(1-\zeta\right)\bm{y},\nonumber\\
    \bm{u_{i}}&=\bm{x_{i}}-\bm{y_{i}}, & \bm{m_{i}}&=\zeta\bm{x_{i}}+\left(1-\zeta\right)\bm{y_{i}},\label{eq:changevarm}\\
    \bm{u_{f}}&=\bm{x_{f}}-\bm{y_{f}}, & \bm{m_{f}}&=\zeta\bm{x_{f}}+\left(1-\zeta\right)\bm{y_{f}},\nonumber
\end{align}
results in
\begin{align}
    \mathcal{K}\left(\Delta \bm{m},\bm{u_{f}},\bm{u_{i}};\Delta t^{+}\right)=G_{0}\left(\Delta t^{+},\Delta \bm{m}|q^{+}\right)\mathcal{J}\left(\bm{u}_{f},\bm{u}_{i};\Delta t^{+}\right),
    \label{eq:functionkappa}
\end{align}
with $\displaystyle \Delta\bm{m}=\bm{m_{f}}-\bm{m_{i}}$,
\begin{equation}
    G_{0}\left(t^{+}_{f}-t^{+}_{i},\Delta\bm{m}|q^{+}\right)=\dfrac{q^{+}}{2\pi i\Delta t^{+}}\exp\left\lbrace\dfrac{iq^{+}}{2\Delta t^{+}}\Delta\bm{m}^{2}\right\rbrace
    \label{eq:functionfree}
\end{equation}
the free propagator and, using the harmonic oscillator approximation,
\begin{align}
\label{eq:functionkho}
    &\mathcal{J}\left(\bm{u}_{f},\bm{u}_{i};\Delta t^{+}\right)=\int^{\bm{u}_{f}}_{\bm{u}_{i}}D\bm{u} \ \exp\left\lbrace\int^{t^{+}_{f}}_{t^{+}_{i}}dt\left[\dfrac{iq^{+}}{2}\zeta\left(1-\zeta\right)\dot{u}^{2}-C_{F}n\sigma\left(\bm{u}\right)\right]\right\rbrace\\
    &\hskip 1cm =\dfrac{1}{2\pi i}\dfrac{q^{+}\zeta\left(1-\zeta\right)\Omega}{\sin\left(\Omega \Delta t^{+}\right)}\exp\left\lbrace i\dfrac{q^{+}\zeta\left(1-\zeta\right)\Omega}{2\sin\left(\Omega \Delta t^{+}\right)}\left[\cos\left(\Omega \Delta t^{+}\right)\left(u_{f}^{2}+u_{i}^{2}\right)-2\bm{u}_{f}\cdot\bm{u}_{i}\right]\right\rbrace,\nonumber 
\end{align}
where we have defined the complex frequency $\displaystyle \Omega^{2}=\dfrac{-iQ^{2}_{s}}{2q^{+}\zeta\left(1-\zeta\right)L^{+}}$.

With these results we can proceed to calculate each term of $\left\langle\overline{\sum} \displaystyle \left|\mathcal{M}\right|^{2}\right\rangle$ separately. In the main text we present full technical details just for the first item, referring to the appendices for additional ones.

\subsubsection{Before-Before}
The first term reads
\begin{equation}
\begin{split}
&\left\langle\overline{\sum}\left|\mathcal{M}_{bef}\right|^{2}\right\rangle=\zeta^{2}\left(1-\zeta\right)^{2}e_{q}^{2}e^{2}\int\dfrac{d^{2}p_{0}}{\left(2\pi\right)^{2}}\dfrac{d^{2}\bar{p}_{0}}{\left(2\pi\right)^{2}}\dfrac{F_{T;L}\left(\bm{p_{0}},\bm{\bar{p}_{0}}\right)}{\left(p^{2}_{0}+\epsilon^{2}\right)\left(\bar{p}^{2}_{0}+\epsilon^{2}\right)}\\
    \times &\int d^{2}x_{0} \ d^{2}x \ d^{2}z_{0} \ d^{2}z
    \int d^{2}\bar{x}_{0} \ d^{2}\bar{x} \ d^{2}\bar{z}_{0} \ d^{2}\bar{z} \ e^{-i{\bm{p}}\cdot\left({\bm{x}}-\bm{\bar{x}}{}\right)}\\
    \times & \ e^{-i{\bm{k}}\cdot\left({\bm{z}}-\bm{\bar{z}}{}\right)}e^{i{\bm{p_{0}}}\cdot\left({\bm{x_{0}}}-{\bm{z_{0}}}\right)}e^{-i\bm{\bar{p}_{0}}{}\cdot\left(\bm{\bar{x}_{0}}{}-\bm{\bar{z}_{0}}{}\right)}\\
    \times &\ Tr\biggl\langle G\left(L^{+},{\bm{x}}; 0,{\bm{x_{0}}}|p^{+}\right)\bar{G}\left(L^{+}, {\bm{z}}; 0, {\bm{z_{0}}}|k^{+}\right)\\
    &\hskip 3cm \times\bar{G}^{\dagger}\left(L^{+}, \bm{\bar{z}}{}; 0, \bm{\bar{z}_{0}}{}|k^{+}\right)G^{\dagger}\left(L^{+},\bm{\bar{x}}{}; 0,\bm{\bar{x}_{0}}{}|p^{+}\right)\biggr\rangle,
    \end{split}
    \label{eq.2.10}
\end{equation}
where 
\begin{equation}
F_{T;L}(\bm{p_{0}},\bm{\bar{p}_{0}}) = \overline{\sum} f_{\lambda;L}^{r,s}(\bm{p_{0}})f_{\lambda;L}^{r,s\ast}(\bm{\bar{p}_{0}})\,,
\end{equation}
with the sum (average) runs over final (initial) helicities only. Explicit expressions for the transverse and longitudinal cases are, respectively,
\begin{equation}
    F_{T}\left(\bm{p_{0},\bm{\bar{p}_{0}}}\right)=\dfrac{2}{\zeta\left(1-\zeta\right)}\left[\zeta^{2}+\left(1-\zeta\right)^{2}\right]\left(\bm{p_{0}}\cdot\bm{\bar{p}_{0}}\right),
\end{equation}
\begin{equation}
    F_{L}\left(\bm{p_{0}},\bm{\bar{p}_{0}}\right)=\dfrac{2}{Q^{2}\zeta\left(1-\zeta\right)}\left[p_{0}^{2}-Q^{2}\zeta\left(1-\zeta\right)\right]\left[\bar{p}_{0}^{2}-Q^{2}\zeta\left(1-\zeta\right)\right].
    \label{eq:FL}
\end{equation}

Now, defining operator\footnote{In order to simplify the notation, in the following we use $\displaystyle X\equiv\left(x^{+},\bm{x}\right)$.} $\displaystyle \mathcal{B}_{p^{+}}\left(X_{1},X_{0}|\bm{r}\right)$ by its action on a function:
\begin{equation}
\mathcal{B}_{p^{+}}\left(X_{1},X_{0}|\bm{r}\right)f\left(\bm{r}\right)\equiv \int^{\bm{x_{1}}{}}_{\bm{x_{0}}{}}D\bm{r}\left(s^{+}\right)\exp\left(\frac{ip^{+}}{2}\int^{x^{+}_{1}}_{x^{+}_{0}}ds^{+}\left[\frac{d\bm{r}}{ds^{+}}\right]^{2}\right)f\left(\bm{r}\right),
   \label{eq.2.11}
\end{equation}
we can write~\eqref{eq.2.10} as
\begin{equation}
    \begin{split}
&\left\langle\overline{\sum}\left|\mathcal{M}_{bef}\right|^{2}\right\rangle=\zeta^{2}\left(1-\zeta\right)^{2}N_{c}e_{q}^{2}e^{2}\int\dfrac{d^{2}p_{0}}{\left(2\pi\right)^{2}}\dfrac{d^{2}\bar{p}_{0}}{\left(2\pi\right)^{2}}\dfrac{F_{T;L}\left(\bm{p_{0}},\bm{\bar{p}_{0}}\right)}{\left(p^{2}_{0}+\epsilon^{2}\right)\left(\bar{p}^{2}_{0}+\epsilon^{2}\right)}\\
    &\times\int d^{2}x_{0} \ d^{2}x \ d^{2}z_{0} \ d^{2}z\int d^{2}\bar{x}_{0} \ d^{2}\bar{x} \ d^{2}\bar{z}_{0} \ d^{2}\bar{z}\\
    &\times e^{-i{\bm{p}}\cdot\left({\bm{x}}-\bm{\bar{x}}{}\right)}e^{-i{\bm{k}}\cdot\left({\bm{z}}-\bm{\bar{z}}{}\right)}e^{i{\bm{p_{0}}}\cdot\left({\bm{x_{0}}}-{\bm{z_{0}}}\right)}e^{-i\bm{\bar{p}_{0}}{}\cdot\left(\bm{\bar{x}_{0}}{}-\bm{\bar{z}_{0}}{}\right)}\\
    &\times\biggl[\mathcal{B}_{p^{+}}\left(X,X_{0}|\bm{r}\right)\mathcal{B}_{k^{+}}\left(Z,Z_{0}|\bm{t}\right)\mathcal{B}_{-k^{+}}\left(\bar{Z},\bar{Z}_{0}|\bm{\bar{t}}\right)\mathcal{B}_{-p^{+}}\left(\bar{X},\bar{X}_{0}|\bm{\bar{r}}\right)S^{\left(4\right)}_{\bm{rt\bar{t}\bar{r}}}\left(L^{+},0\right)\biggr].
    \end{split}
    \label{eq.2.12}
\end{equation}

Using the definitions in~\eqref{eq:functionf},~\eqref{eq:functionfho},~\eqref{eq:functionk},~\eqref{eq:functionfree},~\eqref{eq:functionkho} we can express the path integrals acting on the four-point function as
\begin{equation}
    \begin{split}
&\mathcal{B}_{p^{+}}\left(X,X_{0}|\bm{r}\right)\mathcal{B}_{k^{+}}\left(Z,Z_{0}|\bm{t}\right)\mathcal{B}_{-k^{+}}\left(\bar{Z},\bar{Z}_{0}|\bm{\bar{t}}\right)\mathcal{B}_{-p^{+}}\left(\bar{X},\bar{X}_{0}|\bm{\bar{r}}\right)S^{\left(4\right)}_{\bm{rt\bar{t}\bar{r}}}\left(L^{+},0\right)\\
    =&\int^{\bm{x}{}}_{\bm{x_{0}}{}}D\bm{r}\int^{\bm{z}{}}_{\bm{z_{0}}{}}D\bm{t}\int^{\bm{\bar{z}}{}}_{\bm{\bar{z}_{0}}{}}D\bm{\bar{t}}\int^{\bm{\bar{x}}{}}_{\bm{\bar{x}_{0}}{}}D\bm{\bar{r}} \ \exp\left\lbrace\int^{L^{+}}_{0}d\xi\left[\frac{ip^{+}}{2}\left(\dot{\bm{r}}^{2}-\dot{\bm{\bar{r}}}^{2}\right)+\frac{ik^{+}}{2}\left(\dot{\bm{t}}^{2}-\dot{\bm{\bar{t}}}^{2}\right)\right]\right\rbrace \\ & \hskip 6cm \times S^{\left(4\right)}_{\bm{rt\bar{t}\bar{r}}}\left(L^{+},0\right)\\
=&\mathcal{F}_{p^{+}}\left(\bm{u}_{x},\bm{v}_{x},\bm{u}_{x0},\bm{v}_{x0};L^{+}\right)\mathcal{F}_{k^{+}}\left(\bm{u}_{z},\bm{v}_{z},\bm{u}_{z0},\bm{v}_{z0};L^{+}\right)\\
    &+\frac{Q_{s}^{2}}{2L^{+}}\int^{L^{+}}_{0}ds^{+}\int d^{2}u_{xs}d^{2}v_{xs}d^{2}u_{zs}d^{2}v_{zs} \ \left(\bm{u}_{xs}\cdot\bm{u}_{zs}\right)\\
    &\hskip 0.3cm \times\mathcal{F}_{p^{+}}\left(\bm{u}_{x},\bm{v}_{x},\bm{u}_{xs},\bm{v}_{xs};L^{+}-s^{+}\right)\mathcal{F}_{k^{+}}\left(\bm{u}_{z},\bm{v}_{z},\bm{u}_{zs},\bm{v}_{zs};L^{+}-s^{+}\right)\\
    &\hskip 0.3cm \times\mathcal{K}\left(\bm{m}_{s}-\bm{m}_{0},\bm{\Delta}_{s},\bm{\Delta}_{0};s^{+}\right)\mathcal{K}^{\ast}\left(\bar{\bm{m}}_{s}-\bar{\bm{m}}_{0},\bar{\bm{\Delta}}_{s},\bar{\bm{\Delta}}_{0};s^{+}\right),
    \end{split}
    \label{eq.2.18}
\end{equation}
and the path integral for the two-point function as
\begin{equation}
    \begin{split}
&\mathcal{B}_{p^{+}}\left(X,X_{0}|\bm{r}\right)\mathcal{B}_{k^{+}}\left(Z,Z_{0}|\bm{t}\right)S^{\left(2\right)}_{\bm{rt}}\left(L^{+},0\right)\\
    &=\int^{\bm{x}{}}_{\bm{x_{0}}{}}D\bm{r}\int^{\bm{z}{}}_{\bm{z_{0}}{}}D\bm{t} \ \exp\left\lbrace\int^{L^{+}}_{0}d\xi\left[\frac{ip^{+}}{2}\dot{\bm{r}}^{2}+\frac{ik^{+}}{2}\dot{\bm{t}}^{2}-\frac{Q_{s}^{2}}{4L^{+}}\left(\bm{r}-\bm{t}\right)^{2}\right]\right\rbrace\\
    &=\mathcal{K}\left(\bm{m}-\bm{m}_{0},\bm{\Delta},\bm{\Delta}_{0};L^{+}\right),
    \end{split}
    \label{eq.2.19}
\end{equation}
where we have
made a change of variables analogous to that in~\eqref{eq:changevar}. The intermediate variables $\displaystyle \bm{\Delta}_{s}$, $\displaystyle \bm{m}_{s}$, $\displaystyle \bar{\bm{\Delta}}_{s}$ and $\displaystyle \bar{\bm{m}}_{s}$ in~\eqref{eq.2.18} read
\begin{align}
\bm{\Delta}_{s}=&\bm{v}_{xs}-\bm{v}_{zs}+\dfrac{1}{2}\left(\bm{u}_{xs}-\bm{u}_{zs}\right), & \bm{m}_{s}=&\zeta\left(\bm{v}_{xs}+\dfrac{\bm{u}_{xs}}{2}\right)+\left(1-\zeta\right)\left(\bm{v}_{zs}+\dfrac{\bm{u}_{zs}}{2}\right),\nonumber\\
\bm{\bar{\Delta}}_{s}=&\bm{v}_{xs}-\bm{v}_{zs}-\dfrac{1}{2}\left(\bm{u}_{xs}-\bm{u}_{zs}\right), & \bm{\bar{m}}_{s}=&\zeta\left(\bm{v}_{xs}-\dfrac{\bm{u}_{xs}}{2}\right)+\left(1-\zeta\right)\left(\bm{v}_{zs}-\dfrac{\bm{u}_{zs}}{2}\right).
\label{eq.2.22}
\end{align}
 
Substituting the results of the path integrals~\eqref{eq.2.18} in~\eqref{eq.2.12} we get
\begin{equation}
    \begin{split}
    &\left\langle\overline{\sum}\left|\mathcal{M}_{bef}\right|^{2}\right\rangle=\zeta^{2}\left(1-\zeta\right)^{2}N_{c}e_{q}^{2}e^{2}\int\dfrac{d^{2}p_{0}}{\left(2\pi\right)^{2}}\dfrac{d^{2}\bar{p}_{0}}{\left(2\pi\right)^{2}}\dfrac{F_{T;L}\left(\bm{p_{0}},\bm{\bar{p}_{0}}\right)}{\left(p^{2}_{0}+\epsilon^{2}\right)\left(\bar{p}^{2}_{0}+\epsilon^{2}\right)}\\
    &\times\int d^{2}u_{x0}d^{2}v_{x0}d^{2}u_{z0}d^{2}v_{z0}\int d^{2}u_{x}d^{2}v_{x}d^{2}u_{z}d^{2}v_{z} \ e^{-i{\bm{p}}\cdot\bm{u}_{x}} \ e^{-i{\bm{k}}\cdot\bm{u}_{z}}\\
    &\times e^{i\bm{p_{0}}\cdot\bm{\Delta_{0}}}e^{-i\bm{\bar{p}_{0}}\cdot\bm{\bar{\Delta}_{0}}}\biggl[\mathcal{F}_{p^{+}}\left(\bm{u}_{x},\bm{v}_{x},\bm{u}_{x0},\bm{v}_{x0};L^{+}\right)\mathcal{F}_{k^{+}}\left(\bm{u}_{z},\bm{v}_{z},\bm{u}_{z0},\bm{v}_{z0};L^{+}\right)\biggr.\\
    +&\frac{Q^{2}_{s}}{2L^{+}}\int^{L^{+}}_{0}ds^{+}\int d^{2}u_{xs}d^{2}v_{xs}d^{2}u_{zs}d^{2}v_{zs} \ \left(\bm{u}_{xs}\cdot\bm{u}_{zs}\right)\\
    &\times\mathcal{F}_{p^{+}}\left(\bm{u}_{x},\bm{v}_{x},\bm{u}_{xs},\bm{v}_{xs};L^{+}-s^{+}\right)\mathcal{F}_{k^{+}}\left(\bm{u}_{z},\bm{v}_{z},\bm{u}_{zs},\bm{v}_{zs};L^{+}-s^{+}\right)\\
    &\times \biggl.G_{0}\left(s^{+},\bm{m}_{s}-\bm{m}_{0}|q^{+}\right)\mathcal{J}\left(\bm{\Delta}_{s},\bm{\Delta}_{0};s^{+}\right)\\
    &\times G_{0}^{\ast}\left(s^{+},\bar{\bm{m}}_{s}-\bar{\bm{m}}_{0}|q^{+}\right)\mathcal{J}^{\ast}\left(\bar{\bm{\Delta}}_{s},\bar{\bm{\Delta}}_{0};s^{+}\right)\biggr].
    \end{split}
    \label{eq.2.23}
\end{equation}
We can now use the results derived in appendix~\ref{sec:appC} to simplify the expression by integrating the $\displaystyle \mathcal{F}$'s, to give
\begin{equation}
    \begin{split}
        &\left\langle\overline{\sum}\left|\mathcal{M}_{bef}\right|^{2}\right\rangle=\zeta^{2}\left(1-\zeta\right)^{2}N_{c}e_{q}^{2}e^{2}\int\dfrac{d^{2}p_{0}}{\left(2\pi\right)^{2}}\dfrac{d^{2}\bar{p}_{0}}{\left(2\pi\right)^{2}}\dfrac{F_{T;L}\left(\bm{p_{0}},\bm{\bar{p}_{0}}\right)}{\left(p^{2}_{0}+\epsilon^{2}\right)\left(\bar{p}^{2}_{0}+\epsilon^{2}\right)}\\
        &\times\int d^{2}u_{x}d^{2}u_{z} \ e^{-i\bm{p}\cdot\bm{u}_{x}}e^{-i\bm{k}\cdot\bm{u}_{z}}\\
        &\times \Biggl[\mathcal{P}\left(\bm{u}_{x};L^{+}\right)\mathcal{P}\left(\bm{u}_{z};L^{+}\right)\int d^{2}v_{x0}d^{2}v_{z0} \ \biggl.e^{i\bm{p}_{0}\cdot\bm{\Delta}_{0}}e^{-i\bm{\bar{p}}_{0}\cdot\bm{\bar{\Delta}}_{0}}\biggr|_{\left(\bm{u}_{x0},\bm{u}_{z0}\right)=\left(\bm{u}_{x},\bm{u}_{z}\right)}\Biggr.\\
        +&\frac{Q^{2}_{s}}{2L^{+}}\left(\bm{u}_{x}\cdot\bm{u}_{z}\right)\int^{L^{+}}_{0}ds^{+} \ \mathcal{P}\left(\bm{u}_{x};L^{+}-s^{+}\right)\mathcal{P}\left(\bm{u}_{z};L^{+}-s^{+}\right)\\
        &\times\int d^{2}v_{xs}d^{2}v_{zs}\int d^{2}u_{x0}d^{2}v_{x0}d^{2}u_{z0}d^{2}v_{z0} \ e^{i\bm{p}_{0}\cdot\bm{\Delta}_{0}}e^{-i\bm{\bar{p}}_{0}\cdot\bm{\bar{\Delta}}_{0}}\\
        &\times\Biggl.\biggl.G_{0}\left(s^{+},\bm{m}_{s}-\bm{m}_{0}|q^{+}\right)\mathcal{J}\left(\bm{\Delta}_{s},\bm{\Delta}_{0};s^{+}\right)G_{0}^{\ast}\left(s^{+},\bar{\bm{m}}_{s}-\bar{\bm{m}}_{0}|q^{+}\right)\\
        &\times \mathcal{J}^{\ast}\left(\bar{\bm{\Delta}}_{s},\bar{\bm{\Delta}}_{0};s^{+}\right)\biggr|_{\left(\bm{u}_{xs},\bm{u}_{zs}\right)=\left(\bm{u}_{x},\bm{u}_{z}\right)}\Biggr],
    \end{split}
    \label{eq.2.24}
\end{equation}
with the definition of the broadening function
\begin{equation}
    \mathcal{P}\left(\bm{u};\Delta t^{+}\right)=\exp\left(-\dfrac{Q^{2}_{s}}{4L^{+}}\Delta t^{+}u^{2}\right).
    \label{eq:Pdef}
\end{equation}

Now, integrating $\displaystyle \bm{v}_{x0}, \ \bm{v}_{z0}$ in the first term,  making in the second term the change of variables $\displaystyle\left(\bm{u}_{x0},\bm{v}_{x0},\bm{u}_{z0},\bm{v_{z0}}\right)\to\left(\bm{\Delta}_{0},\bm{m}_{0},\bm{\bar{\Delta}}_{0},\bm{\bar{m}}_{0}\right)$ and $\displaystyle\left(\bm{v}_{xs},\bm{v}_{zs}\right)\to\left(\bm{\Delta}_{s},\bm{m}_{s}\right)$, and using the integral representation for the free propagator (see~\eqref{eq.C.9}), we obtain\footnote{$\displaystyle S_{\perp}=\int d^{2}x$ is the transverse area of the target nucleus.}
\begin{equation}
    \begin{split}
        &\left\langle\overline{\sum}\left|\mathcal{M}_{bef}\right|^{2}\right\rangle=\zeta^{2}\left(1-\zeta\right)^{2}N_{c}e_{q}^{2}e^{2}S_{\perp}\int d^{2}u_{x}d^{2}u_{z} \ e^{-i\bm{p}\cdot\bm{u}_{x}}e^{-i\bm{k}\cdot\bm{u}_{z}}\\
        &\times\Biggl[\mathcal{P}\left(\bm{u}_{x};L^{+}\right)\mathcal{P}\left(\bm{u}_{z};L^{+}\right)\int\dfrac{d^{2}p_{0}}{\left(2\pi\right)^{2}}\dfrac{F_{T;L}\left(\bm{p_{0}},\bm{p_{0}}\right)}{\left(p^{2}_{0}+\epsilon^{2}\right)^{2}}e^{i\bm{p}_{0}\cdot\left(\bm{u}_{x}-\bm{u}_{z}\right)}\Biggr.\\
        +&\frac{Q^{2}_{s}}{2L^{+}}\left(\bm{u}_{x}\cdot\bm{u}_{z}\right)\int^{L^{+}}_{0}ds^{+} \ \mathcal{P}\left(\bm{u}_{x};L^{+}-s^{+}\right)\mathcal{P}\left(\bm{u}_{z};L^{+}-s^{+}\right)\\
        &\Biggl.\times \int d^{2}\Delta_{0}d^{2}\bar{\Delta}_{0}\int\dfrac{d^{2}p_{0}}{\left(2\pi\right)^{2}}\dfrac{d^{2}\bar{p}_{0}}{\left(2\pi\right)}e^{i\bm{p}_{0}\cdot\bm{\Delta}_{0}} e^{-i\bm{\bar{p}}_{0}\cdot\bm{\bar{\Delta}}_{0}}\dfrac{F_{T;L}\left(\bm{p_{0}},\bm{\bar{p}_{0}}\right)}{\left(p^{2}_{0}+\epsilon^{2}\right)\left(\bar{p}^{2}_{0}+\epsilon^{2}\right)}\\
        &\times \int d^{2}\Delta_{s} \ \mathcal{J}\left(\bm{\Delta}_{s},\bm{\Delta}_{0};s^{+}\right)\mathcal{J}^{\ast}\left(\bm{\Delta}_{s}-\bm{u}_{x}+\bm{u}_{z},\bar{\bm{\Delta}}_{0};s^{+}\right)\Biggr].
    \end{split}
    \label{eq.2.25}
\end{equation}

Finally, we can transform this expression to momentum space to read
\begin{equation}
    \begin{split}
        &\left\langle\overline{\sum}\left|\mathcal{M}_{bef}\right|^{2}\right\rangle=\zeta^{2}\left(1-\zeta\right)^{2}N_{c}e_{q}^{2}e^{2}S_{\perp}\Biggl[\int\dfrac{d^{2}p_{0}}{\left(2\pi\right)^{2}}\dfrac{F_{T;L}\left(\bm{p_{0}},\bm{p_{0}}\right)}{\left(p^{2}_{0}+\epsilon^{2}\right)^{2}}\Biggr.\\
        &\times \tilde{\mathcal{P}}\left(\bm{p}-\bm{p_{0}};L^{+}\right)\tilde{\mathcal{P}}\left(\bm{k}+\bm{p_{0}};L^{+}\right)-2\dfrac{Q_{s}^{2}}{L^{+}}\int\dfrac{d^{2}p_{s}}{\left(2\pi\right)^{2}}\left(\bm{p}-\bm{p_{s}}\right)\cdot\left(\bm{k}+\bm{p_{s}}\right)\\
        &\Biggl.\times\int^{L^{+}}_{0}ds^{+}\dfrac{L^{+2}}{\left[Q_{s}^{2}\left(L^{+}-s^{+}\right)\right]^{2}}\tilde{\mathcal{P}}\left(\bm{p}-\bm{p_{s}};L^{+}-s^{+}\right)\tilde{\mathcal{P}}\left(\bm{k}+\bm{p_{s}};L^{+}-s^{+}\right)\\
        &\times \int\dfrac{d^{2}p_{0}}{\left(2\pi\right)^{2}}\dfrac{d^{2}\bar{p}_{0}}{\left(2\pi\right)^{2}}\dfrac{F_{T;L}\left(\bm{p_{0}},\bm{\bar{p}_{0}}\right)}{\left(p^{2}_{0}+\epsilon^{2}\right)\left(\bar{p}^{2}_{0}+\epsilon^{2}\right)}\tilde{\mathcal{J}}\left(\bm{p_{s}},\bm{p_{0}};s^{+}\right)\tilde{\mathcal{J}}^{\ast}\left(\bm{p_{s}},\bm{\bar{p}_{0}};s^{+}\right)\Biggr],
    \end{split}
    \label{eq.2.26}
\end{equation}
with $\tilde{\mathcal{J}}$ and $\tilde{\mathcal{P}}$ given by Fourier transforms of the path integrals~\eqref{eq:functionkho} and~\eqref{eq:Pdef}, respectively, which in the harmonic oscillator approximation are
\begin{equation}
    \tilde{\mathcal{J}}\left(\bm{p},\bm{q};\Delta t^{+}\right)=\exp\left\lbrace\dfrac{i}{2q^{+}\zeta\left(1-\zeta\right)\Omega\sin\left(\Omega \Delta t^{+}\right)}\left[\cos\left(\Omega \Delta t^{+}\right)\left(p^{2}+q^{2}\right)-2\bm{p}\cdot\bm{q}\right]\right\rbrace,
\end{equation}
\begin{equation}
    \tilde{\mathcal{P}}\left(\bm{p};\Delta t^{+}\right)=\dfrac{4\pi}{\Delta t^{+}Q^{2}_{s}}L^{+}\exp\left(\dfrac{-p^{2}}{\Delta t^{+}Q^{2}_{s}}L^{+}\right).
\end{equation}
The interpretation of these functions will be discussed below.

\subsubsection{Before-Inside}
The second term of $\displaystyle \left\langle\overline{\sum}\left|\mathcal{M}\right|^{2}\right\rangle$ in~\eqref{eq.2.9} reads
\begin{equation}
    \begin{split}
    &\left\langle\overline{\sum}\mathcal{M}_{bef}\mathcal{M}_{in}^{\ast}\right\rangle=\frac{-ie_{q}^{2}e^{2}}{2q^{+}}\zeta\left(1-\zeta\right)\int_{0}^{L^{+}}d\bar{y}^{+}e^{iq^{-}\bar{y}^{+}}\int\dfrac{d^{2}p_{0}}{\left(2\pi\right)^{2}}\frac{d^{2}\bar{p}_{0}}{\left(2\pi\right)^{2}}\dfrac{F_{T,L}\left(\bm{p_{0}},\bm{\bar{p}_{0}}\right)}{p^{2}_{0}+\epsilon^{2}}\\
    &\times\int d^{2}x_{0} \ d^{2}x \ d^{2}z_{0} \ d^{2}z\int d^{2}\bar{x}_{1} \ d^{2}\bar{x} \ d^{2}\bar{z}_{1} \ d^{2}\bar{z}\\
    &\times e^{-i{\bm{p}}\cdot\left({\bm{x}}-\bm{\bar{x}}{}\right)}e^{-i{\bm{k}}\cdot\left({\bm{z}}-\bm{\bar{z}}{}\right)}e^{i{\bm{p_{0}}}\cdot\left({\bm{x_{0}}}-{\bm{z_{0}}}\right)}e^{-i\bm{\bar{p}_{0}}\cdot\left(\bm{\bar{x}_{1}}-\bm{\bar{z}_{1}}\right)}\\
    &\times Tr\biggl\langle G\left(L,{\bm{x}}; 0,{\bm{x_{0}}}|p^{+}\right)\bar{G}\left(L, {\bm{z}}; 0, {\bm{z_{0}}}|k^{+}\right)\\
    &\hskip 3cm \times\bar{G}^{\dagger}\left(L, \bm{\bar{z}}{}; \bar{y}^{+}, \bm{\bar{z}_{1}}{}|k^{+}\right)G^{\dagger}\left(L,\bm{\bar{x}}{}; \bar{y}^{+},\bm{\bar{x}_{1}}{}|p^{+}\right)\biggr\rangle.
    \end{split}
    \label{eq.2.27}
\end{equation}

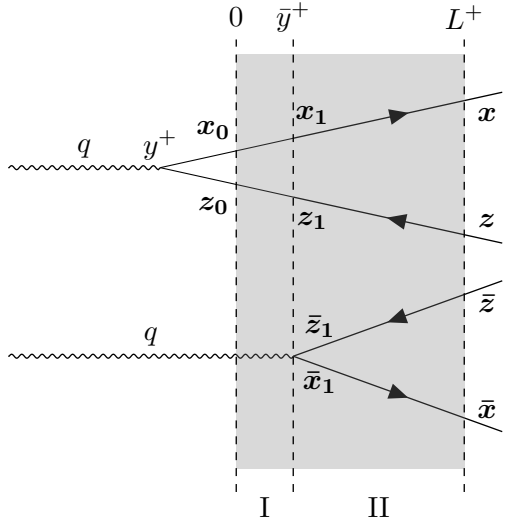
\begin{figure}[ht]
\centering
\begin{tikzpicture}
\begin{feynhand}

\vertex (a) at (-2,0.5);
\vertex (b) at (0,0.5);
\vertex (c) at (4.5,1.5);
\vertex (d) at (4.5,-0.5);
\vertex (m) at (1.75, 0.89);
\vertex (n) at (1.75, 0.11);

\propag [photon] (a) to [edge label = $q$] (b);
\draw (b) to (m);
\propag [fer] (m) to (c);
\draw (b) to (n);
\propag [fer] (d) to (n);

\vertex (e) at (-2,-2);
\vertex (f) at (1.75, -2);
\vertex (g) at (4.5, -1);
\vertex (h) at (4.5, -3);

\propag [photon] (e) to [edge label = $q$] (f);
\propag [fer] (g) to (f);
\propag [fer] (f) to (h);

\node at (1,2.5) {$0$};
\node at (4,2.5) {$L^{+}$};
\node at (1.75, 2.5) {$\bar{y}^{+}$};
\node at (0, 0.8) {$y^{+}$};

\fill[gray, opacity=0.3] (1,-3.5) rectangle (4,2);

\propag [sca] (1.75, -3.8) to (1.75, 2.2);
\propag [sca] (1, -3.8) to (1, 2.2);
\propag [sca] (4, -3.8) to (4, 2.2);

\node at (1.375, -4) {I};
\node at (2.875, -4) {II};

\node at (0.7, 1) {$\bm{x_{0}}$};
\node at (0.7, 0) {$\bm{z_{0}}$};
\node at (2, 1.2) {$\bm{x_{1}}$};
\node at (2, -0.2) {$\bm{z_{1}}$};
\node at (4.3, 1.2) {$\bm{x}$};
\node at (4.3, -0.2) {$\bm{z}$};

\node at (2.1, -1.6) {$\bm{\bar{z}_{1}}$};
\node at (2.1, -2.35) {$\bm{\bar{x}_{1}}$};
\node at (4.3, -1.3) {$\bm{\bar{z}}$};
\node at (4.3, -2.7) {$\bm{\bar{x}}$};

\end{feynhand}
\end{tikzpicture}

\caption{Separation into regions for the Before-Inside contribution.}
\label{Fig :Diagrama 3}
\end{figure}

In order to calculate the trace of the average of the four propagator we need to separate the medium in two regions I and II, see figure~\ref{Fig :Diagrama 3}, and use the composition property of the propagators
\begin{equation}
    G\left(x^{+},\bm{x};y^{+},\bm{y}|p^{+}\right)=\int d^{2}zG\left(x^{+},\bm{x};z^{+},\bm{z}|p^{+}\right)G\left(z^{+},\bm{z};y^{+},\bm{y}|p^{+}\right),
    \label{eq.2.28}
\end{equation}
with $\displaystyle x^{+}>z^{+}>y{+}$.

After the manipulations detailed in appendix~\ref{ap:befin}, the final expression in momentum space reads
\begin{equation}
\begin{split}
    &\left\langle\overline{\sum}\mathcal{M}_{bef}\mathcal{M}^{\ast}_{in}\right\rangle=-\dfrac{iN_{c}e_{q}^{2}e^{2}}{2q^{+}}\zeta\left(1-\zeta\right)S_{\perp}\int^{L^{+}}_{0}d\bar{y}^{+} \ e^{iq^{-}\bar{y}^{+}}\int\dfrac{d^{2}p_{0}}{\left(2\pi\right)^{2}}\dfrac{d^{2}\bar{p}_{0}}{\left(2\pi\right)^{2}}\\
    &\times\dfrac{F_{T;L}\left(\bm{p_{0}},\bm{\bar{p}_{0}}\right)}{p^{2}_{0}+\epsilon^{2}}\Biggl[\tilde{\mathcal{J}}\left(\bm{\bar{p}_{0}},\bm{p_{0}};\bar{y}^{+}\right)\Biggr.\tilde{\mathcal{P}}\left(\bm{p}-\bm{\bar{p}_{0}};L^{+}-\bar{y}^{+}\right)\tilde{\mathcal{P}}\left(\bm{k}+\bm{\bar{p}_{0}};L^{+}-\bar{y}^{+}\right)\\
    &-2\dfrac{Q_{s}^{2}}{L^{+}}\int\dfrac{d^{2}p_{s}}{\left(2\pi\right)^{2}}\left(\bm{p}-\bm{p_{s}}\right)\cdot\left(\bm{k}+\bm{p_{s}}\right)\int\dfrac{d^{2}p_{1}}{\left(2\pi\right)^{2}}\tilde{\mathcal{J}}\left(\bm{p_{1}},\bm{p_{0}};\bar{y}^{+}\right)\\
&\times\int^{L^{+}}_{\bar{y}^{+}}ds^{+}\dfrac{L^{+2}}{\left[Q_{s}^{2}\left(L^{+}-s^{+}\right)\right]^{2}}\tilde{\mathcal{P}}\left(\bm{p}-\bm{p_{s}};L^{+}-s^{+}\right)\tilde{\mathcal{P}}\left(\bm{k}+\bm{p_{s}};L^{+}-s^{+}\right)\\
\Biggl.&\times\tilde{\mathcal{J}}\left(\bm{p_{s}},\bm{p_{1}};s^{+}-\bar{y}^{+}\right)\tilde{\mathcal{J}}^{\ast}\left(\bm{p_{s}},\bm{\bar{p}_{0}};s^{+}-\bar{y}^{+}\right)\Biggr].
\end{split}
\label{eq.2.34}
\end{equation}

\subsubsection{Inside-Inside}
The next term considers the situation when the splitting takes place inside the medium in the amplitude and in the complex conjugate amplitude. With the same procedure as before, we start by
\begin{equation}
    \begin{split}
        &\left\langle\overline{\sum}\left|\mathcal{M}_{in}\right|^{2}\right\rangle=\frac{e_{q}^{2}e^{2}}{\left(2q^{+}\right)^{2}}\int^{L^{+}}_{0}dy^{+}\int^{L^{+}}_{0}d\bar{y}^{+}e^{-iq^{-}\left(y^{+}-\bar{y}^{+}\right)}\int\frac{d^{2}p_{0}}{\left(2\pi\right)^{2}}\frac{d^{2}\bar{p}_{0}}{\left(2\pi\right)^{2}}\\
        &\times F_{T;L}\left(\bm{p_{0}},\bm{\bar{p}_{0}}\right)\int d^{2}x_{0} \ d^{2}z_{0} \ d^{2}\bar{x}_{1} \ d^{2}\bar{z}_{1} \ e^{i\bm{p_{0}}\cdot\left(\bm{x_{0}}{}-\bm{z_{0}}{}\right)}e^{-i\bm{\bar{p}_{0}}\cdot\left(\bm{\bar{x}_{1}}{}-\bm{\bar{z}_{1}}{}\right)}\\
        &\times\int d^{2}x \ d^{2}z \ d^{2}\bar{x} \ d^{2}\bar{z} \ e^{-i\bm{p}\cdot\left(\bm{x}{}-\bm{\bar{x}}{}\right)}e^{-i\bm{k}\cdot\left(\bm{z}{}-\bm{\bar{z}}{}\right)}\\
        &\times
        Tr\biggl\langle G\left(L,\bm{x}{};y^{+},\bm{x_{0}}{}|p^{+}\right)\bar{G}\left(L,\bm{z}{};y^{+},\bm{z_{0}}{}|k^{+}\right)\\
        &\hskip 3cm \times\bar{G}^{\dagger}\left(L,\bm{\bar{z}}{};\bar{y}^{+},\bm{\bar{z}_{1}}{}|k^{+}\right)G^{\dagger}\left(L,\bm{\bar{x}}{};\bar{y}^{+},\bm{\bar{x}_{1}}{}|p^{+}\right)\biggr\rangle.
    \end{split}
    \label{eq.2.35}
\end{equation}

In order to calculate the trace of the average of Wilson lines we have to proceed in the same way as in the previous case, but now we divide the medium in two regions differently depending on which splitting vertex comes first, see figure~\ref{Fig :Diagrama 4}. After the computations detailed in appendix~\ref{ap:inin}, we get the following expression in momentum space:
\begin{equation}
\begin{split}
    &\left\langle\overline{\sum}\left|\mathcal{M}_{in}\right|^{2}\right\rangle=\dfrac{N_{c}e_{q}^{2}e^{2}}{\left(2q^{+}\right)^{2}}S_{\perp}2\mathfrak{Re}\int^{L^{+}}_{0}dy^{+}\int^{L^{+}}_{y^{+}}d\bar{y}^{+} \ e^{-iq^{-}\left(y^{+}-\bar{y}^{+}\right)}\int\dfrac{d^{2}p_{0}}{\left(2\pi\right)^{2}}\dfrac{d^{2}\bar{p}_{0}}{\left(2\pi\right)^{2}}\\
    &\times F_{T;L}\left(\bm{p_{0}},\bm{\bar{p}_{0}}\right)\Biggl[\tilde{\mathcal{P}}\left(\bm{p}-\bm{\bar{p}_{0}};L^{+}-\bar{y}^{+}\right)\tilde{\mathcal{P}}\left(\bm{k}+\bm{\bar{p}_{0}};L^{+}-\bar{y}^{+}\right)\tilde{\mathcal{J}}\left(\bm{\bar{p}_{0}},\bm{p_{0}};\bar{y}^{+}-y^{+}\right)\Biggr.\\
    &\hskip 0.5cm -\dfrac{Q_{s}^{2}}{2L^{+}}4\int\dfrac{d^{2}p_{s}}{\left(2\pi\right)^{2}}\left(\bm{p}-\bm{p_{s}}\right)\cdot\left(\bm{k}+\bm{p_{s}}\right)\int\dfrac{d^{2}p_{1}}{\left(2\pi\right)^{2}}\tilde{\mathcal{J}}\left(\bm{p_{1}},\bm{p_{0}};\bar{y}^{+}-y^{+}\right)\\
    &\times\int^{L^{+}}_{\bar{y}^{+}}ds^{+}\dfrac{L^{+2}}{\left[Q_{s}^{2}\left(L^{+}-s^{+}\right)\right]^{2}}\tilde{\mathcal{P}}\left(\bm{p}-\bm{p_{s}};L^{+}-s^{+}\right)\tilde{\mathcal{P}}\left(\bm{k}+\bm{p_{s}};L^{+}-s^{+}\right)\\
    &\times \Biggl.\tilde{\mathcal{J}}\left(\bm{p_{s}},\bm{p_{1}};s^{+}-\bar{y}^{+}\right)\tilde{\mathcal{J}}^{\ast}\left(\bm{p_{s}},\bm{\bar{p}_{0}};s^{+}-\bar{y}^{+}\right)\Biggr].
\end{split}
\label{eq.2.42}
\end{equation}

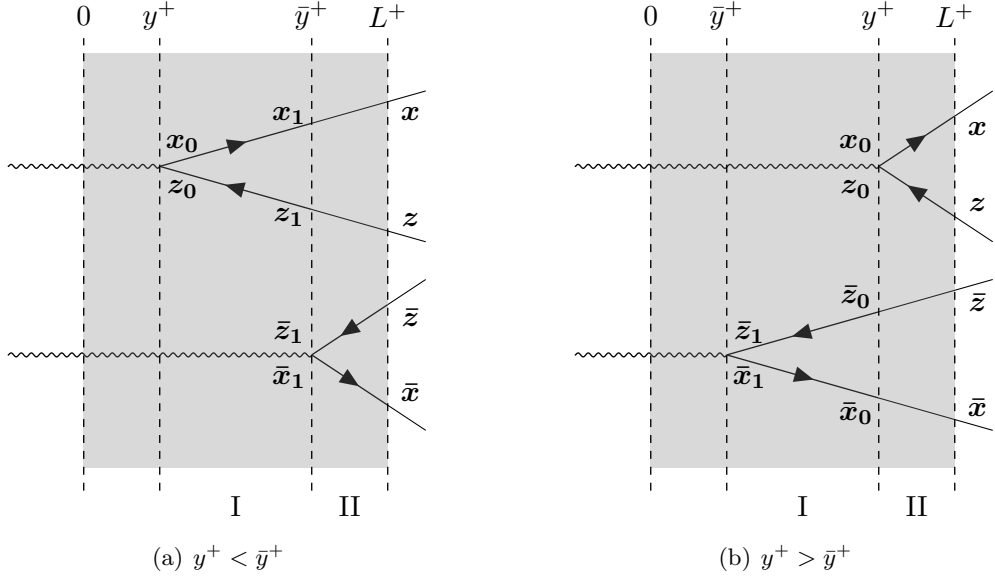
\begin{figure}[ht]
\centering
\subfigure[$y^{+}<\bar{y}^{+}$]{
\begin{tikzpicture}
\begin{feynhand}

\vertex (a) at (-1,0.5);
\vertex (b) at (1,0.5);
\vertex (c) at (4.5,1.5);
\vertex (d) at (4.5,-0.5);
\vertex (m) at (3, 15/14);
\vertex (n) at (3, -1/14);

\propag [photon] (a) to (b);
\draw (m) to (c);
\propag [fer] (b) to (m);
\draw (n) to (d);
\propag [fer] (n) to (b);

\vertex (e) at (-1,-2);
\vertex (f) at (3, -2);
\vertex (g) at (4.5, -1);
\vertex (h) at (4.5, -3);
\vertex (l) at (4,-4/3);
\vertex (r) at (4,-8/3);

\propag [photon] (e) to (f);
\propag [fer] (l) to (f);
\draw (l) to (g);
\propag [fer] (f) to (r);
\draw (r) to (h);

\node at (0,2.5) {$0$};
\node at (4,2.5) {$L^{+}$};
\node at (3, 2.5) {$\bar{y}^{+}$};
\node at (1, 2.5) {$y^{+}$};

\fill[gray, opacity=0.3] (0,-3.5) rectangle (4,2);

\propag [sca] (0, -3.8) to (0, 2.2);
\propag [sca] (4, -3.8) to (4, 2.2);
\propag [sca] (3, -3.8) to (3, 2.2);
\propag [sca] (1, -3.8) to (1, 2.2);

\node at (2, -4) {I};
\node at (3.5, -4) {II};

\node at (1.3, 0.8) {$\bm{x_{0}}$};
\node at (1.3, 0.2) {$\bm{z_{0}}$};
\node at (2.7, 15/14+0.1) {$\bm{x_{1}}$};
\node at (2.7, -1/14-0.1) {$\bm{z_{1}}$};
\node at (4.3, 1.2) {$\bm{x}$};
\node at (4.3, -0.2) {$\bm{z}$};

\node at (2.7,-1.7) {$\bm{\bar{z}_{1}}$};
\node at (2.7, -2.3) {$\bm{\bar{x}_{1}}$};
\node at (4.3, -1.5) {$\bm{\bar{z}}$};
\node at (4.3, -2.5) {$\bm{\bar{x}}$};

\end{feynhand}
\end{tikzpicture}
\label{Fig: Diagrama 4.1}}\hspace{0.1\textwidth}
\subfigure[$y^{+}>\bar{y}^{+}$]{
\begin{tikzpicture}
\begin{feynhand}

\vertex (a) at (-1,-2);
\vertex (b) at (1,-2);
\vertex (c) at (4.5,-1);
\vertex (d) at (4.5,-3);
\vertex (m) at (3, 15/14-2.5);
\vertex (n) at (3, -1/14-2.5);

\propag [photon] (a) to (b);
\draw (m) to (c);
\propag [fer] (m) to (b);
\draw (n) to (d);
\propag [fer] (b) to (n);

\vertex (e) at (-1,0.5);
\vertex (f) at (3, 0.5);
\vertex (g) at (4.5, 1.5);
\vertex (h) at (4.5, -0.5);
\vertex (l) at (4,-4/3+2.5);
\vertex (r) at (4,-8/3+2.5);

\propag [photon] (e) to (f);
\propag [fer] (f) to (l);
\draw (l) to (g);
\propag [fer] (r) to (f);
\draw (r) to (h);

\node at (0,2.5) {$0$};
\node at (4,2.5) {$L^{+}$};
\node at (1, 2.5) {$\bar{y}^{+}$};
\node at (3, 2.5) {$y^{+}$};

\fill[gray, opacity=0.3] (0,-3.5) rectangle (4,2);

\propag [sca] (0, -3.8) to (0, 2.2);
\propag [sca] (4, -3.8) to (4, 2.2);
\propag [sca] (3, -3.8) to (3, 2.2);
\propag [sca] (1, -3.8) to (1, 2.2);

\node at (2, -4) {I};
\node at (3.5, -4) {II};

\node at (1.3, -1.7) {$\bm{\bar{z}_{1}}$};
\node at (1.3, -2.3) {$\bm{\bar{x}_{1}}$};
\node at (2.7, 15/14+0.2-2.5) {$\bm{\bar{z}_{0}}$};
\node at (2.7, -1/14-0.2-2.5) {$\bm{\bar{x}_{0}}$};
\node at (4.3, -1.3) {$\bm{\bar{z}}$};
\node at (4.3, -2.7) {$\bm{\bar{x}}$};

\node at (2.7,0.8) {$\bm{x_{0}}$};
\node at (2.7, 0.2) {$\bm{z_{0}}$};
\node at (4.3, 1) {$\bm{x}$};
\node at (4.3, 0) {$\bm{z}$};

\end{feynhand}
\end{tikzpicture}
\label{Fig: Diagrama 4.2}}

\caption{Separation into regions for the Inside-Inside contribution.}
\label{Fig :Diagrama 4}
\end{figure}

\subsubsection{Before-After}
When the splitting takes place before the medium in the amplitude and after the medium in the complex conjugate, the contribution is
\begin{equation}
    \begin{split}
    &\left\langle\overline{\sum}\mathcal{M}_{bef}\mathcal{M}_{aft}^{\ast}\right\rangle=-e_{q}^{2}e^{2}\zeta^{2}\left(1-\zeta\right)^{2}\left(2\pi\right)^{2}\delta^{\left(2\right)}\left(\bm{p}{}+\bm{k}{}\right)e^{-i\frac{Q^{2}}{2q^{+}}L^{+}}\\
    &\times\int\dfrac{d^{2}p_{0}}{\left(2\pi\right)^{2}}\dfrac{F_{T;L}\left(\bm{p},\bm{p_{0}}\right)}{\left(p^{2}+\epsilon^{2}\right)\left(p^{2}_{0}+\epsilon^{2}\right)}\int d^{2}x_{0} \ d^{2}x \ d^{2}z_{0} \ d^{2}z \ e^{-i\bm{p}\cdot\bm{x}}e^{-i\bm{k}\cdot\bm{z}}e^{i\bm{p_{0}}\cdot\left(\bm{x_{0}}-\bm{z_{0}}\right)}\\
    &\times Tr\Bigl\langle G\left(L,\bm{x}{};0,\bm{x_{0}}{}|p^{+}\right)\bar{G}\left(L,\bm{z}{};0,\bm{z_{0}}{}|k^{+}\right)\Bigr\rangle.
    \end{split}
    \label{eq.2.43}
\end{equation}

Now, solving the path integrals in the trace of the average of two propagators, we get
\begin{equation}
    \begin{split}
    &\left\langle\overline{\sum}\mathcal{M}_{bef}\mathcal{M}_{aft}^{\ast}\right\rangle=-N_{c}e_{q}^{2}e^{2}\zeta^{2}\left(1-\zeta\right)^{2}\left(2\pi\right)^{2}\delta^{\left(2\right)}\left(\bm{p}{}+\bm{k}{}\right)e^{-i\frac{Q^{2}}{2q^{+}}L^{+}}\\
    &\times\int\dfrac{d^{2}p_{0}}{\left(2\pi\right)^{2}}\dfrac{F_{T;L}\left(\bm{p},\bm{p_{0}}\right)}{\left(p^{2}+\epsilon^{2}\right)\left(p^{2}_{0}+\epsilon^{2}\right)}\\
    &\times \int d^{2}x_{0} \ d^{2}z_{0} \ d^{2}x \ d^{2}z \ e^{-i\bm{p}\cdot\bm{x}}e^{-i\bm{k}\cdot\bm{z}}e^{i\bm{p}_{0}\cdot\left(\bm{x}_{0}-\bm{z}_{0}\right)}\mathcal{K}\left(\bm{x},\bm{x_{0}},\bm{z}{},\bm{z_{0}};L^{+}\right).
    \end{split}
    \label{eq.2.44}
\end{equation}

Using the change of variables mentioned in the previous sections and the delta function in the transverse momentum, we get
\begin{equation}
    \begin{split}
        &\left\langle\overline{\sum}\mathcal{M}_{bef}\mathcal{M}_{aft}^{\ast}\right\rangle\\
        &=-N_{c}e_{q}^{2}e^{2}\zeta^{2}\left(1-\zeta\right)^{2}\left(2\pi\right)^{2}\delta^{\left(2\right)}\left(\bm{p}{}+\bm{k}{}\right)e^{-i\frac{Q^{2}}{2q^{+}}L^{+}}\int\dfrac{d^{2}p_{0}}{\left(2\pi\right)^{2}}\dfrac{F_{T;L}\left(\bm{p},\bm{p_{0}}\right)}{\left(p^{2}+\epsilon^{2}\right)\left(p^{2}_{0}+\epsilon^{2}\right)}\\ & \times \int d^{2}\Delta d^{2}m \ d^{2}\Delta_{0}d^{2}m_{0} \ e^{-i\bm{p}\cdot\bm{\Delta}}e^{i\bm{p}_{0}\cdot\bm{\Delta}_{0}}G_{0}\left(\bm{m}-\bm{m}_{0};L^{+}|q^{+}\right)\mathcal{J}\left(\bm{\Delta},\bm{\Delta}_{0};L^{+}\right).
    \end{split}
    \label{eq.2.45}
\end{equation}

Integrating $\displaystyle \bm{m}_{0}$ and $\displaystyle \bm{m}$ with~\eqref{eq.C.9}, we arrive at the following expression in momentum space:
\begin{equation}
    \begin{split}
        &\left\langle\overline{\sum}\mathcal{M}_{bef}\mathcal{M}_{aft}^{\ast}\right\rangle=-N_{c}e_{q}^{2}e^{2}\zeta^{2}\left(1-\zeta\right)^{2}S_{\perp}\left(2\pi\right)^{2}\delta^{\left(2\right)}\left(\bm{p}{}+\bm{k}{}\right)e^{-i\frac{Q^{2}}{2q^{+}}L^{+}}\\
        &\times\int\dfrac{d^{2}p_{0}}{\left(2\pi\right)^{2}}\dfrac{F_{T;L}\left(\bm{p},\bm{p_{0}}\right)}{\left(p^{2}+\epsilon^{2}\right)\left(p^{2}_{0}+\epsilon^{2}\right)}\int d^{2}\Delta d^{2}\Delta_{0} \ e^{-i\bm{p}\cdot\bm{\Delta}}e^{i\bm{p}_{0}\cdot\bm{\Delta}_{0}}\mathcal{J}\left(\bm{\Delta},\bm{\Delta}_{0};L^{+}\right)\\
        &=-N_{c}e_{q}^{2}e^{2}\zeta^{2}\left(1-\zeta\right)^{2}S_{\perp}\left(2\pi\right)^{2}\delta^{\left(2\right)}\left(\bm{p}{}+\bm{k}{}\right)e^{-i\frac{Q^{2}}{2q^{+}}L^{+}}\\
        &\times\int\dfrac{d^{2}p_{0}}{\left(2\pi\right)^{2}}\dfrac{F_{T;L}\left(\bm{p},\bm{p_{0}}\right)}{\left(p^{2}+\epsilon^{2}\right)\left(p^{2}_{0}+\epsilon^{2}\right)}\tilde{\mathcal{J}}\left(\bm{p},\bm{p}_{0};L^{+}\right).
    \end{split}
    \label{eq.2.46}
\end{equation}

\subsubsection{Inside-After}
In this case we have the splittings in and after the medium. Proceeding in the same way as before, we obtain
\begin{equation}
    \begin{split}
    \left\langle\overline{\sum}\mathcal{M}_{in}\mathcal{M}_{aft}^{\ast}\right\rangle=&\frac{-iN_{c}e_{q}^{2}e^{2}}{2q^{+}}\zeta\left(1-\zeta\right)S_{\perp}\left(2\pi\right)^{2}\delta^{\left(2\right)}\left(\bm{p}{}+\bm{k}{}\right)e^{-i\frac{Q^{2}}{2q^{+}}L^{+}}\\
        &\times\int\dfrac{d^{2}p_{0}}{\left(2\pi\right)^{2}}\dfrac{F_{T;L}\left(\bm{p},\bm{p_{0}}\right)}{p^{2}+\epsilon^{2}}\int^{L^{+}}_{0}dy^{+} \ e^{i\frac{Q^{2}}{2q^{+}}y^{+}} \ \tilde{\mathcal{J}}\left(\bm{p},\bm{p}_{0};L^{+}-y^{+}\right).
    \end{split}
    \label{eq.2.47}
\end{equation}

\subsubsection{After-After}
The last term has the splitting after the medium in the amplitude and in the conjugate, corresponding to the vacuum term that reads
\begin{equation}
    \begin{split}
    \left\langle\overline{\sum}\left|\mathcal{M}_{aft}\right|^{2}\right\rangle=&\left(2\pi\right)^{2}\delta^{\left(2\right)}\left(\bm{p}+\bm{k}\right)S_{\perp}N_{c}e_{q}^{2}e^{2}\zeta^{2}\left(1-\zeta\right)^{2}\dfrac{F_{T;L}\left(\bm{p},\bm{p}\right)}{\left(p^{2}+\epsilon^{2}\right)^{2}}.
    \end{split}
    \label{eq.2.48}
\end{equation}

\subsubsection{Interpretation of the results}

A brief discussion of the meaning of the key elements of the results is in order. The interpretation of the functions $\tilde{\mathcal{P}}$ and $\tilde{\mathcal{J}}$ can be seen in figure~\ref{Fig :Diagrama 3}~\cite{Blaizot:2012fh,Apolinario:2014csa}. Consider the first term in the squared bracket in the right hand side of~\eqref{eq.2.34}. We have the colour-correlated propagation of the dipole formed by the $q$ and the $\bar q$ in the amplitude from $0$ to $\bar y^+$ with final transverse coordinates $\bm{{x}_{1}}$ and $\bm{{z}_{1}}$, followed by the colour-correlated propagation from $\bar y^+$ to $L^{+}$ of the dipole formed by the $q$ in the amplitude and in the complex conjugate amplitude, with final transverse coordinates $\bm{{x}}$ and $\bm{\bar{x}}$, and of the dipole formed by the $\bar q$ in the amplitude and in the complex conjugate amplitude, with final transverse coordinates $\bm{{z}}$ and $\bm{\bar{z}}$. Therefore, $\tilde{\mathcal{P}}$ represents the propagation of a dipole formed between amplitude and complex conjugate amplitude, thus the broadening of a parton when traversing the target, while $\tilde{\mathcal{J}}$ represents the propagation of a dipole formed between a parton and an antiparton on a given side of the cut, which cannot be interpreted in terms of broadening of a single parton but as the colour-correlated propagation of a parton pair. The second term in the squared bracket in the right hand side of~\eqref{eq.2.34} corresponds to keeping the latter (now for two dipoles, one in the amplitude and one in the complex conjugate) from $\bar y^+$ to $s^{+}$ where colour coherence is lost through a colour swap; then we have independent broadening of quark and antiquark from $s^{+}$ to $L^{+}$.

\section{Shockwave expansion}
\label{sec:shockwave}

The shockwave expansion consists in the expansion of the expressions in powers of $\displaystyle L^{+}/q^{+}$. Then, at order zero one expects to recover the eikonal shockwave limit, at first order the first non-eikonal correction to the shockwave approximation, and successively. More rigorously, we make the expansion in two dimensionless parameters that appear naturally and independently in the expressions, defined as
\begin{equation}
    \lambda^{2}=Q^{2}\dfrac{L^{+}}{q^{+}}, \ \ \ \ \ \ \ \ \ \ \kappa^{2}=Q_{s}^{2}\dfrac{L^{+}}{q^{+}}\ .
    \label{eq.5.1}
\end{equation}
One needs to change the variables of all the longitudinal integrals for a dimensionless variable, e.g., $\displaystyle y^{+}\to L^{+}t$, which removes all the $L^{+}$ from the limits of the integrals but introduces it in the $\displaystyle \mathcal{P}$ and $\displaystyle \mathcal{J}$ functions. It turns out that $\displaystyle \mathcal{P}$ will not depend on $\displaystyle L^{+}$
\begin{equation}
    \mathcal{P}\left(\bm{x};L^{+}t\right)=e^{-\frac{Q_{s}^{2}}{4L^{+}}\left(L^{+}t\right)x^{2}}=e^{-\frac{Q_{s}^{2}}{4}tx^{2}}.
    \label{eq.5.2}
\end{equation}
On the other hand, $\displaystyle \mathcal{J}$ will only depend on $\displaystyle L^{+}$ through $\displaystyle\kappa$:
\begin{equation}
    \begin{split}
        &\mathcal{J}\left(\bm{x},\bm{y};L^{+}t\right)\\
        &=\dfrac{1}{2\pi i}\dfrac{q^{+}\zeta\left(1-\zeta\right)\Omega}{\sin\left(\Omega L^{+}t\right)}\exp\left\lbrace i\dfrac{q^{+}\zeta\left(1-\zeta\right)\Omega}{2\sin\left(\Omega L^{+}t\right)}\left[\cos\left(\Omega L^{+}t\right)\left(x^{2}+y^{2}\right)-2\bm{x}\cdot\bm{y}\right]\right\rbrace\\
        &=\dfrac{1}{2\pi i}\dfrac{Q_{s}^{2}\zeta\left(1-\zeta\right)\omega}{\kappa\sin\left(\omega\kappa t\right)}\exp\left\lbrace i\dfrac{Q_{s}^{2}\zeta\left(1-\zeta\right)\omega}{2\kappa\sin\left(\omega \kappa t\right)}\left[\cos\left(\omega \kappa t\right)\left(x^{2}+y^{2}\right)-2\bm{x}\cdot\bm{y}\right]\right\rbrace,
    \end{split}
    \label{eq.5.3}
\end{equation}
where $\displaystyle \Omega^{2}=\dfrac{-iQ_{s}^{2}}{2q^{+}\zeta\left(1-\zeta\right)L^{+}}$ and we define $\omega^{2}=\dfrac{-i}{2\zeta\left(1-\zeta\right)}$, so  $\displaystyle\Omega L^{+}=\omega\kappa$ and $\displaystyle q^{+}\Omega=Q_{s}^{2}\omega/\kappa$. With the expression for $\displaystyle \mathcal{J}$ in terms of $\displaystyle \kappa$, we expand it in this variable but, because the expression is divergent when $\displaystyle \kappa\to 0$, we found more convenient to expand it in momentum space and then return to position space. Writing the expansion as (only even terms are non-zero)
\begin{equation}
    \mathcal{J}\left(\bm{x},\bm{y};L^{+}t\right)=\sum_{j=0}^{\infty}\kappa^{2j}\mathcal{J}^{\left(2j\right)}\left(\bm{x},\bm{y};t\right),
    \label{eq.5.4}
\end{equation}
we get the following expressions for the first three terms:
\begin{equation}
    \mathcal{J}^{\left(0\right)}\left(\bm{x},\bm{y};t\right)=\delta^{\left(2\right)}\left(\bm{x}-\bm{y}\right)e^{-\frac{tQ_{s}^{2}}{16}\left(\bm{x}+\bm{y}\right)^{2}},
    \label{eq.5.5}
\end{equation}
\begin{equation}
    \begin{split}
        \mathcal{J}^{\left(2\right)}\left(\bm{x},\bm{y};t\right)=&\dfrac{- it}{2\zeta\left(1-\zeta\right)} \ e^{-\frac{tQ_{s}^{2}}{16}\left(\bm{x}+\bm{y}\right)^{2}}\left[\dfrac{t}{4}-\dfrac{1}{Q_{s}^{2}}\nabla^{2}-\dfrac{t^{2}Q_{s}^{2}}{192}\left(\bm{x}+\bm{y}\right)^{2}\right]\delta^{\left(2\right)}\left(\bm{x}-\bm{y}\right),
    \end{split}
    \label{eq.5.6}
\end{equation}
\begin{equation}
    \begin{split}
        &\mathcal{J}^{\left(4\right)}\left(\bm{x},\bm{y};t\right)=\dfrac{t^2}{360\zeta^{2}\left(1-\zeta\right)^{2}} \ e^{-\frac{tQ_{s}^{2}}{16}\left(\bm{x}+\bm{y}\right)^{2}}\Biggl[-\dfrac{45}{Q_{s}^{4}}\nabla^{4}+30\dfrac{t}{Q_{s}^{2}}\nabla^{2}\\
        &-\dfrac{15}{32}t^{2}\left(8+\left(\bm{x}+\bm{y}\right)^{2}\nabla^{2}\right)+\dfrac{21}{128}t^{3}Q_{s}^{2}\left(\bm{x}+\bm{y}\right)^{2}-\dfrac{5}{4096}t^{4}Q_{s}^{4}\left(\bm{x}+\bm{y}\right)^{4}\Biggr]\delta^{\left(2\right)}\left(\bm{x}-\bm{y}\right).
    \end{split}
    \label{eq.5.7}
\end{equation}

Now it is straightforward to obtain the contribution of each term up to the desired order in $\displaystyle \kappa$ and $\displaystyle\lambda$ (i.e., in $\displaystyle L^{+}/q^{+}$).
As a consistency check, we note  that at zero order (eikonal) only the terms before-before, before-after and after-after will contribute. The terms before-inside and after-inside, having a prefactor $\displaystyle \lambda^{2}=Q^{2}L^{+}/q^{+}$, will begin to contribute at first order (next-to-eikonal). The last term, inside-inside, has a prefactor $\displaystyle\lambda^{4}=Q^{4}\left(L^{+}/q^{+}\right)^{2}$, so its first contribution is at second order (next-to-nex-to-eikonal): each splitting vertex inside the target adds a factor $\displaystyle\lambda^{2}$ -- a non-eikonal contribution.

\subsection{$\displaystyle 0^{th}$-order (Eikonal)}
We show explicitly the calculation for the eikonal contribution of the term before-before. In position space, we take~\eqref{eq.2.25} that, with the change of longitudinal variable $\displaystyle s^{+}\to L^{+}t$ and using~\eqref{eq.5.2}, reads
\begin{equation}
    \begin{split}
        &\left\langle\overline{\sum}\left|\mathcal{M}_{bef}\right|^{2}\right\rangle=\zeta^{2}\left(1-\zeta\right)^{2}N_{c}e_{q}^{2}e^{2}S_{\perp}\int d^{2}u_{x}d^{2}u_{z} \ e^{-i\bm{p}\cdot\bm{u}_{x}}e^{-i\bm{k}\cdot\bm{u}_{z}}\\
        &\times\Biggl[e^{-\frac{Q_{s}^{2}}{4}\left(u_{x}^{2}+u_{z}^{2}\right)}\int\dfrac{d^{2}p_{0}}{\left(2\pi\right)^{2}}\dfrac{F_{T;L}\left(\bm{p_{0}},\bm{p_{0}}\right)}{\left(p^{2}_{0}+\epsilon^{2}\right)^{2}}e^{i\bm{p}_{0}\cdot\left(\bm{u}_{x}-\bm{u}_{z}\right)}\Biggr.\\
        &+\frac{Q^{2}_{s}}{2}\left(\bm{u}_{x}\cdot\bm{u}_{z}\right)\int^{1}_{0}dt \ e^{-\frac{Q_{s}^{2}\left(1-t\right)}{4}\left(u_{x}^{2}+u_{z}^{2}\right)}\int d^{2}\Delta_{0}d^{2}\bar{\Delta}_{0}\int\dfrac{d^{2}p_{0}}{\left(2\pi\right)^{2}}\dfrac{d^{2}\bar{p}_{0}}{\left(2\pi\right)}e^{i\bm{p}_{0}\cdot\bm{\Delta}_{0}}e^{-i\bm{\bar{p}}_{0}\cdot\bm{\bar{\Delta}}_{0}}\\
        &\times\Biggl.\dfrac{F_{T;L}\left(\bm{p_{0}},\bm{\bar{p}_{0}}\right)}{\left(p^{2}_{0}+\epsilon^{2}\right)\left(\bar{p}_{0}^{2}+\epsilon^{2}\right)}\int d^{2}\Delta_{s} \ \mathcal{J}\left(\bm{\Delta}_{s},\bm{\Delta}_{0};L^{+}t\right)\mathcal{J}^{\ast}\left(\bm{\Delta}_{s}-\bm{u}_{x}+\bm{u}_{z},\bar{\bm{\Delta}}_{0};L^{+}t\right)\Biggr].
    \end{split}
    \label{eq.5.8}
\end{equation}

Then, at zero-order in the parameters $\displaystyle \kappa$ and $\lambda$, taking $\mathcal{J}\to \mathcal{J}^{\left(0\right)}$ and using \eqref{eq.5.5}, this term can be written as\footnote{The superscript $\displaystyle(0)$ in the left hand part of the expression indicates that this term correspond to the term $\displaystyle\left(L/q^{+}\right)^{0}$ in the expansion.}
\begin{equation}
    \begin{split}
        &\left\langle\overline{\sum}\left|\mathcal{M}_{bef}\right|^{2}\right\rangle^{\left(0\right)}=\zeta^{2}\left(1-\zeta\right)^{2}N_{c}e_{q}^{2}e^{2}S_{\perp}\int d^{2}u_{x}d^{2}u_{z}e^{-i\bm{p}\cdot\bm{u}_{x}}e^{-i\bm{k}\cdot\bm{u}_{z}}\\
        \times&\Biggl[e^{-\frac{Q_{s}^{2}}{4}\left(u_{x}^{2}+u_{z}^{2}\right)}\int\dfrac{d^{2}p_{0}}{\left(2\pi\right)^{2}}\dfrac{F_{T;L}\left(\bm{p_{0}},\bm{p_{0}}\right)}{\left(p^{2}_{0}+\epsilon^{2}\right)^{2}}e^{i\bm{p}_{0}\cdot\left(\bm{u}_{x}-\bm{u}_{z}\right)}\Biggr.\\
        &+\dfrac{Q^{2}_{s}}{2}\left(\bm{u}_{x}\cdot\bm{u}_{z}\right)\int^{1}_{0}dt \ e^{-\frac{Q_{s}^{2}\left(1-t\right)}{4}\left(u_{x}^{2}+u_{z}^{2}\right)}\int d^{2}\Delta_{0}d^{2}\bar{\Delta}_{0}\\
        &\times \int\dfrac{d^{2}p_{0}}{\left(2\pi\right)^{2}}\dfrac{d^{2}\bar{p}_{0}}{\left(2\pi\right)}e^{i\bm{p}_{0}\cdot\bm{\Delta}_{0}}e^{-i\bm{\bar{p}}_{0}\cdot\bm{\bar{\Delta}}_{0}}\dfrac{F_{T;L}\left(\bm{p_{0}},\bm{\bar{p}_{0}}\right)}{\left(p_{0}^{2}+\epsilon^{2}\right)\left(\bar{p}_{0}^{2}+\epsilon^{2}\right)}\int d^{2}\Delta_{s}\delta^{\left(2\right)}\left(\bm{\Delta}_{s}-\bm{\Delta}_{0}\right)\\
        \Biggl.&\times \ e^{-\frac{tQ_{s}^{2}}{16}\left(\bm{\Delta}_{s}+\bm{\Delta}_{0}\right)^{2}}\delta^{\left(2\right)}\left(\bm{\Delta}_{s}-\bm{u}_{x}+\bm{u}_{z}-\bm{\bar{\Delta}}_{0}\right)e^{-\frac{tQ_{s}^{2}}{16}\left(\bm{\Delta}_{s}-\bm{u}_{x}+\bm{u}_{z}+\bm{\bar{\Delta}}_{0}\right)^{2}}\Biggr]
    \end{split}
    \label{eq.5.10}
\end{equation}
which, integrating $\displaystyle \bm{\Delta}_{s}$, $\displaystyle \bm{\bar{\Delta}}_{0}$, yields
\begin{equation}
    \begin{split}
        &\left\langle\overline{\sum}\left|\mathcal{M}_{bef}\right|^{2}\right\rangle^{\left(0\right)}=\zeta^{2}\left(1-\zeta\right)^{2}N_{c}e_{q}^{2}e^{2}S_{\perp}\int d^{2}u_{x}d^{2}u_{z} \ e^{-i\bm{p}\cdot\bm{u}_{x}}e^{-i\bm{k}\cdot\bm{u}_{z}}\\
        &\times\Biggl[e^{-\frac{Q_{s}^{2}}{4}\left(u_{x}^{2}+u_{z}^{2}\right)}\int\dfrac{d^{2}p_{0}}{\left(2\pi\right)^{2}}\dfrac{F_{T;L}\left(\bm{p_{0}},\bm{p_{0}}\right)}{\left(p^{2}_{0}+\epsilon^{2}\right)^{2}}e^{i\bm{p}_{0}\cdot\left(\bm{u}_{x}-\bm{u}_{z}\right)}+\dfrac{Q^{2}_{s}}{2}\left(\bm{u}_{x}\cdot\bm{u}_{z}\right)\Biggr.\\
        &\times\int^{1}_{0}dt \ e^{-\frac{Q_{s}^{2}\left(1-t\right)}{4}\left(u_{x}^{2}+u_{z}^{2}\right)}\int d^{2}\Delta_{0}\int\dfrac{d^{2}p_{0}}{\left(2\pi\right)^{2}}\dfrac{d^{2}\bar{p}_{0}}{\left(2\pi\right)^{2}}e^{i\bm{p}_{0}\cdot\bm{\Delta}_{0}}e^{-i\bm{\bar{p}}_{0}\cdot\left(\bm{\Delta}_{0}-\bm{u}_{x}+\bm{u}_{z}\right)}\\
        &\times \Biggl.\dfrac{F_{T;L}\left(\bm{p_{0}},\bm{\bar{p}_{0}}\right)}{\left(p_{0}^{2}+\epsilon^{2}\right)\left(\bar{p}_{0}^{2}+\epsilon^{2}\right)} \,e^{-\frac{tQ_{s}^{2}}{4}\Delta_{0}^{2}} \ e^{-\frac{tQ_{s}^{2}}{4}\left(\bm{\Delta}_{0}-\bm{u}_{x}+\bm{u}_{z}\right)^{2}}\Biggr].
    \end{split}
    \label{eq.5.11}
\end{equation}

Analogously we compute the eikonal contribution coming from the before-after term. From~\eqref{eq.2.46} we have
\begin{equation}
    \begin{split}
        &\left\langle\overline{\sum}\mathcal{M}_{bef}\mathcal{M}_{aft}^{\ast}\right\rangle=-N_{c}e_{q}^{2}e^{2}\zeta^{2}\left(1-\zeta\right)^{2}S_{\perp}\left(2\pi\right)^{2}\delta^{\left(2\right)}\left(\bm{p}+\bm{k}\right)e^{-i\frac{\lambda^{2}}{2}}\\
        &\times\int\dfrac{d^{2}p_{0}}{\left(2\pi\right)^{2}}\dfrac{F_{T;L}\left(\bm{p},\bm{p_{0}}\right)}{\left(p^{2}+\epsilon^{2}\right)\left(p^{2}_{0}+\epsilon^{2}\right)}\int d^{2}\Delta d^{2}\Delta_{0} \ e^{-i\bm{p}\cdot\bm{\Delta}}e^{i\bm{p}_{0}\cdot\bm{\Delta}_{0}}\mathcal{J}\left(\bm{\Delta},\bm{\Delta}_{0};L^{+}\right),
    \end{split}
    \label{eq.5.12}
\end{equation}
whose expansion to zeroth order zero in $\displaystyle\lambda^{2}$, $\displaystyle\kappa^{2}$ gives
\begin{equation}
    \begin{split}
        &\left\langle\overline{\sum}\mathcal{M}_{bef}\mathcal{M}_{aft}^{\ast}\right\rangle^{\left(0\right)}=-N_{c}e_{q}^{2}e^{2}\zeta^{2}\left(1-\zeta\right)^{2}S_{\perp}\left(2\pi\right)^{2}\delta^{\left(2\right)}\left(\bm{p}{}+\bm{k}{}\right)\\
        &\times\int\dfrac{d^{2}p_{0}}{\left(2\pi\right)^{2}}\dfrac{F_{T;L}\left(\bm{p},\bm{p_{0}}\right)}{\left(p^{2}+\epsilon^{2}\right)\left(p^{2}_{0}+\epsilon^{2}\right)}\int d^{2}\Delta d^{2}\Delta_{0} \ e^{-i\bm{p}\cdot\bm{\Delta}}e^{i\bm{p}_{0}\cdot\bm{\Delta}_{0}}\mathcal{J}^{\left(0\right)}\left(\bm{\Delta},\bm{\Delta}_{0};L^{+}\right)\\
        &=-N_{c}e_{q}^{2}e^{2}\zeta^{2}\left(1-\zeta\right)^{2}S_{\perp}\left(2\pi\right)^{2}\delta^{\left(2\right)}\left(\bm{p}+\bm{k}\right)\int\dfrac{d^{2}p_{0}}{\left(2\pi\right)^{2}}\dfrac{F_{T;L}\left(\bm{p},\bm{p_{0}}\right)}{\left(p^{2}+\epsilon^{2}\right)\left(p_{0}^{2}+\epsilon^{2}\right)}\\
        &\times \int d^{2}\Delta \ e^{-i\left(\bm{p}-\bm{p}_{0}\right)\cdot\bm{\Delta}} \ e^{-\frac{Q_{s}^{2}}{4}\Delta^{2}}.
    \end{split}
    \label{eq.5.13}
\end{equation}

Finally, the after-after term~\eqref{eq.2.48} is purely eikonal.

\subsection{$\displaystyle1^{st}$-order (Next-to-Eikonal)}

For the first-order correction we have the contributions from the before-before and the before-after terms as in the previous case, plus the new first contributions coming from the terms before-inside and after-inside. We anticipate that the crossed terms (before-after, before-inside, after-inside) are purely imaginary, giving a null contribution to the cross section, while the term before-before is identically zero; hence, the first non-eikonal corrections (in the shockwave sense) will come from the next-to-next-to-eikonal contributions.
To illustrate these points, we first show the calculation for the before-inside term, which yields a purely imaginary expression (the calculation for the rest of crossing terms is identical), and later the before-before term resulting in zero.

Starting from~\eqref{eq.2.33} for the before-inside term, we change the integration variables in the longitudinal integrals to a dimensionless integral variable to remove $L^{+}$ from the limits of the integrals, $\displaystyle\bar{y}^{+}\to L^{+}t$, $s^{+}\to L^{+}t^{\prime}$, giving
\begin{equation}
    \begin{split}
        &\left\langle\overline{\sum}\mathcal{M}_{bef}\mathcal{M}_{in}^{\ast}\right\rangle=\dfrac{-iN_{c}e_{q}^{2}e^{2}}{2Q^{2}}\zeta\left(1-\zeta\right)S_{\perp}\lambda^{2}\int^{1}_{0}dt \ e^{-i\frac{\lambda^{2}}{2}t}\int\dfrac{d^{2}p_{0}}{\left(2\pi\right)^{2}}\dfrac{d^{2}\bar{p}_{0}}{\left(2\pi\right)^{2}}\\
        &\times\dfrac{F_{T;L}\left(\bm{p_{0}},\bm{\bar{p}_{0}}\right)}{p^{2}_{0}+\epsilon^{2}}\int d^{2}u_{x}d^{2}u_{z} \ e^{-i\bm{p}\cdot\bm{u}_{x}} e^{-i\bm{k}\cdot\bm{u}_{z}}\Biggl[\mathcal{P}\left(\bm{u}_{x};L^{+}\left(1-t\right)\right)\mathcal{P}\left(\bm{u}_{z};L^{+}\left(1-t\right)\right)\\
        &\times\int d^{2}\bar{\Delta}_{1}\int d^{2}\Delta_{0} \ e^{i\bm{p}_{0}\cdot\bm{\Delta}_{0}}e^{-i\bm{\bar{p}}_{0}\cdot\bm{\bar{\Delta}}_{1}} \ \mathcal{J}\left(\bm{\bar{\Delta}}_{1}+\bm{u}_{x}-\bm{u}_{z},\bm{\Delta}_{0};L^{+}t\right)\Biggr.\\
        &+\dfrac{Q_{s}^{2}}{2}\left(\bm{u}_{x}\cdot\bm{u}_{z}\right)\int^{1}_{t}dt^{\prime} \ \mathcal{P}\left(\bm{u}_{x};L^{+}\left(1-t^{\prime}\right)\right)\mathcal{P}\left(\bm{u}_{z};L^{+}\left(1-t^{\prime}\right)\right)\\
        &\times\int d^{2}\Delta_{s}d^{2}\Delta_{0}d^{2}\bar{\Delta}_{1}d^{2}\Delta_{1} \ e^{i\bm{p}_{0}\cdot\bm{\Delta}_{0}}e^{-i\bm{\bar{p}}_{0}\cdot\bm{\bar{\Delta}}_{1}}\\
        &\times \Biggl.\mathcal{J}\left(\bm{\Delta}_{s},\bm{\Delta}_{1};L^{+}\left(t^{\prime}-t\right)\right)\mathcal{J}^{\ast}\left(\bm{\Delta}_{s}-\bm{u}_{x}+\bm{u}_{z},\bm{\bar{\Delta}}_{1};L^{+}\left(t^{\prime}-t\right)\right)\mathcal{J}\left(\bm{\Delta}_{1},\bm{\Delta}_{0};L^{+}t\right)\Biggr].
    \end{split}
    \label{eq.5.15}
\end{equation}

The dependence of the term before-inside of the two parameters $\displaystyle \lambda^{2}$, $\displaystyle \kappa^{2}$ is, schematically,
\begin{equation}
    \left\langle\overline{\sum}\mathcal{M}_{bef}\mathcal{M}^{\ast}_{in}\right\rangle\thicksim\lambda^{2} e^{-i\frac{\lambda^{2}}{2}t}\left[\mathcal{J}\left(\kappa^{2}\right)+\mathcal{J}^{3}\left(\kappa^{2}\right)\right].
    \label{eq.5.16}
\end{equation}
Then, the first term in the expansion is $\propto \displaystyle\lambda^{2}\left[\mathcal{J}^{\left(0\right)}+\mathcal{J}^{\left(0\right)3}\right]$, so it will not contribute to the eikonal cross section as expected since it is an inside-splitting term, as discussed earlier. Computing the first non-eikonal correction, we get
\begin{equation}
    \begin{split}
        &\left\langle\overline{\sum}\mathcal{M}_{bef}\mathcal{M}_{in}^{\ast}\right\rangle^{\left(1\right)}=\dfrac{-iN_{c}e_{q}^{2}e^{2}}{2Q^{2}}\zeta\left(1-\zeta\right)S_{\perp}\lambda^{2}\int^{1}_{0}dt\int\dfrac{d^{2}p_{0}}{\left(2\pi\right)^{2}}\dfrac{d^{2}\bar{p}_{0}}{\left(2\pi\right)^{2}}\dfrac{F_{T;L}\left(\bm{p_{0}},\bm{\bar{p}_{0}}\right)}{p^{2}_{0}+\epsilon^{2}}\\
        &\times\int d^{2}u_{x}d^{2}u_{z} \ e^{-i\bm{p}\cdot\bm{u}_{x}} e^{-i\bm{k}\cdot\bm{u}_{z}}\Biggl[e^{-\frac{Q_{s}^{2}\left(1-t\right)}{4}\left(u_{x}^{2}+u_{z}^{2}\right)}\int d^{2}\bar{\Delta}_{1}d^{2}\Delta_{0} \ e^{i\bm{p}_{0}\cdot\bm{\Delta}_{0}}e^{-i\bm{\bar{p}}_{0}\cdot\bm{\bar{\Delta}}_{1}}\\
        &\times\delta^{\left(2\right)}\left(\bm{\bar{\Delta}_{1}}+\bm{u_{x}}-\bm{u_{z}}-\bm{\Delta_{0}}\right)e^{-\frac{tQ_{s}^{2}}{16}\left(\bm{\bar{\Delta}_{1}}+\bm{u_{x}}-\bm{u_{z}}+\bm{\Delta_{0}}\right)^{2}}\\
        &+\dfrac{Q_{s}^{2}}{2}\left(\bm{u}_{x}\cdot\bm{u}_{z}\right)\int^{1}_{t}dt^{\prime} \ e^{-\frac{Q_{s}^{2}\left(1-t^{\prime}\right)}{4}\left(u_{x}^{2}+u_{z}^{2}\right)}\int d^{2}\Delta_{s}d^{2}\Delta_{0}d^{2}\bar{\Delta}_{1}d^{2}\Delta_{1} \ e^{i\bm{p}_{0}\cdot\bm{\Delta}_{0}}e^{-i\bm{\bar{p}}_{0}\cdot\bm{\bar{\Delta}}_{1}}\\
        &\times\ \delta^{\left(2\right)}\left(\bm{\Delta_{s}}-\bm{\Delta_{1}}\right)\ \Biggl.e^{-\frac{\left(t^{\prime}-t\right)Q_{s}^{2}}{16}\left(\bm{\Delta_{s}}+\bm{\Delta_{1}}\right)^{2}}\delta^{\left(2\right)}\left(\bm{\Delta}_{s}-\bm{u}_{x}+\bm{u}_{z}-\bar{\bm{\Delta}}_{1}\right)\\
        &\times e^{-\frac{\left(t^{\prime}-t\right)Q_{s}^{2}}{16}\left(\bm{\Delta}_{s}-\bm{u}_{x}+\bm{u}_{z}+\bar{\bm{\Delta}}_{1}\right)^{2}}\delta^{\left(2\right)}\left(\bm{\Delta}_{1}-\bm{\Delta}_{0}\right)e^{-\frac{tQ_{s}^{2}}{16}\left(\bm{\Delta}_{1}+\bm{\Delta}_{0}\right)^{2}}\Biggr].
    \end{split}
    \label{eq.5.17}
\end{equation}
Integrating the delta functions we obtain
\begin{equation}
    \begin{split}
        &\left\langle\overline{\sum}\mathcal{M}_{bef}\mathcal{M}_{in}^{\ast}\right\rangle^{\left(1\right)}=\dfrac{-iN_{c}e_{q}^{2}e^{2}}{2Q^{2}}\zeta\left(1-\zeta\right)S_{\perp}\lambda^{2}\int^{1}_{0}dt\int\dfrac{d^{2}p_{0}}{\left(2\pi\right)^{2}}\dfrac{d^{2}\bar{p}_{0}}{\left(2\pi\right)^{2}}\dfrac{F_{T;L}\left(\bm{p_{0}},\bm{\bar{p}_{0}}\right)}{p^{2}_{0}+\epsilon^{2}}\\
        &\times\int d^{2}u_{x}d^{2}u_{z} \ e^{-i\bm{p}\cdot\bm{u}_{x}} e^{-i\bm{k}\cdot\bm{u}_{z}}\\
        &\times \Biggl[e^{-\frac{Q_{s}^{2}\left(1-t\right)}{4}\left(u_{x}^{2}+u_{z}^{2}\right)}\int d^{2}\Delta_{0} \ e^{i\bm{p}_{0}\cdot\bm{\Delta}_{0}}e^{-i\bm{\bar{p}}_{0}\cdot\left(\bm{\Delta}_{0}-\bm{u}_{x}+\bm{u}_{z}\right)} \ e^{-\frac{tQ_{s}^{2}}{4}\Delta_{0}^{2}}\Biggr.\\
        &\ \ +\dfrac{Q_{s}^{2}}{2}\left(\bm{u}_{x}\cdot\bm{u}_{z}\right)\int^{1}_{t}dt^{\prime} \ e^{-\frac{Q_{s}^{2}\left(1-t^{\prime}\right)}{4}\left(u_{x}^{2}+u_{z}^{2}\right)}\int d^{2}\Delta_{0} \ e^{i\bm{p}_{0}\cdot\bm{\Delta}_{0}}e^{-i\bm{\bar{p}}_{0}\cdot\left(\bm{\Delta}_{0}-\bm{u}_{x}+\bm{u}_{z}\right)}\\
        &\ \ \times \Biggl.e^{-\frac{\left(t^{\prime}-t\right)Q_{s}^{2}}{4}\Delta_{0}^{2}}e^{-\frac{\left(t^{\prime}-t\right)Q_{s}^{2}}{4}\left(\bm{\Delta}_{0}-\bm{u}_{x}+\bm{u}_{z}\right)^{2}}e^{-\frac{tQ_{s}^{2}}{4}\Delta_{0}^{2}}\Biggr].
    \end{split}
    \label{eq.5.18}
\end{equation}
At this point we do not need to solve the remaining integrals. We only need to realize that the complex conjugate of this expression reads the same but with opposite sign, which is easy to check by taking the complex conjugate and then performing the change of variables $\displaystyle \left(\bm{u}_{x},\bm{u}_{z},\bm{\Delta}_{0}\right)\to\left(-\bm{u}_{x},-\bm{u}_{z},-\bm{\Delta}_{0}\right)$. This implies that this contribution is purely imaginary and, therefore, it will not contribute to the cross section at first order in the shockwave expansion. The same happens with the remaining crossing terms.

Let us return to the before-before term and prove that it is zero at first order. From~\eqref{eq.5.8} the first-order term in $\displaystyle\lambda^{2}$ and $\displaystyle\kappa^{2}$ reads
\begin{equation}
    \begin{split}
        &\left\langle\overline{\sum}\left|\mathcal{M}_{bef}\right|^{2}\right\rangle^{\left(1\right)}=\zeta^{2}\left(1-\zeta\right)^{2}N_{c}e_{q}^{2}e^{2}S_{\perp}\int d^{2}u_{x}d^{2}u_{z} \ e^{-i\bm{p}\cdot\bm{u}_{x}}e^{-i\bm{k}\cdot\bm{u}_{z}}\frac{Q^{2}_{s}}{2}\left(\bm{u}_{x}\cdot\bm{u}_{z}\right)\\
        &\times\int^{1}_{0}dt \ e^{-\frac{Q_{s}^{2}\left(1-t\right)}{4}\left(u_{x}^{2}+u_{z}^{2}\right)}\int d^{2}\Delta_{0}d^{2}\bar{\Delta}_{0}\int\dfrac{d^{2}p_{0}}{\left(2\pi\right)^{2}}\dfrac{d^{2}\bar{p}_{0}}{\left(2\pi\right)}e^{i\bm{p}_{0}\cdot\bm{\Delta}_{0}}e^{-i\bm{\bar{p}}_{0}\cdot\bm{\bar{\Delta}}_{0}}\\
        &\times\dfrac{F_{T;L}\left(\bm{p_{0}},\bm{\bar{p}_{0}}\right)}{\left(p^{2}_{0}+\epsilon^{2}\right)\left(\bar{p}^{2}_{0}+\epsilon^{2}\right)}\int d^{2}\Delta_{s}\Biggl[\kappa^{2}\mathcal{J}^{\left(2\right)}\left(\bm{\Delta}_{s},\bm{\Delta}_{0};t\right)\mathcal{J}^{\left(0\right)\ast}\left(\bm{\Delta}_{s}-\bm{u}_{x}+\bm{u}_{z},\bar{\bm{\Delta}}_{0};t\right)\Biggr.\\
        &\Biggl.+\ \mathcal{J}^{\left(0\right)}\left(\bm{\Delta}_{s},\bm{\Delta}_{0};t\right)\kappa^{2}\mathcal{J}^{\left(2\right)\ast}\left(\bm{\Delta}_{s}-\bm{u}_{x}+\bm{u}_{z},\bar{\bm{\Delta}}_{0};t\right)\Biggr]\\
        &=\left\langle\overline{\sum}\left|\mathcal{M}_{bef}\right|^{2}\right\rangle^{\left(20\right)}+\left\langle\overline{\sum}\left|\mathcal{M}_{bef}\right|^{2}\right\rangle^{\left(02\right)},\\
    \end{split}
    \label{eq.5.19}
\end{equation}
where we split the expression according to the two addends inside the square brackets. The first one, using~\eqref{eq.5.5} and~\eqref{eq.5.6}, reads
\begin{equation}
    \begin{split}
        &\left\langle\overline{\sum}\left|\mathcal{M}_{bef}\right|^{2}\right\rangle^{\left(20\right)}=\zeta^{2}\left(1-\zeta\right)^{2}N_{c}e_{q}^{2}e^{2}S_{\perp}\kappa^{2}\\
        &\times\int d^{2}u_{x}d^{2}u_{z}e^{-i\bm{p}\cdot\bm{u}_{x}}e^{-i\bm{k}\cdot\bm{u}_{z}}\frac{Q^{2}_{s}}{2}\left(\bm{u}_{x}\cdot\bm{u}_{z}\right)\\
        &\times\int^{1}_{0}dt \ e^{-\frac{Q_{s}^{2}\left(1-t\right)}{4}\left(u_{x}^{2}+u_{z}^{2}\right)}\int d^{2}\Delta_{0}d^{2}\bar{\Delta}_{0}\int\dfrac{d^{2}p_{0}}{\left(2\pi\right)^{2}}\dfrac{d^{2}\bar{p}_{0}}{\left(2\pi\right)}e^{i\bm{p}_{0}\cdot\bm{\Delta}_{0}}e^{-i\bm{\bar{p}}_{0}\cdot\bm{\bar{\Delta}}_{0}}\\
        &\times\dfrac{F_{T;L}\left(\bm{p_{0}},\bm{\bar{p}_{0}}\right)}{\left(p^{2}_{0}+\epsilon^{2}\right)\left(\bar{p}^{2}_{0}+\epsilon^{2}\right)}\int d^{2}\Delta_{s}\dfrac{-it}{2\zeta\left(1-\zeta\right)}\Biggl[\dfrac{t}{4}\delta^{\left(2\right)}\left(\bm{\Delta}_{s}-\bm{\Delta}_{0}\right)\\
        &\hskip 2cm -\dfrac{1}{Q_{s}^{2}}\nabla^{2}\delta^{\left(2\right)}\left(\bm{\Delta}_{s}-\bm{\Delta}_{0}\right)-\dfrac{t^{2}Q_{s}^{2}}{192}\left(\bm{\Delta}_{s}+\bm{\Delta}_{0}\right)^{2}\delta^{\left(2\right)}\left(\bm{\Delta_{s}}-\bm{\Delta_{0}}\right)\Biggr]\\
        &\times e^{-\frac{tQ_{s}^{2}}{16}\left(\bm{\Delta}_{s}+\bm{\Delta}_{0}\right)^{2}}\delta^{\left(2\right)}\left(\bm{\Delta}_{s}-\bm{u}_{x}+\bm{u}_{z}-\bar{\bm{\Delta}}_{0}\right)e^{-\frac{tQ_{s}^{2}}{16}\left(\bm{\Delta}_{s}-\bm{u}_{x}+\bm{u}_{z}+\bar{\bm{\Delta}}_{0}\right)^{2}},\\
    \end{split}
    \label{eq.5.20}
\end{equation}
which, integrating $\displaystyle \bm{\bar{\Delta}}_{0}$ and $\displaystyle \bm{\Delta}_{0}$, gives
\begin{equation}
    \begin{split}
        &\left\langle\overline{\sum}\left|\mathcal{M}_{bef}\right|^{2}\right\rangle^{\left(20\right)}=-i\dfrac{N_{c}e_{q}^{2}e^{2}}{2}\zeta\left(1-\zeta\right)S_{\perp}\kappa^{2}\int d^{2}u_{x}d^{2}u_{z} \ e^{-i\bm{p}\cdot\bm{u}_{x}}e^{-i\bm{k}\cdot\bm{u}_{z}}\\
        &\times\frac{Q^{2}_{s}}{2}\left(\bm{u}_{x}\cdot\bm{u}_{z}\right)\int^{1}_{0}dt \ e^{-\frac{Q_{s}^{2}\left(1-t\right)}{4}\left(u_{x}^{2}+u_{z}^{2}\right)}\int \ d^{2}\Delta_{s}e^{-\frac{tQ_{s}^{2}}{4}\Delta_{s}^{2}}e^{-\frac{tQ_{s}^{2}}{4}\left(\bm{\Delta}_{s}-\bm{u}_{x}+\bm{u}_{z}\right)^{2}}\\&\times\int\dfrac{d^{2}p_{0}}{\left(2\pi\right)^{2}}\dfrac{d^{2}\bar{p}_{0}}{\left(2\pi\right)}e^{i\bm{p}_{0}\cdot\bm{\Delta}_{s}}e^{-i\bm{\bar{p}}_{0}\cdot\left(\bm{\Delta}_{s}-\bm{u}_{x}+\bm{u}_{z}\right)}\left(\dfrac{1}{p^{2}_{0}+\epsilon^{2}}\right)\left(\dfrac{F_{T;L}\left(\bm{p_{0}},\bm{\bar{p}_{0}}\right)}{\bar{p}^{2}_{0}+\epsilon^{2}}\right)\\
        &\times t\Biggl[\dfrac{t}{4}-\dfrac{1}{Q_{s}^{2}}\left[-p_{0}^{2}-\dfrac{tQ_{s}^{2}}{4}\left(1+2i\bm{p}_{0}\cdot\bm{\Delta}_{s}\right)+\dfrac{t^{2}Q_{s}^{4}}{16}\Delta_{s}^{2}\right]-\dfrac{t^{2}Q_{s}^{2}}{48}\Delta_{s}^{2}\Biggr].
    \end{split}
    \label{eq.5.22}
\end{equation}

With the same procedure we obtain the other term:
\begin{equation}
    \begin{split}
        &\left\langle\overline{\sum}\left|\mathcal{M}_{bef}\right|^{2}\right\rangle^{\left(02\right)}=i\dfrac{N_{c}e_{q}^{2}e^{2}}{2}\zeta\left(1-\zeta\right)S_{\perp}\kappa^{2}\int d^{2}u_{x}d^{2}u_{z}e^{-i\bm{p}\cdot\bm{u}_{x}}e^{-i\bm{k}\cdot\bm{u}_{z}}\\
        &\times\frac{Q^{2}_{s}}{2}\left(\bm{u}_{x}\cdot\bm{u}_{z}\right)\int^{1}_{0}dt \ e^{-\frac{Q_{s}^{2}\left(1-t\right)}{4}\left(u_{x}^{2}+u_{z}^{2}\right)}\int \ d^{2}\bar{\Delta}_{0}\ t \ e^{-\frac{tQ_{s}^{2}}{4}\left(\bm{\bar{\Delta}}_{0}+\bm{u}_{x}-\bm{u}_{z}\right)^{2}}e^{-\frac{tQ_{s}^{2}}{4}\bar{\Delta}_{0}^{2}}\\
        &\times\int\dfrac{d^{2}p_{0}}{\left(2\pi\right)^{2}}\dfrac{d^{2}\bar{p}_{0}}{\left(2\pi\right)}e^{i\bm{p}_{0}\cdot\left(\bm{\bar{\Delta}}_{0}+\bm{u}_{x}-\bm{u}_{z}\right)}e^{-i\bm{\bar{p}}_{0}\cdot\bm{\bar{\Delta}}_{0}}\dfrac{F_{T;L}\left(\bm{p_{0}},\bm{\bar{p}_{0}}\right)}{\left(p^{2}_{0}+\epsilon^{2}\right)\left(\bar{p}^{2}_{0}+\epsilon^{2}\right)}\\
        &\times\Biggl[\dfrac{t}{4}-\dfrac{1}{Q_{s}^{2}}\left[-p_{0}^{2}-\dfrac{tQ_{s}^{2}}{4}\left(1+2i\bm{p}_{0}\cdot\left(\bm{\bar{\Delta}}_{0}+\bm{u}_{x}-\bm{u}_{z}\right)\right)+\dfrac{t^{2}Q_{s}^{4}}{16}\left(\bm{\bar{\Delta}}_{0}+\bm{u}_{x}-\bm{u}_{z}\right)^{2}\right]\Biggr.\\
        &\hskip 0.5cm \Biggl.-\dfrac{t^{2}Q_{s}^{2}}{48}\left(\bm{\bar{\Delta}}_{0}^{2}+\bm{u}_{x}-\bm{u}_{z}\right)^{2}\Biggr] .\\
    \end{split}
    \label{eq.5.23}
\end{equation}
Making the change of variables $\displaystyle \bm{\bar{\Delta}}_{0}\to\bm{\Delta}_{s}-\bm{u}_{x}+\bm{u}_{z}$ in the last expression, we recover  the same result as for the first term,~\eqref{eq.5.22}, but with an overall minus sign. Hence, when we add the two expressions they cancel each other resulting in a zero contribution for the before-before term to the cross section at first-order:
\begin{equation}
    \left\langle\overline{\sum}\left|\mathcal{M}_{bef}\right|^{2}\right\rangle^{\left(1\right)}=\left\langle\overline{\sum}\left|\mathcal{M}_{bef}\right|^{2}\right\rangle^{\left(20\right)}+\left\langle\overline{\sum}\left|\mathcal{M}_{bef}\right|^{2}\right\rangle^{\left(02\right)}=0.
    \label{eq.5.24}
\end{equation}

The vanishing of the first order non-eikonal corrections coming from the relaxation of the shockwave approximation  can checked directly in the corresponding expressions in~\cite{Altinoluk:2022jkk} for homogeneous isotropic models for the target averages as the ones that we employ -- see also the discussions in~\cite{Agostini:2024xqs}. This null result was also found for single gluon production in proton-nucleus collisions in~\cite{Altinoluk:2014oxa} for the same kind of next-to-eikonal corrections.

\subsection{$\displaystyle 2^{nd}$-order (Next-to-Next-to-Eikonal)}
Following the same procedure we obtain the second-order contributions in the shockwave expansion, i.e., at next-to-next-to-eikonal accuracy, for each term. Using the same procedure and defining ${\bf q} = {\bf p} + {\bf k}$ and the relative momentum as ${\bf P} = ({\bf p} - {\bf k})/2$, we find
\begin{equation}
    \begin{split}
        &\left\langle\overline{\sum}\left|\mathcal{M}_{bef}\right|^{2}\right\rangle^{\left(2\right)}=\zeta^{2}\left(1-\zeta\right)^{2}N_{c}e_{q}^{2}e^{2}S_{\perp}\dfrac{Q_{s}^{4}}{360\zeta^{2}\left(1-\zeta\right)^{2}}\left(\dfrac{L^{+}}{q^{+}}\right)^{2}\int d^{2}\Delta_{b} \ e^{-i\bm{q}\cdot\bm{\Delta_{b}}}\\
        &\times\int d^{2}r \ d^{2}\bar{r} \ e^{-i\bm{P}\cdot\left(\bm{r}-\bm{\bar{r}}\right)}\int \dfrac{d^{2}p_{0}}{\left(2\pi\right)^{2}}\dfrac{d^{2}\bar{p}_{0}}{\left(2\pi\right)^{2}}e^{i\bm{p_{0}}\cdot\bm{r}}e^{-i\bm{\bar{p}_{0}}\cdot\bm{\bar{r}}}\dfrac{F_{T;L}\left(\bm{p_{0}},\bm{\bar{p}_{0}}\right)}{\left(p^{2}_{0}+\epsilon^{2}\right)\left(\bar{p}_{0}^{2}+\epsilon^{2}\right)}\\
        &\times\dfrac{Q_{s}^{2}}{2}\left[\Delta_{b}^{2}-\dfrac{\left(\bm{r}-\bm{\bar{r}}\right)^{2}}{4}\right]\int^{1}_{0}dt \ e^{-\frac{Q_{s}^{2}\left(1-t\right)}{2}\Delta_{b}^{2}}e^{-\frac{Q_{s}^{2}\left(1-t\right)}{8}\left(\bm{r}-\bm{\bar{r}}\right)^{2}}e^{-\frac{Q_{s}^{2}t}{4}r^{2}}e^{-\frac{Q_{s}^{2}t}{4}\bar{r}^{2}} t^{2}\\
        &\times \Biggl\lbrace t\dfrac{\bar{p}_{0}^{2}}{Q_{s}^{2}}\left(120-45i\bm{\bar{p}_{0}}\cdot\bm{\bar{r}}-45i\bm{\bar{p}_{0}}\cdot\bm{r}\right)-t^{2}\dfrac{15}{2}\Bigl[-8+3\left(\bm{\bar{p}_{0}}\cdot\bm{\bar{r}}\right)^{2}+2\bar{p}_{0}^{2}\bar{r}^{2}+6i\bm{\bar{p}_{0}}\cdot\bm{r}\Bigr.\Biggr.\\
        &\Bigl.+\bm{\bar{p}_{0}}\cdot\bm{\bar{r}}\left(14i+3\bm{\bar{p}_{0}}\cdot\bm{r}\right)+3\bar{p}_{0}^{2}\left(\bm{r}\cdot\bm{\bar{r}}\right)+\bar{p}_{0}^{2}r^{2}\Bigr]+\dfrac{3}{4}it^{3}Q_{s}^{2}\Bigl[\bar{r}^{2}\left(32i+15\bm{\bar{p}_{0}}\cdot\bm{\bar{r}}+5\bm{\bar{p}_{0}}\cdot\bm{r}\right)\Bigr.\\
        &\Biggl.\Bigl.+\left(25i+15\bm{\bar{p}_{0}}\cdot\bm{\bar{r}}\right)\left(\bm{r}\cdot\bm{\bar{r}}\right)+5r^{2}\left(i+\bm{\bar{p}_{0}}\cdot\bm{\bar{r}}\right)\Bigr]+\dfrac{5}{8}t^{4}Q_{s}^{4}\bar{r}^{2}\left(2\bar{r}^{2}+3\bm{r}\cdot\bm{\bar{r}}+r^{2}\right)\Biggr\rbrace
    \end{split}
\end{equation}
for the before-before contribution,
\begin{equation}
    \begin{split}
        &\left\langle\overline{\sum}\mathcal{M}_{bef}\mathcal{M}_{in\left(fact\right)}^{\ast}\right\rangle^{\left(2\right)}=-\dfrac{N_{c}e_{q}^{2}e^{2}}{2Q^{2}}\zeta\left(1-\zeta\right)S_{\perp}\int d^{2}\Delta_{b} \ e^{-i\bm{q}\cdot\bm{\Delta_{b}}}\int d^{2}r \ d^{2}\bar{r} \\
        &\times\ e^{-i\bm{P}\cdot\left(\bm{r}-\bm{\bar{r}}\right)}\int\dfrac{d^{2}p_{0}}{\left(2\pi\right)^{2}}\dfrac{d^{2}\bar{p}_{0}}{\left(2\pi\right)^{2}} \ e^{i\bm{p_{0}}\cdot\bm{r}} \ e^{-i\bm{\bar{p}_{0}}\cdot\bm{\bar{r}}} \ \dfrac{F_{T;L}\left(\bm{p_{0}},\bm{\bar{p}_{0}}\right)}{p_{0}^{2}+\epsilon^{2}}\int^{1}_{0}dt \ e^{-\frac{Q_{s}^{2}\left(1-t\right)}{2}\Delta_{b}^{2}} \ \\
        &\times\ e^{-\frac{Q_{s}^{2}\left(1-t\right)}{8}\left(\bm{r}-\bm{\bar{r}}\right)^{2}}e^{-\frac{Q_{s}^{2}t}{4}r^{2}}t\Biggl\lbrace\dfrac{\lambda^{4}}{2}+\dfrac{\lambda^{2}\kappa^{2}}{2\zeta\left(1-\zeta\right)}\biggl[\dfrac{t}{2}+\dfrac{\bar{p}_{0}^{2}}{Q_{s}^{2}}-i\dfrac{t}{2}\bm{\bar{p}_{0}}\cdot\bm{r}-\dfrac{t^{2}Q_{s}^{2}}{12}r^{2}\biggr]\Biggr\rbrace
    \end{split}
\end{equation}
for the factorisable part of the before-inside contribution,
\begin{equation}
    \begin{split}
        &\left\langle\overline{\sum}\mathcal{M}_{bef}\mathcal{M}_{in\left(Nfact\right)}^{\ast}\right\rangle^{\left(2\right)}=-\dfrac{N_{c}e_{q}^{2}e^{2}}{2Q^{2}}\zeta\left(1-\zeta\right)S_{\perp}\int d^{2}\Delta_{b} \ e^{-i\bm{q}\cdot\bm{\Delta_{b}}}\int d^{2}r \ d^{2}\bar{r} \\
        &\times e^{-i\bm{P}\cdot\left(\bm{r}-\bm{\bar{r}}\right)}\int\dfrac{d^{2}p_{0}}{\left(2\pi\right)^{2}}\dfrac{d^{2}\bar{p}_{0}}{\left(2\pi\right)^{2}} \ e^{i\bm{p_{0}}\cdot\bm{r}} \ e^{-i\bm{\bar{p}_{0}}\cdot\bm{\bar{r}}} \ \dfrac{F_{T;L}\left(\bm{p_{0}},\bm{\bar{p}_{0}}\right)}{p_{0}^{2}+\epsilon^{2}}\dfrac{Q_{s}^{2}}{2}\left[\Delta^{2}_{b}-\dfrac{\left(\bm{r}-\bm{\bar{r}}\right)^{2}}{4}\right]\\
        &\times\int^{1}_{0}dt\int^{1}_{t}dt^{\prime} \ e^{-\frac{Q_{s}^{2}\left(1-t^{\prime}\right)}{2}\Delta_{b}^{2}} \ e^{-\frac{Q_{s}^{2}\left(1-t^{\prime}\right)}{8}\left(\bm{r}-\bm{\bar{r}}\right)^{2}} \ e^{-\frac{Q_{s}^{2}t^{\prime}}{4}r^{2}} \ e^{-\frac{Q_{s}^{2}\left(t^{\prime}-t\right)}{4}\bar{r}^{2}}\Biggl\lbrace\dfrac{\lambda^{4}}{2}t\Biggr.\\
        &\Biggl.+\dfrac{\lambda^{2}\kappa^{2}}{2\zeta\left(1-\zeta\right)}\biggl[t^{\prime2}-\dfrac{t^{2}}{2}+t\dfrac{\bar{p}_{0}^{2}}{Q_{s}^{2}}+\dfrac{i}{2}\left(t^{2}-t^{\prime2}\right)\bm{\bar{p}_{0}}\cdot\bm{\bar{r}}+\dfrac{1}{24}\left(-2t^{3}+9tt^{\prime2}-7t^{\prime3}\right)Q_{s}^{2}\bar{r}^{2}\biggr.\Biggr.\\
        &\Biggl.\biggl.-i\dfrac{t^{\prime2}}{2}\bm{\bar{p}_{0}}\cdot\bm{r}+\dfrac{1}{24}\left(3t-5t^{\prime}\right)t^{\prime2}Q_{s}^{2}r^{2}+\dfrac{1}{8}t^{\prime2}\left(t^{\prime}-t\right)Q_{s}^{2}\left(\bm{r}-\bm{\bar{r}}\right)^{2}\biggr]\Biggr\rbrace
    \end{split}
\end{equation}
for the non-factorisable part of the before-inside contribution,
\begin{equation}
    \begin{split}
        &\left\langle\overline{\sum}\left|\mathcal{M}_{in}\right|^{2}\right\rangle^{\left(2\right)}=\dfrac{N_{c}e_{q}^{2}e^{2}}{2Q^{4}}S_{\perp}\lambda^{4}\int d^{2}\Delta_{b} \ e^{-i\bm{q}\cdot\bm{\Delta_{b}}}\int d^{2}r \ d^{2}\bar{r} \ e^{-i\bm{P}\cdot\left(\bm{r}-\bm{\bar{r}}\right)}\\
        &\times\int\dfrac{d^{2}p_{0}}{\left(2\pi\right)^{2}}\dfrac{d^{2}\bar{p}_{0}}{\left(2\pi\right)^{2}} \ e^{i\bm{p_{0}}\cdot\bm{r}} \ e^{-i\bm{\bar{p}_{0}}\cdot\bm{\bar{r}}} \ F_{T;L}\left(\bm{p_{0}},\bm{\bar{p}_{0}}\right)\\
        &\times\int^{1}_{0}dt\int^{1}_{t}dt^{\prime}\Biggl\lbrace e^{-\frac{Q_{s}^{2}\left(1-t^{\prime}\right)}{2}\Delta_{b}^{2}} \ e^{-\frac{Q_{s}^{2}\left(1-t^{\prime}\right)}{8}\left(\bm{r}-\bm{\bar{r}}\right)^{2}} \ e^{-\frac{Q_{s}^{2}\left(t^{\prime}-t\right)}{4}r^{2}}+\dfrac{Q_{s}^{2}}{4}\left[\Delta_{b}^{2}-\dfrac{\left(\bm{r}-\bm{\bar{r}}\right)^{2}}{4}\right] \Biggr.\\
        &\Biggl.\times \int^{1}_{t^{\prime}}dt^{\prime\prime} \ e^{-\frac{Q_{s}^{2}\left(1-t^{\prime\prime}\right)}{2}\Delta_{b}^{2}} \ e^{-\frac{Q_{s}^{2}\left(1-t^{\prime\prime}\right)}{8}\left(\bm{r}-\bm{\bar{r}}\right)^{2}} \ e^{-\frac{Q_{s}^{2}\left(t^{\prime\prime}-t\right)}{4}r^{2}} \ e^{-\frac{Q_{s}^{2}\left(t^{\prime\prime}-t^{\prime}\right)}{4}\bar{r}^{2}}\Biggr\rbrace
    \end{split}
\end{equation}
for the inside-inside contribution,
\begin{equation}
    \begin{split}
        &\left\langle\overline{\sum}\mathcal{M}_{bef}\mathcal{M}_{aft}^{\ast}\right\rangle^{\left(2\right)}=\zeta^{2}\left(1-\zeta\right)^{2}N_{c}e_{q}^{2}e^{2}S_{\perp}\int d^{2}\Delta_{b} \ e^{-i\bm{q}\cdot\bm{\Delta_{b}}}\int d^{2}r \ d^{2}\bar{r} \ e^{-i\bm{P}\cdot\left(\bm{r}-\bm{\bar{r}}\right)}\\
        &\times\int\dfrac{d^{2}p_{0}}{\left(2\pi\right)}\dfrac{d^{2}\bar{p}_{0}}{\left(2\pi\right)^{2}} \ e^{i\bm{p_{0}}\cdot\bm{r}} \ e^{-i\bm{\bar{p}_{0}}\cdot\bm{\bar{r}}}\dfrac{F_{T;L}\left(\bm{p_{0}},\bm{\bar{p}_{0}}\right)}{\left(p^{2}_{0}+\epsilon^{2}\right)\left(\bar{p}^{2}_{0}+\epsilon^{2}\right)} \ e^{-\frac{Q_{s}^{2}}{4}\bar{r}^{2}}\\
        &\times\Biggl\lbrace\dfrac{\lambda^{2}}{8}+\dfrac{\lambda^{2}\kappa^{2}}{4\zeta\left(1-\zeta\right)}\biggl[\dfrac{1}{2}+\dfrac{P^{2}}{Q_{s}^{2}}+\dfrac{i}{2}\bm{P}\cdot\bm{\bar{r}}-\dfrac{Q_{s}^{2}}{12}\bar{r}^{2}\biggr]\Biggr.\\
        &+\dfrac{\kappa^{4}}{360\zeta^{2}\left(1-\zeta\right)^{2}}\biggl[\dfrac{75}{4}+45\dfrac{P^{4}}{Q_{s}^{4}}+75\dfrac{P^{2}}{Q_{s}^{2}}+i45\dfrac{P^{2}}{Q_{s}^{2}}\bm{P}\cdot\bm{\bar{r}}-\dfrac{45}{4}\left(\bm{P}\cdot\bm{\bar{r}}\right)^{2}\biggr.\\
        &\Biggl.\biggl.-\dfrac{15}{2}P^{2}\bar{r}^{2}+i\dfrac{165}{4}\bm{P}\cdot\bm{\bar{r}}-i\dfrac{15}{4}Q_{s}^{2}\bar{r}^{2}\bm{P}\cdot\bm{\bar{r}}-\dfrac{27}{4}Q_{s}^{2}\bar{r}^{2}+\dfrac{5}{16}Q_{s}^{2}\bar{r}^{4}\biggr]\Biggr\rbrace
    \end{split}
\end{equation}
for the before-after contribution and, finally,
\begin{equation}
    \begin{split}
        &\left\langle\overline{\sum}\mathcal{M}_{in}\mathcal{M}_{aft}^{\ast}\right\rangle^{\left(2\right)}=-\dfrac{N_{c}e_{q}^{2}e^{2}}{2Q^{2}}\zeta\left(1-\zeta\right)S_{\perp}\int d^{2}\Delta_{b} \ e^{-i\bm{q}\cdot\bm{\Delta_{b}}}\int d^{2}r \ d^{2}\bar{r} \ e^{-i\bm{P}\cdot\left(\bm{r}-\bm{\bar{r}}\right)}\\
        &\times\int\dfrac{d^{2}p_{0}}{\left(2\pi\right)}\dfrac{d^{2}\bar{p}_{0}}{\left(2\pi\right)^{2}} \ e^{i\bm{p_{0}}\cdot\bm{r}} \ e^{-i\bm{\bar{p}_{0}}\cdot\bm{\bar{r}}}\left(\dfrac{F_{T;L}\left(\bm{p_{0}},\bm{\bar{p}_{0}}\right)}{p_{0}^{2}+\epsilon^{2}}\right)\int^{1}_{0}dt\left(1-t\right)e^{-\frac{Q_{s}^{2}\left(1-t\right)}{4}\bar{r}^{2}}\\
        &\times\Biggl\lbrace\dfrac{\lambda^{4}}{2}+\dfrac{\lambda^{2}\kappa^{2}}{2\zeta\left(1-\zeta\right)}\biggl[\dfrac{P^{2}}{Q_{s}^{2}}+\dfrac{\left(1-t\right)}{2}+\dfrac{i}{2}\left(1-t\right)\bm{P}\cdot\bm{\bar{r}}-\dfrac{1}{12}\left(1-t\right)^{2}Q_{s}^{2}\bar{r}^{2}\biggr]\Biggr\rbrace
    \end{split}
\end{equation}
for the inside-after contribution.

\section{Correlation limit}
\label{sec:correl}
The correlation limit corresponds to the kinematic region in which the relative momentum is much larger than the momentum imbalance, $\displaystyle\left|\bm{P}\right|\gg\left|\bm{q}\right|$, $\displaystyle \bm{P}=\left(\bm{p}-\bm{k}\right)/2$, $\displaystyle\bm{q}=\bm{p}+\bm{q}$, and than the saturation scale $\displaystyle\left|\bm{P}\right|\gg Q_{s}$. On the other hand, in this process two types of correction may arise, density corrections in powers of $Q_s/\left|\bm{P}\right|$ and kinematic twists in powers of $\left|\bm{q}\right|/\left|\bm{P}\right|$. Our procedure is somewhat different to the standard method for obtaining the correlation limit, based on an expansion of the integrand of the expressions for values of the transverse position variable conjugate to $|\bm{P}|$ much smaller than those of the transverse position variable conjugate to $|\bm{q}|$. Instead, we make an expansion in powers of $1/|\bm{P}|$ and, in the harmonic oscillator approximation that we use, $|\bm{q}|\sim Q_s$ -- thus the expansion becomes effectively one in a single parameter $Q_s/|\bm{P}|$.

In this section we limit ourselves to the expression for a longitudinal photon, providing those corresponding to a transverse photon in appendices.

\subsection{$\displaystyle 0^{th}-order$ (Eikonal)}

The calculation of the correlation limit for the longitudinal photon is performed in appendix~\ref{app:0corrlim}. The final result reads
\begin{equation}
    \left\langle\overline{\sum}\left|\mathcal{M}\right|^{2}\right\rangle^{\left(0\right)}=2N_{c}e_{q}^{2}e^{2}\dfrac{\zeta\left(1-\zeta\right)}{Q^{2}}S_{\perp}\int d^{2}\Delta_{b} e^{-i\bm{q}\cdot\bm{\Delta_{b}}}\left[\dfrac{16a^{4}}{\Delta_{b}^{2}Q_{s}^{2}}\left(1-e^{-\frac{Q_{s}^{2}\Delta_{b}^{2}}{2}}\right)x^{6}+\mathcal{O}\left(x^{8}\right)\right],
    \label{eq:0corrlimexpr}
\end{equation}
with $\displaystyle x=Q_{s}/|\mathbf{P}|$ and $\displaystyle a^{2}=\epsilon^{2}/Q_{s}^{2}$.
Introducing the hard partonic cross section
\begin{equation}
    H_{\gamma^{\ast}_{L}g\to q\bar{q}}=\alpha_{s}\alpha_{em}e_{q}^{2}\zeta^{2}\left(1-\zeta\right)^{2}\dfrac{8\epsilon^{2}P^{2}}{\left(P^{2}+\epsilon^{2}\right)^{4}},
\end{equation}
we get
\begin{equation}
    \begin{split}
        \left\langle\overline{\sum}\left|\mathcal{M}\right|^{2}\right\rangle^{\left(0\right)}&=\dfrac{2\zeta\left(1-\zeta\right)}{Q^{2}\zeta^{2}\left(1-\zeta\right)^{2}}\dfrac{N_{c}S_{\perp}}{\alpha_{s}}4\pi H_{\gamma^{\ast}_{L}g\to q\bar{q}}\dfrac{Q_{s}^{6}}{8\epsilon^{2}}\dfrac{\left(1+x^{2}a^{2}\right)^{4}}{x^{6}}\\
        &\hskip 0.5cm \times\int d^{2}\Delta_{b} \ e^{-i\bm{q}\cdot\bm{\Delta_{b}}}\left[\dfrac{16a^{4}}{\Delta_{b}^{2}Q_{s}^{2}}\left(1-e^{-\frac{Q_{s}^{2}\Delta_{b}^{2}}{2}}\right)x^{6}+\mathcal{O}\left(x^{8}\right)\right]\\
        &=H_{\gamma^{\ast}_{L}g\to q\bar{q}}\left[\dfrac{16\pi N_{c}}{\alpha_{s}}S_{\perp}\int d^{2}\Delta_{b} \ \dfrac{e^{-i\bm{q}\cdot\bm{\Delta_{b}}}}{\Delta_{b}^{2}}\left(1-e^{-\frac{Q_{s}^{2}\Delta_{b}^{2}}{2}}\right)+\mathcal{O}\left(\dfrac{Q_{s}^{2}}{P^{2}}\right)\right].
    \end{split}
\end{equation}
This result coincides with that in~\cite{Dominguez:2011wm}.

\subsection{$\displaystyle 2^{nd}-order$ (Next-to-Next-to-Eikonal)}
With a similar calculation but more cumbersome than in the previous case, we obtain the correlation limit for the second term in the shockwave expansion:
\begin{equation}
    \begin{split}
        \left\langle\overline{\sum}\left|\mathcal{M}\right|^{2}\right\rangle^{\left(2\right)}=&H_{\gamma^{\ast}_{L}g\to q\bar{q}}\dfrac{P^{2}}{Q_{s}^{2}}\left(\dfrac{L}{q^{+}}\right)^{2}\dfrac{2}{Q_{s}^{2}Q^{2}\zeta^{3}\left(1-\zeta\right)^{3}}\Biggl\lbrace\dfrac{\pi N_{c}}{\alpha_{s}}S_{\perp}\int d^{2}\Delta_{b}\dfrac{e^{-i\bm{q}\cdot\bm{\Delta_{b}}}}{\Delta_{b}^{10}}\Biggl.\\
        &\times\biggl[1920\left(1-e^{-\frac{Q_{s}^{2}\Delta_{b}^{2}}{2}}\right)-960\Delta_{b}^{2}Q_{s}^{2}+240\Delta_{b}^{4}Q_{s}^{4}-40\Delta_{b}^{6}Q_{s}^{6}\biggr.\\
        &\hskip 0.5cm \Biggl.\biggl.+5\Delta_{b}^{8}Q_{s}^{8}-\dfrac{1}{2}\Delta_{b}^{10}Q_{s}^{10}-\dfrac{\zeta\left(1-\zeta\right)}{24}\Delta_{b}^{10}Q_{s}^{8}Q^{2}\biggr]+\mathcal{O}\left(\dfrac{Q_{s}^{2}}{P^{2}}\right)\Biggr\rbrace.
    \end{split}
\end{equation}
We note that this equation does not diverge: the expansion of the exponential inside the squared brackets exactly cancels the first terms until order $Q_s^{10}$.

\section{Conclusions and outlook}
\label{sec:conclu}

In this work we have computed the non-eikonal corrections to dijet production in DIS off a nucleus that stem solely from the relaxation of the shockwave approximation. For both longitudinally and transversely polarized photons, we provide general expressions in terms of two two-dimensional path integrals. In order to proceed further, we use a given model, GBW or, equivalently, the harmonic oscillator approximation, for the target averages of Wilson lines.

We then expand the general expressions order by order beyond the shockwave limit which provides the eikonal results. This shockwave expansion is performed up to next-to-next-to-eikonal accuracy. We recover the known results at eikonal accuracy. We observe that next-to-eikonal corrections to this observable vanish for the GBW model that we employ for target averages, as previously found for single gluon production in proton-nucleus collisions~\cite{Altinoluk:2014oxa}.

We then proceed to calculate the back-to-back of correlation limit or our expressions. Our expansion includes both density corrections in $Q_s/|\bm{P}|$ and kinematic twists $|\bm{q}|/|\bm{P}|$. We observe that both corrections become mixed for the GBW model that we employ.

Several avenues, left for future studies, could be the outlook of this work. On the one hand,  further investigation of the TMDs beyond eikonal order could be done on the basis of our results, although restricted to the GBW model for target averages. On the other hand, a numerical implementation with comparison of the all-order expressions with those at next-to-next-to-eikonal accuracy could be useful for phenomenological applications, as well as the comparison with other non-eikonal corrections and eikonal results at higher order in the strong coupling.

\acknowledgments

We thank Tolga Altinoluk and Guillaume Beuf for useful discussions. This research was supported by European Research Council project ERC-2018-ADG-835105 YoctoLHC, by Xunta de Galicia (CIGUS Network of Research Centres), by European Union ERDF, and by the Spanish Research State Agency under projects PID20231527\-62NB---I00 and CEX2023-001318-M financed by MICIU/AEI/10.13039/501100011033. The
work of Adrián Romero González is supported by grant PID2023‐152762NB‐I00 funded by
MICIU/AEI/10.13039/\-501100011033 and by ERDF/EU. This project received funding from the École Polytechnique Foundation.

\appendix

\section{Useful properties}
\label{sec:appC}
From~\eqref{eq:functionfho} we can calculate an integral that appears recurrently in the main calculations. Let $\displaystyle f\left(\bm{u}_{i},\bm{v}_{i}\right)$ be a function of the initial variables. Then
\begin{equation}
    \begin{split}
        \mathcal{A}_f\left(\bm{u_{i}},\bm{v_{i}};\Delta t^{+}\right)=&\int d^{2}u_{f}d^{2}u_{i}d^{2}v_{f}d^{2}v_{i} \ e^{-i\bm{p}\cdot\bm{u_{f}}}\mathcal{F}_{p^{+}}\left(\bm{u_{f}},\bm{u_{i}},\bm{v_{f}},\bm{v_{i}};\Delta t^{+}\right)f\left(\bm{u_{i}},\bm{v_{i}}\right)\\
        =&\int d^{2}u_{f}d^{2}u_{i}d^{2}v_{f}d^{2}v_{i} \ e^{-i\bm{p}\cdot\bm{u_{f}}}\left(\dfrac{p^{+}}{2\pi i\Delta t^{+}}\right)^{2}\exp\left\lbrace\dfrac{ip^{+}}{\Delta t^{+}}\Delta\bm{u}\cdot\Delta\bm{v}\right\rbrace\\
        &\times \exp\left\lbrace-\dfrac{Q_{s}^{2}}{12L^{+}}\Delta t^{+}\left(u_{f}^{2}+u_{i}^{2}+\bm{u_{f}}\cdot\bm{u_{i}}\right)\right\rbrace f\left(\bm{u_{i}},\bm{v_{i}}\right).
    \end{split}
    \label{eq.C.1}
\end{equation}
Integrating $\displaystyle \bm{v_{f}}$ and then $\displaystyle \bm{u_{i}}$, we get
\begin{equation}
    \begin{split}
        \mathcal{A}_f\left(\bm{u_{i}},\bm{v_{i}};\Delta t^{+}\right)&=\int d^{2}u_{f}d^{2}u_{i}d^{2}v_{i} \ e^{-i\bm{p}\cdot\bm{u_{f}}}\delta^{\left(2\right)}\left(\Delta\bm{u}\right)\exp\left\lbrace-\dfrac{ip^{+}}{\Delta t^{+}}\Delta\bm{u}\cdot\bm{v_{i}}\right\rbrace\\
        &\times\exp\left\lbrace-\dfrac{Q^{2}_{s}}{12L^{+}}\Delta t^{+}\left(u_{f}^{2}+u_{i}^{2}+\bm{u}_{f}\cdot\bm{u}_{i}\right)\right\rbrace f\left(\bm{u_{i}},\bm{v_{i}}\right)\\
        &=\int d^{2}u_{f} \ e^{-i\bm{p}\cdot\bm{u_{f}}}\exp\left(-\dfrac{Q^{2}_{s}}{4L^{+}}\Delta t^{+}u^{2}_{f}\right)\int d^{2}v_{i} \ \biggl.f\left(\bm{u_{i}},\bm{v_{i}}\right)\biggr|_{\bm{u_{i}}=\bm{u_{f}}}\\
        &=\int d^{2}u_{f} \ e^{-i\bm{p}\cdot\bm{u_{f}}} \ \mathcal{P}\left(\bm{u_{f}};\Delta t^{+}\right)\int d^{2}v_{i} \ \biggl.f\left(\bm{u_{i}},\bm{v_{i}}\right)\biggr|_{\bm{u_{i}}=\bm{u_{f}}},
    \end{split}
    \label{eq.C.2}
\end{equation}
where we have defined the broadening function $\displaystyle \mathcal{P}$.

We also make use of the normalization of the free propagator, i.e.,
\begin{equation}
\begin{split}
    &\int d^{2}m \ G_{0}\left(s,\bm{m}_{f}-\bm{m}|q^{+}\right)=\int d^{2}m\int\dfrac{d^{2}l}{\left(2\pi\right)^{2}}e^{-i\frac{l^{2}}{2q^{+}}s}e^{i\bm{l}\cdot\left(\bm{m}_{f}-\bm{m}\right)}\\
    &=\int\dfrac{d^{2}l}{\left(2\pi\right)^{2}}e^{-i\frac{l^{2}}{2q^{+}}s}e^{i\bm{l}\cdot\bm{m}_{f}}\left(2\pi\right)^{2}\delta^{\left(2\right)}\left(\bm{l}\right)=1.
    \end{split}
    \label{eq.C.9}
\end{equation}

\section{Calculation of the Before-Inside contribution}
\label{ap:befin}
Using~\eqref{eq.2.28} we can write the trace in~\eqref{eq.2.27} as
\begin{equation}
    \begin{split}
        &Tr\biggl\langle G\left(L^{+},{\bm{x}}; 0,{\bm{x_{0}}}|p^{+}\right)\bar{G}\left(L^{+}, {\bm{z}}; 0, {\bm{z_{0}}}|k^{+}\right)\\
        &\hskip 3cm \times\bar{G}^{\dagger}\left(L^{+}, \bm{\bar{z}}{}; \bar{y}^{+}, \bm{\bar{z}_{1}}{}|k^{+}\right)G^{\dagger}\left(L^{+},\bm{\bar{x}}{}; \bar{y}^{+},\bm{\bar{x}_{1}}{}|p^{+}\right)\biggr\rangle\\
        =&\int d^{2}x_{1}d^{2}z_{1}Tr\biggl\langle G\left(L^{+},{\bm{x}}; \bar{y}^{+},{\bm{x_{1}}}|p^{+}\right)G\left(\bar{y}^{+}, {\bm{x_{1}}}; 0, {\bm{x_{0}}}|p^{+}\right)\bar{G}\left(\bar{y}^{+},{\bm{z_{1}}}; 0,{\bm{z_{0}}}|k^{+}\right)\biggr.\\
        &\times\biggl.\bar{G}\left(L^{+},{\bm{z}}; \bar{y}^{+},{\bm{z_{1}}}|k^{+}\right)\bar{G}^{\dagger}\left(L^{+},\bm{\bar{z}}{}; \bar{y}^{+},\bm{\bar{z}_{1}}{}|k^{+}\right)G^{\dagger}\left(L^{+},\bm{\bar{x}}{}; \bar{y}^{+},\bm{\bar{x}_{1}}{}|p^{+}\right)\biggr\rangle\\
        =&\int d^{2}x_{1}d^{2}z_{1}\mathcal{B}_{p^{+}}\left(X,X_{1}|\bm{r}\right)\mathcal{B}_{p^{+}}\left(X_{1},X_{0}|\bm{u}\right)\mathcal{B}_{k^{+}}\left(Z_{1},Z_{0}|\bm{v}\right)\mathcal{B}_{k^{+}}\left(Z,Z_{1}|\bm{t}\right)\\
        &\times \mathcal{B}_{-k^{+}}\left(\bar{Z},\bar{Z}_{1}|\bm{\bar{t}}\right){B}_{-p^{+}}\left(\bar{X},\bar{X}_{1}|\bm{\bar{r}}\right)Tr\Bigl\langle U\left(L^{+},\bar{y}^{+};\bm{r}\right)U\left(\bar{y}^{+},0;\bm{u}\right)U^{\dagger}\left(\bar{y}^{+},0;\bm{v}\right)\\
        &\hskip 3cm \times U^{\dagger}\left(L^{+},\bar{y}^{+},\bm{t}\right)U\left(L^{+},\bar{y}^{+};\bm{\bar{t}}\right)U^{\dagger}\left(L^{+},\bar{y}^{+};\bm{\bar{r}}\right)\Bigr\rangle.
    \end{split}
    \label{eq.2.29}
\end{equation}
From~\eqref{eq.2.13} we know that only the correlations of fields at the same +-coordinate can contribute to the average of Wilson lines, so using this locality of the averages we can split this trace of Wilson into two traces, one for Wilson lines in the first region and the other for the second one:
\begin{equation}
    \begin{split}
        &Tr\Bigl\langle U\left(L^{+},\bar{y}^{+};\bm{r}\right)U\left(\bar{y}^{+},0;\bm{u}\right)U^{\dagger}\left(\bar{y}^{+},0;\bm{v}\right)U^{\dagger}\left(L^{+},\bar{y}^{+};\bm{t}\right)\\
        &\hskip 2cm\times U\left(L^{+},\bar{y}^{+};\bm{\bar{t}}\right)U^{\dagger}\left(L^{+},\bar{y}^{+};\bm{\bar{r}}\right)\Bigr\rangle\\
        =&\frac{1}{N_{c}}Tr\Bigl\langle U\left(\bar{y}^{+},0;\bm{u}\right)U^{\dagger}\left(\bar{y}^{+},0;\bm{v}\right)\Bigr\rangle Tr\Bigl\langle U^{\dagger}\left(L^{+},\bar{y}^{+};\bm{t}\right)U\left(L^{+},\bar{y}^{+};\bm{\bar{t}}\right)\\
        &\hskip 2cm \times U^{\dagger}\left(L^{+},\bar{y}^{+};\bm{\bar{r}}\right)U\left(L^{+},\bar{y}^{+};\bm{r}\right)\Bigr\rangle
        =N_{c} \ S^{\left(2\right)}_{\bm{uv}}\left(\bar{y}^{+},0\right)S^{\left(4\right)}_{\bm{rt\bar{t}\bar{r}}}\left(L^{+},\bar{y}^{+}\right).
    \end{split}
    \label{eq.2.30}
\end{equation}
Using this expression and solving the path integrals we get
\begin{equation}
    \begin{split}
        &Tr\biggl\langle G\left(L^{+},{\bm{x}}; 0,{\bm{x_{0}}}|p^{+}\right)\bar{G}\left(L^{+}, {\bm{z}}; 0, {\bm{z_{0}}}|k^{+}\right)\\
        &\hskip 2cm \times \bar{G}^{\dagger}\left(L^{+}, \bm{\bar{z}}{}; \bar{y}^{+}, \bm{\bar{z}_{1}}{}|k^{+}\right)G^{\dagger}\left(L^{+},\bm{\bar{x}}{}; \bar{y}^{+},\bm{\bar{x}_{1}}{}|p^{+}\right)\biggr\rangle\\
        =&N_{c}\int d^{2}x_{1}d^{2}z_{1}\biggl[\mathcal{B}_{p^{+}}\left(X,X_{1}|\bm{r}\right)\mathcal{B}_{k^{+}}\left(Z,Z_{1}|\bm{t}\right)\mathcal{B}_{-k^{+}}\left(\bar{Z},\bar{Z}_{1}|\bm{\bar{t}}\right)\\
        &\hskip 3cm \times{B}_{-p^{+}}\left(\bar{X},\bar{X}_{1}|\bm{\bar{r}}\right)S^{\left(4\right)}_{\bm{rt\bar{t}\bar{r}}}\left(L^{+},\bar{y}^{+}\right)\biggr]\\
        &\hskip 1cm \times\biggl[\mathcal{B}_{p^{+}}\left(X_{1},X_{0}|\bm{u}\right)\mathcal{B}_{k^{+}}\left(Z_{1},Z_{0}|\bm{v}\right)S^{\left(2\right)}_{\bm{uv}}\left(\bar{y}^{+},0\right)\biggr]\\
        =&N_{c}\int d^{2}x_{1}d^{2}z_{1}\biggl[\mathcal{F}_{p^{+}}\left(\bm{u_{x}},\bm{v_{x}},\bm{u_{x1}},\bm{v_{x1}};L^{+}-\bar{y}^{+}\right)\mathcal{F}_{k^{+}}\left(\bm{u_{z}},\bm{v_{z}},\bm{u_{z1}},\bm{v_{z1}};L^{+}-\bar{y}^{+}\right)\biggr.
            \end{split}
\end{equation}
\begin{equation}
    \begin{split}
        +&\frac{Q^{2}_{s}}{2L^{+}}\int^{L^{+}}_{\bar{y}^{+}}ds^{+}\int d^{2}u_{xs}d^{2}v_{xs}d^{2}u_{zs}d^{2}v_{zs}\left(\bm{u_{xs}}\cdot\bm{u_{zs}}\right)\mathcal{F}_{p^{+}}\left(\bm{u_{x}},\bm{v_{x}},\bm{u_{xs}},\bm{v_{xs}};L^{+}-s^{+}\right)\\
        \times&\biggl.\mathcal{F}_{k^{+}}\left(\bm{u_{z}},\bm{v_{z}},\bm{u_{zs}},\bm{v_{zs}};L^{+}-s^{+}\right)\mathcal{K}\left(\bm{m_{s}}-\bm{m_{1}},\bm{\Delta_{s}},\bm{\Delta_{1}};s^{+}-\bar{y}^{+}\right)\\
        \times &\mathcal{K}^{\ast}\left(\bm{\bar{m}_{s}}-\bm{\bar{m}_{1}},\bm{\bar{\Delta}_{s}},\bm{\bar{\Delta}_{1}};s^{+}-\bar{y}^{+}\right)\biggr] \biggl[\mathcal{K}\left(\bm{m_{1}}-\bm{m_{0}},\bm{\Delta_{1}},\bm{\Delta_{0}};\bar{y}^{+}\right)\biggr].\notag
    \end{split}
    \label{eq.2.31}
\end{equation}
Plugging this expression into~\eqref{eq.2.27} and using~\eqref{eq.C.2}, we get
\begin{equation}
    \begin{split}
        &\left\langle\overline{\sum}\mathcal{M}_{bef}\cdot\mathcal{M}_{in}^{\ast}\right\rangle=\dfrac{-iN_{c}e_{q}^{2}e^{2}}{2q^{+}}\zeta\left(1-\zeta\right)\int^{L^{+}}_{0}d\bar{y}^{+} \ e^{iq^{-}\bar{y}^{+}}\int\dfrac{d^{2}p_{0}}{\left(2\pi\right)^{2}}\dfrac{d^{2}\bar{p}_{0}}{\left(2\pi\right)^{2}}\\
        &\times \left(\dfrac{F_{T;L}\left(\bm{p_{0}},\bm{\bar{p}_{0}}\right)}{p^{2}_{0}+\epsilon^{2}}\right)\int d^{2}u_{x}d^{2}u_{z} \ e^{-i\bm{p}\cdot\bm{u_{x}}} e^{-i\bm{k}\cdot\bm{u_{z}}}\Biggl[\mathcal{P}\left(\bm{u}_{x};L^{+}-\bar{y}^{+}\right)\mathcal{P}\left(\bm{u}_{z};L^{+}-\bar{y}^{+}\right)\\
        &\times\int d^{2}v_{x1}d^{2}v_{z1}\int d^{2}\Delta_{0}d^{2}m_{0} \ e^{i\bm{p}_{0}\cdot\bm{\Delta_{0}}}e^{-i\bm{\bar{p}}_{0}\cdot\bm{\bar{\Delta}_{1}}} \ G_{0}\left(\bm{m_{1}}-\bm{m_{0}};\bar{y}^{+}|q^{+}\right)\\
        &\hskip 4cm \times \mathcal{J}\left(\bm{\Delta_{1}},\bm{\Delta_{0}};\bar{y}^{+}\right)\biggr|_{\left(\bm{u_{x1}},\bm{u_{z1}}\right)=\left(\bm{u_{x}},\bm{u_{z}}\right)}\Biggr.\\
        &\biggl.+\dfrac{Q_{s}^{2}}{2L^{+}}\left(\bm{u_{x}}\cdot\bm{u_{z}}\right)\int^{L^{+}}_{\bar{y}^{+}}ds^{+}\mathcal{P}\left(\bm{u_{x}};L^{+}-s^{+}\right)\mathcal{P}\left(\bm{u_{z}};L^{+}-s^{+}\right)\int d^{2}v_{xs}d^{2}v_{zs}\\
        &\times \int d^{2}\Delta_{0}d^{2}m_{0}d^{2}\bar{\Delta}_{1}d^{2}\bar{m}_{1}d^{2}\Delta_{1}d^{2}m_{1} \ e^{i\bm{p}_{0}\cdot\bm{\Delta_{0}}}e^{-i\bm{\bar{p}}_{0}\cdot\bm{\bar{\Delta}_{1}}} \ G_{0}\left(\bm{m_{s}}-\bm{m_{1}};s^{+}-\bar{y}^{+}|q^{+}\right)\\
        &\times\Biggl.\mathcal{J}\left(\bm{\Delta_{s}},\bm{\Delta_{1}};s^{+}-\bar{y}^{+}\right)G^{\ast}_{0}\left(\bm{\bar{m}_{s}}-\bm{\bar{m}_{1}};s^{+}-\bar{y}^{+}|q^{+}\right)\mathcal{J}^{\ast}\left(\bm{\bar{\Delta}_{s}},\bm{\bar{\Delta}_{1}};s^{+}-\bar{y}^{+}\right)\\
        &\times G_{0}\left(\bm{m_{1}}-\bm{m_{0}};\bar{y}^{+}|q^{+}\right)\biggl.\mathcal{J}\left(\bm{\Delta_{1}},\bm{\Delta_{0}};\bar{y}^{+}\right)\biggr|_{\left(\bm{u_{xs}},\bm{u_{zs}}\right)=\left(\bm{u_{x}},\bm{u_{z}}\right)}\Biggr].
    \end{split}
    \label{eq.2.32}
\end{equation}
With the change of integration variables $\displaystyle\left(\bm{v_{x1}},\bm{v_{z1}}\right)\to\left(\bm{\bar{\Delta}_{1}},\bm{\bar{m}_{1}}\right)$ in the first term and $\displaystyle \left(\bm{v_{xs}},\bm{v_{zs}}\right)\to\left(\bm{\Delta}_{s},\bm{m}_{s}\right)$ in the second, and using~\eqref{eq.C.9}, we get
\begin{equation}
    \begin{split}
        &\left\langle\overline{\sum}\mathcal{M}_{bef}\cdot\mathcal{M}_{in}^{\ast}\right\rangle=\dfrac{-iN_{c}e_{q}^{2}e^{2}}{2q^{+}}\zeta\left(1-\zeta\right)S_{\perp}\int^{L^{+}}_{0}d\bar{y}^{+} \ e^{iq^{-}\bar{y}^{+}}\int\dfrac{d^{2}p_{0}}{\left(2\pi\right)^{2}}\dfrac{d^{2}\bar{p}_{0}}{\left(2\pi\right)^{2}}\\
        &\times\left(\dfrac{F_{T;L}\left(\bm{p_{0}},\bm{\bar{p}_{0}}\right)}{p^{2}_{0}+\epsilon^{2}}\right)\int d^{2}u_{x}d^{2}u_{z}e^{-i\bm{p}\cdot\bm{u_{x}}}e^{-i\bm{k}\cdot\bm{u_{z}}}\Biggl[\mathcal{P}\left(\bm{u_{x}};L^{+}-\bar{y}^{+}\right)\mathcal{P}\left(\bm{u_{z}};L^{+}-\bar{y}^{+}\right)\\
        &\times\int d^{2}\bar{\Delta}_{1}\int d^{2}\Delta_{0} \ e^{i\bm{p}_{0}\cdot\bm{\Delta}_{0}}e^{-i\bm{\bar{p}}_{0}\cdot\bm{\bar{\Delta}}_{1}} \ \mathcal{J}\left(\bm{\bar{\Delta}_{1}}+\bm{u_{x}}-\bm{u_{z}},\bm{\Delta_{0}};\bar{y}^{+}\right)\Biggr.\\
        &+\dfrac{Q_{s}^{2}}{2L^{+}}\left(\bm{u_{x}}\cdot\bm{u_{z}}\right)\int^{L^{+}}_{\bar{y}^{+}}ds^{+}\mathcal{P}\left(\bm{u_{x}};L^{+}-s^{+}\right)\mathcal{P}\left(\bm{u_{z}};L^{+}-s^{+}\right) \\
        &\times \Biggl.\int d^{2}\Delta_{s}d^{2}\Delta_{0}d^{2}\bar{\Delta}_{1}d^{2}\Delta_{1}\ e^{i\bm{p}_{0}\cdot\bm{\Delta_{0}}}e^{-i\bm{\bar{p}_{0}}\cdot\bm{\bar{\Delta}_{1}}} \mathcal{J}\left(\bm{\Delta_{s}},\bm{\Delta_{1}};s^{+}-\bar{y}^{+}\right)\\
        &\times \mathcal{J}^{\ast}\left(\bm{\Delta_{s}}-\bm{u_{x}}+\bm{u_{z}},\bm{\bar{\Delta}_{1}};s^{+}-\bar{y}^{+}\right)\mathcal{J}\left(\bm{\Delta_{1}},\bm{\Delta_{0}};\bar{y}^{+}\right)\Biggr].
    \end{split}
    \label{eq.2.33}
\end{equation}

\section{Calculation of the Inside-Inside contribution}
\label{ap:inin}

The calculation proceeds in the following way. Once we have the separations in regions shown in figure~\ref{Fig :Diagrama 4}, we can compute the trace as before using~\eqref{eq.2.28} again\footnote{Note that in the abbreviated notation we have $\displaystyle X_{1}=\left(\bar{y}^{+},\bm{x_{1}}{}\right)$, $\displaystyle Z_{1}=\left(\bar{y}^{+},\bm{z_{1}}{}\right)$, $\displaystyle \bar{X}_{0}=\left(y^{+},\bm{\bar{x}_{0}}{}\right)$, $\displaystyle \bar{Z}_{0}=\left(y^{+},\bm{\bar{z}_{0}}{}\right)$.}:
\begin{equation}
    \begin{split}
        &Tr\biggl\langle G\left(L^{+},\bm{x}{};y^{+},\bm{x_{0}}{}|p^{+}\right)\bar{G}\left(L^{+},\bm{z}{};y^{+},\bm{z_{0}}{}|k^{+}\right)\\
        &\times\bar{G}^{\dagger}\left(L^{+},\bm{\bar{z}}{};\bar{y}^{+},\bm{\bar{z}_{1}}{}|k^{+}\right) G^{\dagger}\left(L^{+},\bm{\bar{x}}{};\bar{y}^{+},\bm{\bar{x}_{1}}{}|p^{+}\right)\biggr\rangle\\
        =&\Theta\left(\bar{y}^{+}-y^{+}\right)N_{c}\int d^{2}x_{1}d^{2}z_{1}\biggl[\mathcal{B}_{p^{+}}\left(X_{1},X_{0}|\bm{u}\right)\mathcal{B}_{k^{+}}\left(Z_{1},Z_{0}|\bm{v}\right)S^{\left(2\right)}_{\bm{uv}}\left(\bar{y}^{+},y^{+}\right)\biggr]\\
        &\times\biggl[\mathcal{B}_{p^{+}}\left(X,X_{1}|\bm{r}\right)\mathcal{B}_{k^{+}}\left(Z,Z_{1}|\bm{t}\right)\mathcal{B}_{-k^{+}}\left(\bar{Z},\bar{Z}_{1}|\bm{\bar{t}}\right){B}_{-p^{+}}\left(\bar{X},\bar{X}_{1}|\bm{\bar{r}}\right)S^{\left(4\right)}_{\bm{rt\bar{t}\bar{r}}}\left(L^{+},\bar{y}^{+}\right)\biggr]\\
        &+\Theta\left(y^{+}-\bar{y}^{+}\right)N_{c}\int d^{2}\bar{z}_{0}d^{2}\bar{x}_{0}\biggl[\mathcal{B}_{-p^{+}}\left(\bar{X}_{0},\bar{X}_{1}|\bm{u}\right)\mathcal{B}_{-k^{+}}\left(\bar{Z}_{0},\bar{Z}_{1}|\bm{v}\right)S^{\left(2\right)}_{\bm{vu}}\left(y^{+},\bar{y}^{+}\right)\biggr]\\
        &\times\biggl[\mathcal{B}_{p^{+}}\left(X,X_{0}|\bm{r}\right)\mathcal{B}_{k^{+}}\left(Z,Z_{0}|\bm{t}\right)\mathcal{B}_{-k^{+}}\left(\bar{Z},\bar{Z}_{0}|\bm{\bar{t}}\right){B}_{-p^{+}}\left(\bar{X},\bar{X}_{0}|\bm{\bar{r}}\right)S^{\left(4\right)}_{\bm{rt\bar{t}\bar{r}}}\left(L^{+},y^{+}\right)\biggr]\\
        =&N_{c}\int d^{2}x_{1}d^{2}z_{1}\Theta\left(\bar{y}^{+}-y^{+}\right)\mathcal{K}\left(\bm{x_{1}}{},\bm{x_{0}}{},\bm{z_{1}}{},\bm{z_{0}}{};\bar{y}^{+}-y^{+}\right)\\
        &\times\Biggl[\mathcal{F}_{p^{+}}\left(\bm{x},\bm{x_{1}},\bm{\bar{x}},\bm{\bar{x}_{1}};L^{+}-\bar{y}^{+}\right)\Biggr.\mathcal{F}_{k^{+}}\left(\bm{z},\bm{z_{1}},\bm{\bar{z}},\bm{\bar{z}_{1}};L^{+}-\bar{y}^{+}\right)\\
        &+\frac{Q^{2}_{s}}{2L^{+}}\int^{L^{+}}_{\bar{y}^{+}}ds^{+}\int d^{2}x_{s}d^{2}\bar{x}_{s}d^{2}z_{s}d^{2}\bar{z}_{s}\left(\bm{x_{s}}-\bm{\bar{x}_{s}}\right)\cdot\left(\bm{z_{s}}-\bm{\bar{z}_{s}}\right)\\
        &\times\mathcal{F}_{p^{+}}\left(\bm{x},\bm{x_{s}},\bm{\bar{x}},\bm{\bar{x}_{s}};L^{+}-s^{+}\right)\Biggl.\mathcal{F}_{k^{+}}\left(\bm{z},\bm{z_{s}},\bm{\bar{z}},\bm{\bar{z}_{s}};L^{+}-s^{+}\right)\\
        &\times\mathcal{K}\left(\bm{x_{s}},\bm{x_{1}},\bm{z_{s}},\bm{z_{1}};s^{+}-\bar{y}^{+}\right)\mathcal{K}^{\ast}\left(\bm{\bar{x}_{s}},\bm{\bar{x}_{1}},\bm{\bar{z}_{s}},\bm{\bar{z}_{1}};s^{+}-\bar{y}^{+}\right)\Biggr]\\
        &+N_{c}\int d^{2}\bar{z}_{0}d^{2}\bar{x}_{0}\Theta\left(y^{+}-\bar{y}^{+}\right)\mathcal{K}^{\ast}\left(\bm{\bar{x}_{0}}{},\bm{\bar{x}_{1}}{},\bm{\bar{z}_{0}}{},\bm{\bar{z}_{1}}{};y^{+}-\bar{y}^{+}\right)\\
        &\times\Biggl[\mathcal{F}_{p^{+}}\left(\bm{x},\bm{x_{0}},\bm{\bar{x}},\bm{\bar{x}_{0}};L^{+}-y^{+}\right)\Biggr.\mathcal{F}_{k^{+}}\left(\bm{z},\bm{z_{0}},\bm{\bar{z}},\bm{\bar{z}_{0}};L^{+}-y^{+}\right)\\
        & +\frac{Q^{2}}{2L^{+}}\int^{L^{+}}_{y^{+}}ds^{+}\int d^{2}x_{s}d^{2}\bar{x}_{s}d^{2}z_{s}d^{2}\bar{z}_{s}\left(\bm{x_{s}}-\bm{\bar{x}_{s}}\right)\cdot\left(\bm{z_{s}}-\bm{\bar{z}_{s}}\right)\\
        &\Biggr.\times\mathcal{F}_{p^{+}}\left(\bm{x},\bm{x_{s}},\bm{\bar{x}},\bm{\bar{x}_{s}};L^{+}-s^{+}\right)\mathcal{F}_{k^{+}}\left(\bm{z},\bm{z_{s}},\bm{\bar{z}},\bm{\bar{z}_{s}};L^{+}-s^{+}\right)\\
        &\times \mathcal{K}\left(\bm{x_{s}},\bm{x_{0}},\bm{z_{s}},\bm{z_{0}};s^{+}-y^{+}\right)\mathcal{K}^{\ast}\left(\bm{\bar{x}_{s}},\bm{\bar{x}_{0}},\bm{\bar{z}_{s}},\bm{\bar{z}_{0}};s^{+}-y^{+}\right)\Biggr].
    \end{split}
    \label{eq.2.36}
\end{equation}

Then,
\begin{equation}
    \begin{split}
        &\left\langle\overline{\sum}\left|\mathcal{M}_{in}\right|^{2}\right\rangle=\frac{N_{c}e_{q}^{2}e^{2}}{\left(2q^{+}\right)^{2}}\int^{L^{+}}_{0}dy^{+}\int^{L^{+}}_{0}d\bar{y}^{+}e^{-iq^{-}\left(y^{+}-\bar{y}^{+}\right)}\int\frac{d^{2}p_{0}}{\left(2\pi\right)^{2}}\frac{d^{2}\bar{p}_{0}}{\left(2\pi\right)^{2}}\\
        &\times F_{T;L}\left(\bm{p_{0}},\bm{\bar{p}_{0}}\right)\int d^{2}x_{0}d^{2}z_{0}d^{2}\bar{x}_{1}d^{2}\bar{z}_{1} \ e^{i\bm{p_{0}}\cdot\left(\bm{x_{0}}{}-\bm{z_{0}}{}\right)}e^{-i\bm{\bar{p}_{0}}\cdot\left(\bm{\bar{x}_{1}}{}-\bm{\bar{z}_{1}}{}\right)}\\
        &\times \int d^{2}xd^{2}zd^{2}\bar{x}d^{2}\bar{z} \ e^{-i\bm{p}\cdot\left(\bm{x}{}-\bm{\bar{x}}{}\right)}e^{-i\bm{k}\cdot\left(\bm{z}{}-\bm{\bar{z}}{}\right)}\\
        &\Biggl\lbrace\int d^{2}x_{1}d^{2}z_{1}\Theta\left(\bar{y}^{+}-y^{+}\right)\mathcal{K}\left(\bm{x_{1}}{},\bm{x_{0}}{},\bm{z_{1}}{},\bm{z_{0}}{};\bar{y}^{+}-y^{+}\right)\Biggr.\Biggr.\\
        &\times\biggl[\mathcal{F}_{p^{+}}\left(\bm{x},\bm{x_{1}},\bm{\bar{x}},\bm{\bar{x}_{1}};L^{+}-\bar{y}^{+}\right)\biggr.\mathcal{F}_{k^{+}}\left(\bm{z},\bm{z_{1}},\bm{\bar{z}},\bm{\bar{z}_{1}};L^{+}-\bar{y}^{+}\right)\\
        &+\frac{Q^{2}_{s}}{2L^{+}}\int^{L^{+}}_{\bar{y}^{+}}ds^{+}\int d^{2}x_{s}d^{2}\bar{x}_{s}d^{2}z_{s}d^{2}\bar{z}_{s}\left(\bm{x_{s}}-\bm{\bar{x}_{s}}\right)\cdot\left(\bm{z_{s}}-\bm{\bar{z}_{s}}\right)\\
        &\times\biggl.\mathcal{F}_{p^{+}}\left(\bm{x},\bm{x_{s}},\bm{\bar{x}},\bm{\bar{x}_{s}};L^{+}-s^{+}\right)\mathcal{F}_{k^{+}}\left(\bm{z},\bm{z_{s}},\bm{\bar{z}},\bm{\bar{z}_{s}};L^{+}-s^{+}\right)\\
        &\times\mathcal{K}\left(\bm{x_{s}},\bm{x_{1}},\bm{z_{s}},\bm{z_{1}};s^{+}-\bar{y}^{+}\right)\mathcal{K}^{\ast}\left(\bm{\bar{x}_{s}},\bm{\bar{x}_{1}},\bm{\bar{z}_{s}},\bm{\bar{z}_{1}};s^{+}-\bar{y}^{+}\right)\biggr]\\
        &+\int d^{2}\bar{z}_{0}d^{2}\bar{x}_{0}\Theta\left(y^{+}-\bar{y}^{+}\right)\mathcal{K}^{\ast}\left(\bm{\bar{x}_{0}}{},\bm{\bar{x}_{1}}{},\bm{\bar{z}_{0}}{},\bm{\bar{z}_{1}}{};y^{+}-\bar{y}^{+}\right)\\
        &\times \biggl[\mathcal{F}_{p^{+}}\left(\bm{x},\bm{x_{0}},\bm{\bar{x}},\bm{\bar{x}_{0}};L^{+}-y^{+}\right)\biggr.\mathcal{F}_{k^{+}}\left(\bm{z},\bm{z_{0}},\bm{\bar{z}},\bm{\bar{z}_{0}};L^{+}-y^{+}\right)\\
        &+\frac{Q^{2}_{s}}{2L^{+}}\int^{L^{+}}_{y^{+}}ds^{+}\int d^{2}x_{s}d^{2}\bar{x}_{s}d^{2}z_{s}d^{2}\bar{z}_{s}\left(\bm{x_{s}}-\bm{\bar{x}_{s}}\right)\cdot\left(\bm{z_{s}}-\bm{\bar{z}_{s}}\right)\\
        &\times \mathcal{F}_{p^{+}}\left(\bm{x},\bm{x_{s}},\bm{\bar{x}},\bm{\bar{x}_{s}};L^{+}-s\right)\Biggr.\biggr.\mathcal{F}_{k^{+}}\left(\bm{z},\bm{z_{s}},\bm{\bar{z}},\bm{\bar{z}_{s}};L^{+}-s\right)\\
        &\times \mathcal{K}\left(\bm{x_{s}},\bm{x_{0}},\bm{z_{s}},\bm{z_{0}};s^{+}-y^{+}\right)\mathcal{K}^{\ast}\left(\bm{\bar{x}_{s}},\bm{\bar{x}_{0}},\bm{\bar{z}_{s}},\bm{\bar{z}_{1}};s^{+}-y^{+}\right)\biggr]\Biggr\rbrace.
    \end{split}
    \label{eq.2.37}
\end{equation}

The second term, corresponding to the splitting ordering given by $\displaystyle \Theta\left(y^{+}-\bar{y}^{+}\right)$, is the complex conjugate of the first term. Then we can write the above expression using only the term in which $\displaystyle \bar{y}^{+}>y^{+}$ (see figure~\ref{Fig: Diagrama 4.1}):
\begin{equation}
    \begin{split}
        &\left\langle\overline{\sum}\left|\mathcal{M}_{in}\right|^{2}\right\rangle=\frac{N_{c}e_{q}^{2}e^{2}}{\left(2q^{+}\right)^{2}}\\
        &\times2 \ \mathfrak{Re}\Biggl\lbrace\int^{L^{+}}_{0}dy^{+}\int^{L^{+}}_{0}d\bar{y}^{+}e^{-iq^{-}\left(y^{+}-\bar{y}^{+}\right)}\Theta\left(\bar{y}^{+}-y^{+}\right)\int\frac{d^{2}p_{0}}{\left(2\pi\right)^{2}}\frac{d^{2}\bar{p}_{0}}{\left(2\pi\right)^{2}}\Biggr.\\
        &\times F_{T;L}\left(\bm{p_{0}},\bm{\bar{p}_{0}}\right)\int d^{2}x_{0}d^{2}z_{0}d^{2}\bar{x}_{1}d^{2}\bar{z}_{1} \ e^{i\bm{p_{0}}\cdot\left(\bm{x_{0}}{}-\bm{z_{0}}{}\right)}e^{-i\bm{\bar{p}_{0}}\cdot\left(\bm{\bar{x}_{1}}{}-\bm{\bar{z}_{1}}{}\right)}\\
        &\times \int d^{2}xd^{2}zd^{2}\bar{x}d^{2}\bar{z} \ e^{-i\bm{p}\cdot\left(\bm{x}{}-\bm{\bar{x}}{}\right)}e^{-i\bm{k}\cdot\left(\bm{z}{}-\bm{\bar{z}}{}\right)}\int d^{2}x_{1}d^{2}z_{1}\mathcal{K}\left(\bm{x_{1}}{},\bm{x_{0}}{},\bm{z_{1}}{},\bm{z_{0}}{};\bar{y}^{+}-y^{+}\right)
        \end{split}
\end{equation}
\begin{equation}
    \begin{split}
        &\times \biggl[\mathcal{F}_{p^{+}}\left(\bm{x},\bm{x_{1}},\bm{\bar{x}},\bm{\bar{x}_{1}};L^{+}-\bar{y}^{+}\right)\mathcal{F}_{k^{+}}\left(\bm{z},\bm{z_{1}},\bm{\bar{z}},\bm{\bar{z}_{1}};L^{+}-\bar{y}^{+}\right)\biggr.\\
        &+\frac{Q^{2}_{s}}{2L^{+}}\int^{L^{+}}_{\bar{y}^{+}}ds^{+}\int d^{2}x_{s}d^{2}\bar{x}_{s}d^{2}z_{s}d^{2}\bar{z}_{s} \left(\bm{x_{s}}-\bm{\bar{x}_{s}}\right)\cdot\left(\bm{z_{s}}-\bm{\bar{z}_{s}}\right)\mathcal{F}_{p^{+}}\left(\bm{x},\bm{x_{s}},\bm{\bar{x}},\bm{\bar{x}_{s}};L^{+}-s^{+}\right)\Biggl.\biggl.\\
        &\times\mathcal{F}_{k^{+}}\left(\bm{z},\bm{z_{s}},\bm{\bar{z}},\bm{\bar{z}_{s}};L^{+}-s^{+}\right)\mathcal{K}\left(\bm{x_{s}},\bm{x_{1}},\bm{z_{s}},\bm{z_{1}};s^{+}-\bar{y}^{+}\right)\mathcal{K}^{\ast}\left(\bm{\bar{x}_{s}},\bm{\bar{x}_{1}},\bm{\bar{z}_{s}},\bm{\bar{z}_{1}};s^{+}-\bar{y}^{+}\right)\biggr]\Biggr\rbrace.\notag
    \end{split}
    \label{eq.2.38}
\end{equation}
Writing everything in terms of the appropriate variables to solve the path integrals, see~\ref{sec:long_squ} (the definition of the variables is analogous to that in ~\eqref{eq:changevar},~\eqref{eq:changevarm},~\eqref{eq.2.22}), we get
\begin{equation}
    \begin{split}
        &\left\langle\overline{\sum}\left|\mathcal{M}_{in}\right|^{2}\right\rangle=\frac{N_{c}e_{q}^{2}e^{2}}{\left(2q^{+}\right)^{2}}\\
        &\times 2 \ \mathfrak{Re}\Biggl\lbrace\int^{L^{+}}_{0}dy^{+}\int^{L^{+}}_{0}d\bar{y}^{+}e^{-iq^{-}\left(y^{+}-\bar{y}^{+}\right)}\Theta\left(\bar{y}^{+}-y^{+}\right)\int\frac{d^{2}p_{0}}{\left(2\pi\right)^{2}}\frac{d^{2}\bar{p}_{0}}{\left(2\pi\right)^{2}}\Biggr.\\
        &\times F_{T;L}\left(\bm{p_{0}},\bm{\bar{p}_{0}}\right)\int d^{2}\Delta_{0} \ d^{2}m_{0} \ e^{i\bm{p_{0}}\cdot\bm{\Delta}_{0}}\int d^{2}u_{x1}d^{2}v_{x1}d^{2}u_{z1}d^{2}v_{z1} \ e^{-i\bm{\bar{p}_{0}}\cdot\bar{\bm{\Delta}}_{1}}\\
        &\times \int d^{2}u_{x}d^{2}v_{x}d^{2}u_{z}d^{2}v_{z} \ e^{-i\bm{p}\cdot\bm{u}_{x}}e^{-i\bm{k}\cdot\bm{u}_{z}}\mathcal{K}\left(\bm{m}_{1}-\bm{m}_{0},\bm{\Delta}_{1},\bm{\Delta}_{0};\bar{y}^{+}-y^{+}\right)\\
        &\times \biggl[\mathcal{F}_{p^{+}}\left(\bm{u}_{x},\bm{u}_{x1},\bm{v}_{x},\bm{v}_{x1};L^{+}-\bar{y}^{+}\right)\biggr.\mathcal{F}_{k^{+}}\left(\bm{u}_{z},\bm{u}_{z1},\bm{v}_{z},\bm{v}_{z1};L^{+}-\bar{y}^{+}\right)\\
        & +\frac{Q^{2}_{s}}{2L^{+}}\int^{L^{+}}_{\bar{y}^{+}}ds^{+}\int d^{2}u_{xs}d^{2}v_{xs}d^{2}u_{zs}d^{2}v_{zs}\left(\bm{u}_{xs}\cdot\bm{u}_{zs}\right)\\
        &\times\mathcal{F}_{p^{+}}\left(\bm{u}_{x},\bm{u}_{xs},\bm{v}_{x},\bm{v}_{xs};L^{+}-s^{+}\right)\Biggl.\biggl.\mathcal{F}_{k^{+}}\left(\bm{u}_{z},\bm{u}_{zs},\bm{v}_{z},\bm{v}_{zs};L^{+}-s^{+}\right)\\
        &\times\mathcal{K}\left(\bm{m}_{s}-\bm{m}_{1},\bm{\Delta}_{s},\bm{\Delta}_{1};s^{+}-\bar{y}^{+}\right)\mathcal{K}^{\ast}\left(\bar{\bm{m}}_{s}-\bar{\bm{m}}_{1},\bar{\bm{\Delta}}_{s},\bar{\bm{\Delta}}_{1};s^{+}-\bar{y}^{+}\right)\biggr]\Biggr\rbrace.
    \end{split}
    \label{eq.2.39}
\end{equation}
Using~\eqref{eq.C.2}, we can write
\begin{equation}
    \begin{split}
        &\left\langle\overline{\sum}\left|\mathcal{M}_{in}\right|^{2}\right\rangle=\frac{N_{c}e_{q}^{2}e^{2}}{\left(2q^{+}\right)^{2}}\\
        &\times2 \ \mathfrak{Re}\Biggl\lbrace\int^{L^{+}}_{0}dy^{+}\int^{L^{+}}_{0}d\bar{y}^{+}e^{-iq^{-}\left(y^{+}-\bar{y}^{+}\right)}\Theta\left(\bar{y}^{+}-y^{+}\right)\int\frac{d^{2}p_{0}}{\left(2\pi\right)^{2}}\frac{d^{2}\bar{p}_{0}}{\left(2\pi\right)^{2}}\Biggr.\\
        &\times F_{T;L}\left(\bm{p_{0}},\bm{\bar{p}_{0}}\right)\int d^{2}u_{x}d^{2}u_{z} \ e^{-i\bm{p}\cdot\bm{u}_{x}}e^{-i\bm{k}\cdot\bm{u}_{z}}\int d^{2}\Delta_{0} \ d^{2}m_{0} \ e^{i\bm{p_{0}}\cdot\bm{\Delta}_{0}}\\
        &\times\biggl[\mathcal{P}\left(\bm{u}_{x};L^{+}-\bar{y}^{+}\right)\mathcal{P}\left(\bm{u}_{z};L^{+}-\bar{y}^{+}\right)\biggr.\int d^{2}v_{x1}d^{2}v_{z1} \ e^{-i\bm{\bar{p}}_{0}\cdot\bm{\bar{\Delta}}_{1}} \\
        &\hskip 1cm \times \ G_{0}\left(\bm{m}_{1}-\bm{m}_{0};\bar{y}^{+}-y^{+}|q^{+}\right)\biggl.\mathcal{J}\left(\bm{\Delta}_{1},\bm{\Delta}_{0};\bar{y}^{+}-y^{+}\right)\biggr|_{\left(\bm{u}_{x1},\bm{u}_{z1}\right)=\left(\bm{u}_{x},\bm{u}_{z}\right)}
        \end{split}
\end{equation}
\begin{equation}
    \begin{split}
        &+\frac{Q^{2}_{s}}{2L^{+}}\left(\bm{u}_{x}\cdot\bm{u}_{z}\right)\int^{L^{+}}_{\bar{y}^{+}}ds^{+}\mathcal{P}\left(\bm{u}_{x};L^{+}-s^{+}\right)\mathcal{P}\left(\bm{u}_{z};L^{+}-s^{+}\right)\int d^{2}v_{xs}d^{2}v_{zs}\\
        &\times \int d^{2}\Delta_{1}d^{2}m_{1}d^{2}\bar{\Delta}_{1}d^{2}\bar{m}_{1} \ e^{-i\bm{\bar{p}}_{0}\cdot\bm{\bar{\Delta}}_{1}}G_{0}\left(\bm{m}_{s}-\bm{m}_{1};s^{+}-\bar{y}^{+}|q^{+}\right)\\
        &\times \mathcal{J}\left(\bm{\Delta}_{s},\bm{\Delta}_{1};s^{+}-\bar{y}^{+}\right)G^{\ast}_{0}\left(\bm{\bar{m}}_{s}-\bm{\bar{m}}_{1};s^{+}-\bar{y}^{+}|q^{+}\right)\mathcal{J}^{\ast}\left(\bm{\bar{\Delta}}_{s},\bm{\bar{\Delta}}_{1};s^{+}-\bar{y}^{+}\right)\\
        &\times\Biggl.\biggl.G_{0}\left(\bm{m}_{1}-\bm{m}_{0};\bar{y}^{+}-y^{+}|q^{+}\right)\biggl.\mathcal{J}\left(\bm{\Delta}_{1},\bm{\Delta}_{0};\bar{y}^{+}-y^{+}\right)\biggr|_{\left(\bm{u}_{xs},\bm{u}_{zs}\right)=\left(\bm{u}_{x},\bm{u}_{z}\right)}\biggr]\Biggr\rbrace.\notag
    \end{split}
    \label{eq.2.40}
\end{equation}

With the change of variables $\displaystyle \left(\bm{v}_{x1},\bm{v}_{z1}\right)\to\left(\bm{\bar{\Delta}}_{1},\bm{\bar{m}}_{1}\right)$ in the first term and $\displaystyle\left(\bm{v}_{xs},\bm{v}_{zs}\right)\to\left(\bm{\Delta}_{s},\bm{m}_{s}\right)$ in the second, and then integrating all the $\displaystyle \bm{m}$'s with~\eqref{eq.C.9}, we arrive at
\begin{equation}
    \begin{split}
        &\left\langle\overline{\sum}\left|\mathcal{M}_{in}\right|^{2}\right\rangle=\frac{N_{c}e_{q}^{2}e^{2}}{\left(2q^{+}\right)^{2}}S_{\perp}\\
        &\times2 \ \mathfrak{Re}\Biggl\lbrace\int^{L^{+}}_{0}dy^{+}\int^{L^{+}}_{0}d\bar{y}^{+}e^{-iq^{-}\left(y^{+}-\bar{y}^{+}\right)}\Theta\left(\bar{y}^{+}-y^{+}\right)\int\frac{d^{2}p_{0}}{\left(2\pi\right)^{2}}\frac{d^{2}\bar{p}_{0}}{\left(2\pi\right)^{2}}F_{T;L}\left(\bm{p_{0}},\bm{\bar{p}_{0}}\right)\Biggr.\\
        &\times \int d^{2}u_{x}d^{2}u_{z} \ e^{-i\bm{p}\cdot\bm{u}_{x}}e^{-i\bm{k}\cdot\bm{u}_{z}}\int d^{2}\Delta_{0} \ e^{i\bm{p_{0}}\cdot\bm{\Delta}_{0}}\int d^{2}\bar{\Delta}_{1} \ e^{-i\bm{\bar{p}}_{0}\cdot\bm{\bar{\Delta}}_{1}}\\      &\times\biggl[\mathcal{P}\left(\bm{u}_{x};L^{+}-\bar{y}^{+}\right)\mathcal{P}\left(\bm{u}_{z};L^{+}-\bar{y}^{+}\right)\biggr.\mathcal{J}\left(\bm{\bar{\Delta}}_{1}+\bm{u}_{x}-\bm{u}_{z},\bm{\Delta}_{0};\bar{y}^{+}-y^{+}\right)\\
        &+\frac{Q^{2}_{s}}{2L^{+}}\left(\bm{u}_{x}\cdot\bm{u}_{z}\right)\int^{L^{+}}_{\bar{y}^{+}}ds^{+}\mathcal{P}\left(\bm{u}_{x};L^{+}-s^{+}\right)\mathcal{P}\left(\bm{u}_{z};L^{+}-s^{+}\right)\\
        &\times\Biggl.\biggl.\int d^{2}\Delta_{s}\int d^{2}\Delta_{1}\mathcal{J}\left(\bm{\Delta}_{s},\bm{\Delta}_{1};s^{+}-\bar{y}^{+}\right)\mathcal{J}^{\ast}\left(\bm{\Delta}_{s}-\bm{u}_{x}+\bm{u}_{z},\bm{\bar{\Delta}}_{1};s^{+}-\bar{y}^{+}\right)\\
        &\hskip 4cm \times \mathcal{J}\left(\bm{\Delta}_{1},\bm{\Delta}_{0};\bar{y}^{+}-y^{+}\right)\biggr]\Biggr\rbrace.
    \end{split}
    \label{eq.2.41}
\end{equation}

\section{Calculation of the correlation limit of the eikonal contribution from a longitudinal photon}
\label{app:0corrlim}
We show explicitly the calculation of the correlation limit for the non-factorizable part of the before-before term for a longitudinal photon, second addend inside the square brackets in \eqref{eq.5.11} with $\displaystyle F_{T;L}\left(\bm{p_{0}},\bm{\bar{p}_{0}}\right)$ replaced by~\eqref{eq:FL}, which is the most involved one. Writing it in terms of  variables $\displaystyle \bm{P}=\left(\bm{p}-\bm{k}\right)/2$, $\displaystyle\bm{q}=\bm{p}+\bm{q}$, and with the change of integration variables
\begin{equation}
    \displaystyle\left(\bm{u_{x}},\bm{u_{z}},\bm{\Delta_{0}}\right)\to\left(\bm{\Delta_{b}},\bm{r},\bm{\bar{r}}\right)=\left(\left(\bm{u_{x}}+\bm{u_{z}}\right)/2,\bm{\Delta_{0}},\bm{\Delta_{0}}-\bm{u_{x}}+\bm{u_{z}}\right),
\end{equation}
we have
\begin{equation}
    \begin{split}
        &\left\langle\sum\left|\mathcal{M}_{bef}\right|^{2}_{\left(Nfact\right)}\right\rangle^{\left(0\right)}=\dfrac{\zeta\left(1-\zeta\right)}{Q^{2}}2N_{c}e_{q}^{2}e^{2}S_{\perp}\int d^{2}\Delta_{b} \ e^{-i\bm{q}\cdot\bm{\Delta_{b}}}\int d^{2}r \ d^{2}\bar{r} \\
        &\times e^{-i\bm{P}\cdot\left(\bm{r}-\bm{\bar{r}}\right)}\dfrac{Q_{s}^{2}}{2}\left[\Delta_{b}^{2}-\dfrac{\left(\bm{r}-\bm{\bar{r}}\right)^{2}}{4}\right]\int^{1}_{0}dt \ e^{-\frac{Q_{s}^{2}\left(1-t\right)}{2}\Delta_{b}^{2}} \ e^{-\frac{Q_{s}^{2}\left(1-t\right)}{8}\left(\bm{r}-\bm{\bar{r}}\right)^{2}} \\
        &\times e^{-\frac{Q_{s}^{2}t}{4}r^{2}} \ e^{-\frac{Q_{s}^{2}t}{4}\bar{r}}\int\dfrac{d^{2}p_{0}}{\left(2\pi\right)^{2}}\dfrac{d^{2}\bar{p}_{0}}{\left(2\pi\right)^{2}}e^{i\bm{p_{0}}\cdot\bm{r}} \ e^{-i\bm{\bar{p}_{0}}\cdot\bm{\bar{r}}}\left(\dfrac{p_{0}^{2}-\epsilon^{2}}{p^{2}_{0}+\epsilon^{2}}\right)\left(\dfrac{\bar{p}_{0}^{2}-\epsilon^{2}}{\bar{p}_{0}^{2}+\epsilon^{2}}\right)\\
        &=\dfrac{\zeta\left(1-\zeta\right)}{Q^{2}}2N_{c}e_{q}^{2}e^{2}S_{\perp}\int d^{2}\Delta_{b} \ e^{-i\bm{q}\cdot\bm{\Delta_{b}}}\int d^{2}r \ d^{2}\bar{r} \ e^{-i\bm{P}\cdot\left(\bm{r}-\bm{\bar{r}}\right)}\int^{1}_{0}dt \ e^{-\frac{Q_{s}^{2}\left(1-t\right)}{2}\Delta_{b}^{2}}\\
        &\times\left[\dfrac{Q_{s}^{2}\Delta_{b}^{2}}{2}+\dfrac{\partial}{\partial T}\right] \ e^{-\frac{Q_{s}^{2}T}{8}\left(\bm{r}-\bm{\bar{r}}\right)^{2}} \ e^{-\frac{Q_{s}^{2}t}{4}r^{2}} \ e^{-\frac{Q_{s}^{2}t}{4}\bar{r}}\\
        &\times\int\dfrac{d^{2}p_{0}}{\left(2\pi\right)^{2}}\dfrac{d^{2}\bar{p}_{0}}{\left(2\pi\right)^{2}}e^{i\bm{p_{0}}\cdot\bm{r}} \ e^{-i\bm{\bar{p}_{0}}\cdot\bm{\bar{r}}}\left(\dfrac{p_{0}^{2}-\epsilon^{2}}{p^{2}_{0}+\epsilon^{2}}\right)\left(\dfrac{\bar{p}_{0}^{2}-\epsilon^{2}}{\bar{p}_{0}^{2}+\epsilon^{2}}\right),
    \end{split}
    \label{eq:G2}
\end{equation}
with $T=1-t$.
Then we expand the exponentials with $\displaystyle\bm{r},\bm{\bar{r}}$ variables and rewrite them as derivatives of $\displaystyle e^{-i\bm{P}\cdot\left(\bm{r}-\bm{\bar{r}}\right)}$:
\begin{equation}
    \begin{split}
        &\left\langle\sum\left|\mathcal{M}_{bef}\right|^{2}_{\left(Nfact\right)}\right\rangle^{\left(0\right)}=\dfrac{\zeta\left(1-\zeta\right)}{Q^{2}}2N_{c}e_{q}^{2}e^{2}S_{\perp}\int d^{2}\Delta_{b} \ e^{-i\bm{q}\cdot\bm{\Delta_{b}}}\int d^{2}r \ d^{2}\bar{r} \\
        &\times e^{-i\bm{P}\cdot\left(\bm{r}-\bm{\bar{r}}\right)}\int^{1}_{0}dt \ e^{-\frac{Q_{s}^{2}\left(1-t\right)}{2}\Delta_{b}^{2}}\left[\dfrac{Q_{s}^{2}\Delta_{b}^{2}}{2}+\dfrac{\partial}{\partial T}\right]\sum_{ijk=0}^{\infty}\dfrac{1}{n!j!k!}\left(-\dfrac{Q_{s}^{2}T}{8}\right)^{n}\left(-\dfrac{Q_{s}^{2}t}{4}\right)^{j+k}\\
        &\times\left(\bm{r}-\bm{\bar{r}}\right)^{2n}r^{2j}\bar{r}^{2k}\int\dfrac{d^{2}p_{0}}{\left(2\pi\right)^{2}}\dfrac{d^{2}\bar{p}_{0}}{\left(2\pi\right)^{2}}e^{i\bm{p_{0}}\cdot\bm{r}} \ e^{-i\bm{\bar{p}_{0}}\cdot\bm{\bar{r}}}\left(\dfrac{p_{0}^{2}-\epsilon^{2}}{p^{2}_{0}+\epsilon^{2}}\right)\left(\dfrac{\bar{p}_{0}^{2}-\epsilon^{2}}{\bar{p}_{0}^{2}+\epsilon^{2}}\right)\\
        &=\dfrac{\zeta\left(1-\zeta\right)}{Q^{2}}2N_{c}e_{q}^{2}e^{2}S_{\perp}\int d^{2}\Delta_{b} \ e^{-i\bm{q}\cdot\bm{\Delta_{b}}}\int^{1}_{0}dt \ e^{-\frac{Q_{s}^{2}\left(1-t\right)}{2}\Delta_{b}^{2}}\\
        &\times\left[\dfrac{Q_{s}^{2}\Delta_{b}^{2}}{2}+\dfrac{\partial}{\partial T}\right]\sum_{ijk=0}^{\infty}\dfrac{1}{n!j!k!}\left(\dfrac{Q_{s}^{2}T}{8}\right)^{n}\left(\dfrac{Q_{s}^{2}t}{4}\right)^{j+k}\int d^{2}r \ d^{2}\bar{r} \ \nabla^{2n}_{\bm{P}}\\
        &\times\left[\left(\nabla^{2j}_{\bm{P}}e^{-i\bm{P}\cdot\bm{r}}\right)\left(\nabla^{2k}_{\bm{P}}e^{i\bm{P}\cdot\bm{r}}\right)\right]\int\dfrac{d^{2}p_{0}}{\left(2\pi\right)^{2}}\dfrac{d^{2}\bar{p}_{0}}{\left(2\pi\right)^{2}}e^{i\bm{p_{0}}\cdot\bm{r}} \ e^{-i\bm{\bar{p}_{0}}\cdot\bm{\bar{r}}}\left(\dfrac{p_{0}^{2}-\epsilon^{2}}{p^{2}_{0}+\epsilon^{2}}\right)\left(\dfrac{\bar{p}_{0}^{2}-\epsilon^{2}}{\bar{p}_{0}^{2}+\epsilon^{2}}\right).
    \end{split}
\end{equation}
As stated in the main text, our procedure is somewhat different to the standard method for obtaining the correlation limit, based on an expansion of the integrand in~\eqref{eq:G2} for $\left|\bm{r}-\bm{\bar{r}}\right|\ll |\bm{\Delta}_b|$. Instead, we make an expansion in powers of $1/|\bm{P}|$, and in the harmonic oscillator approximation that we use, $|\bm{q}|\sim Q_s$, thus the expansion becomes effectively one in a single parameter $Q_s/|\bm{P}|$. Now, integrating $\bm{r},\bm{\bar{r}}$ and then $\displaystyle\bm{p_{0}},\bm{\bar{p}_{0}}$, we get
\begin{equation}
    \begin{split}
        &\left\langle\sum\left|\mathcal{M}_{bef}\right|^{2}_{\left(Nfact\right)}\right\rangle^{\left(0\right)}=\dfrac{\zeta\left(1-\zeta\right)}{Q^{2}}2N_{c}e_{q}^{2}e^{2}S_{\perp}\int d^{2}\Delta_{b} \ e^{-i\bm{q}\cdot\bm{\Delta_{b}}}\int^{1}_{0}dt \ e^{-\frac{Q_{s}^{2}\left(1-t\right)}{2}\Delta_{b}^{2}}\\
        \\
        &\times\left[\dfrac{Q_{s}^{2}\Delta_{b}^{2}}{2}+\dfrac{\partial}{\partial T}\right]\sum_{ijk=0}^{\infty}\dfrac{1}{n!j!k!}\left(\dfrac{Q_{s}^{2}T}{8}\right)^{n}\left(\dfrac{Q_{s}^{2}t}{4}\right)^{j+k} \\
        &\times\nabla^{2n}_{\bm{P}}\left[\nabla^{2j}_{\bm{P}}\left(\dfrac{P^{2}-\epsilon^{2}}{P^{2}+\epsilon^{2}}\right)\nabla^{2k}_{\bm{P}}\left(\dfrac{P^{2}-\epsilon^{2}}{P^{2}+\epsilon^{2}}\right)\right].
    \end{split}
\end{equation}
In order to take the limit $\displaystyle\left|\bm{P}\right|\gg Q_{s}$ we define the dimensionless variable $\displaystyle x=\dfrac{Q_{s}}{\left|\bm{P}\right|}\ll 1$. We can rewrite the above expression using
\begin{equation}
    \nabla^{2}_{\bm{P}}f\left(P\right)=\left(\dfrac{1}{P}\dfrac{\partial}{\partial P}+\dfrac{\partial^{2}}{\partial P^{2}}\right)f\left(P\right)=\dfrac{x^{4}}{Q_{s}^{2}}\left(\dfrac{1}{x}\dfrac{\partial}{\partial x}+\dfrac{\partial^{2}}{\partial x^{2}}\right)f\left(Q_{s}/x\right)\equiv\dfrac{1}{Q_{s}^{2}}\hat{\mathcal{L}}_{x}f\left(Q_{s}/x\right),
\end{equation}
to obtain
\begin{equation}
    \begin{split}
        &\left\langle\sum\left|\mathcal{M}_{bef}\right|^{2}_{\left(Nfact\right)}\right\rangle^{\left(0\right)}=\dfrac{\zeta\left(1-\zeta\right)}{Q^{2}}2N_{c}e_{q}^{2}e^{2}S_{\perp}\int d^{2}\Delta_{b} \ e^{-i\bm{q}\cdot\bm{\Delta_{b}}}\int^{1}_{0}dt \ e^{-\frac{Q_{s}^{2}\left(1-t\right)}{2}\Delta_{b}^{2}}\\
        &\times\left[\dfrac{Q_{s}^{2}\Delta_{b}^{2}}{2}+\dfrac{\partial}{\partial T}\right]\sum_{ijk=0}^{\infty}\dfrac{1}{n!j!k!}\left(\dfrac{T}{8}\right)^{n}\left(\dfrac{t}{4}\right)^{j+k} \\
        &\times \hat{\mathcal{L}}_{x}^{n}\left[\hat{\mathcal{L}}^{j}_{x}\left(\dfrac{1-x^{2}\epsilon^{2}/Q_{s}^{2}}{1+x^{2}\epsilon^{2}/Q_{s}^{2}}\right)\hat{\mathcal{L}}^{k}_{x}\left(\dfrac{1-x^{2}\epsilon^{2}/Q_{s}^{2}}{1+x^{2}\epsilon^{2}/Q_{s}^{2}}\right)\right].
    \end{split}
\end{equation}
Defining
\begin{equation}
    f^{njk}\left(x\right)=\hat{\mathcal{L}}_{x}^{n}\left[\hat{\mathcal{L}}^{j}_{x}\left(\dfrac{1-x^{2}\epsilon^{2}/Q_{s}^{2}}{1+x^{2}\epsilon^{2}/Q_{s}^{2}}\right)\hat{\mathcal{L}}^{k}_{x}\left(\dfrac{1-x^{2}\epsilon^{2}/Q_{s}^{2}}{1+x^{2}\epsilon^{2}/Q_{s}^{2}}\right)\right],
\end{equation}
we can compute the required terms (cancellations occur between different contributions to the cross section), up to order $\displaystyle x^{6}$, in the above expansion\footnote{We define the dimensionless parameter $\displaystyle a^{2}=\epsilon^{2}/Q_{s}^{2}$.}:
\begin{equation}
    f^{000}\left(x\right)=1-4a^{2}x^{2}+8a^{4}x^{4}-12a^{6}x^{6}+\mathcal{O}\left(x^{8}\right),
\end{equation}
\begin{equation}
    f^{100}\left(x\right)=-16a^{2}x^{4}+128a^{4}x^{6}+\mathcal{O}\left(x^{8}\right),
\end{equation}
\begin{equation}
    f^{010}\left(x\right)=f^{001}\left(x\right)=-8a^{2}x^{4}+48a^{4}x^{6}+\mathcal{O}\left(x^{8}\right),
\end{equation}
\begin{equation}
    f^{110}\left(x\right)=f^{101}\left(x\right)=-128a^{2}x^{6}+\mathcal{O}\left(x^{8}\right),
\end{equation}
\begin{equation}
    f^{011}\left(x\right)=\mathcal{O}\left(x^{8}\right),
\end{equation}
\begin{equation}
    f^{200}\left(x\right)=-256a^2x^6+\mathcal{O}\left(x^{8}\right),
\end{equation}
\begin{equation}
    f^{020}\left(x\right)=f^{002}\left(x\right)=-128a^2x^6+\mathcal{O}\left(x^{8}\right).
\end{equation}
Additional terms are of increasingly higher order in $\displaystyle x$. In fact, it can be proven that for $\displaystyle j,k\geq1$
\begin{equation}
    f^{njk}\left(x\right)=\mathcal{O}\left(x^{2n+2\left(j+1\right)+2\left(k+1\right)}\right),
\end{equation}
while
\begin{equation}
    f^{nj0}\left(x\right)=f^{n0j}\left(x\right)=\mathcal{O}\left(x^{2n+2\left(j+1\right)}\right),
\end{equation}
provided that $\displaystyle n\neq0$ or $j\neq0$. With this in mind, we can write the complete expression up to order $\displaystyle x^{6}$:
\begin{equation}
    \begin{split}
        &\left\langle\sum\left|\mathcal{M}_{bef}\right|^{2}_{\left(Nfact\right)}\right\rangle^{\left(0\right)}=\dfrac{\zeta\left(1-\zeta\right)}{Q^{2}}2N_{c}e_{q}^{2}e^{2}S_{\perp}\int d^{2}\Delta_{b} \ e^{-i\bm{q}\cdot\bm{\Delta_{b}}}\int^{1}_{0}dt \ e^{-\frac{Q_{s}^{2}\left(1-t\right)}{2}\Delta_{b}^{2}}\\
        &\times\left[\dfrac{Q_{s}^{2}\Delta_{b}^{2}}{2}+\dfrac{\partial}{\partial T}\right]\Biggl[1-4a^{2}x^{2}+8a^{4}x^{4}-12a^{6}x^{6}+\dfrac{T}{8}\Bigl(-16a^{2}x^{4}+128a^{4}x^{6}\Bigr)\Biggr.\\
        &\Biggl.+2\dfrac{t}{4}\Bigl(-8a^{2}x^{4}+48a^{4}x^{6}\Bigr)+2\dfrac{T}{8}\dfrac{t}{4}\Bigl(-128a^{2}x^{6}\Bigr)+\dfrac{1}{2!}\left(\dfrac{T}{8}\right)^{2}\left(-256a^{2}x^{6}\right)\\
        &+2\dfrac{1}{2!}\left(\dfrac{t}{4}\right)^{2}\left(-128a^{2}x^{6}\right)+\mathcal{O}\left(x^{8}\right)\Biggr]
    \end{split}
\end{equation}
and, integrating $t$, we finally get
\begin{equation}
    \begin{split}
        &\left\langle\sum\left|\mathcal{M}_{bef}\right|^{2}_{\left(Nfact\right)}\right\rangle^{\left(0\right)}=\dfrac{\zeta\left(1-\zeta\right)}{Q^{2}}2N_{c}e_{q}^{2}e^{2}S\int d^{2}\Delta_{b} \ e^{-i\bm{q}\cdot\bm{\Delta_{b}}}\Biggl\lbrace1-e^{-\frac{Q_{s}^{2}\Delta_{b}^{2}}{2}}\\
        &-4a^2\left(1-e^{-\frac{Q_{s}^{2}\Delta_{b}^{2}}{2}}\right)x^{2}\Biggr.+2a^2\left[-2+e^{-\frac{Q_{s}^{2}\Delta_{b}^{2}}{2}}+4a^2\left(1-e^{-\frac{Q_{s}^{2}\Delta_{b}^{2}}{2}}\right)\right]x^{4}\\
        &\Biggl.+\left[2a^{2}\left(-4+12a^2-6a^4+\left(1-8a^2+6a^4\right)e^{-\frac{Q_{s}^{2}\Delta_{b}^{2}}{2}}\right)\right.\\
        &\Biggl.\left.+\dfrac{16a^{4}}{\Delta_{b}^{2}Q_{s}^{2}}\left(1-e^{-\frac{Q_{s}^{2}\Delta_{b}^{2}}{2}}\right)\right]x^{6}+\mathcal{O}\left(x^{8}\right)\Biggr\rbrace.
    \end{split}
\end{equation}

Proceeding in the same way for all terms that contribute to eikonal order, we get~\eqref{eq:0corrlimexpr}.

\section{$\displaystyle 0^{th}$-order term in the shockwave expansion for a transverse photon}
Only the before-before, the before-after and the after-after contribute at order zero (eikonal). The first two for a transverse photon are, respectively,
\begin{equation}
    \begin{split}
        &\left\langle\overline{\sum}\left|\mathcal{M}_{bef}\right|^{2}\right\rangle^{\left(0\right)}=2N_{c}e_{q}^{2}e^{2}\zeta\left(1-\zeta\right)\left[\zeta^{2}+\left(1-\zeta\right)^{2}\right]S_{\perp}\int d^{2}u_{x}d^{2}u_{z}\\
        &\times e^{-i\bm{p}\cdot\bm{u}_{x}}e^{-i\bm{k}\cdot\bm{u}_{z}}\Biggl[e^{-\frac{Q_{s}^{2}}{4}\left(u_{x}^{2}+u_{z}^{2}\right)}\int\dfrac{d^{2}p_{0}}{\left(2\pi\right)^{2}}\dfrac{p^{2}_{0}}{\left(p^{2}_{0}+\epsilon^{2}\right)^{2}}e^{i\bm{p}_{0}\cdot\left(\bm{u}_{x}-\bm{u}_{z}\right)}+\dfrac{Q^{2}_{s}}{2}\left(\bm{u}_{x}\cdot\bm{u}_{z}\right)\Biggr.\\
        &\times\int^{1}_{0}dt \ e^{-\frac{Q_{s}^{2}\left(1-t\right)}{4}\left(u_{x}^{2}+u_{z}^{2}\right)}\int d^{2}\Delta_{0}\int\dfrac{d^{2}p_{0}}{\left(2\pi\right)^{2}}\dfrac{d^{2}\bar{p}_{0}}{\left(2\pi\right)^{2}}e^{i\bm{p}_{0}\cdot\bm{\Delta}_{0}}e^{-i\bm{\bar{p}}_{0}\cdot\left(\bm{\Delta}_{0}-\bm{u}_{x}+\bm{u}_{z}\right)}\\
        &\Biggl.\times\dfrac{\bm{p_{0}}\cdot\bm{\bar{p}_{0}}}{\left(p^{2}_{0}+\epsilon^{2}\right)\left(\bar{p}^{2}_{0}+\epsilon^{2}\right)}\  e^{-\frac{tQ_{s}^{2}}{4}\Delta_{0}^{2}} \ e^{-\frac{tQ_{s}^{2}}{4}\left(\bm{\Delta}_{0}-\bm{u}_{x}+\bm{u}_{z}\right)^{2}}\Biggr]
    \end{split}
\end{equation}
and
\begin{equation}
    \begin{split}
        &\left\langle\overline{\sum}\mathcal{M}_{bef}\mathcal{M}_{aft}^{\ast}\right\rangle^{\left(0\right)}=-2N_{c}e_{q}^{2}e^{2}\zeta\left(1-\zeta\right)\left[\zeta^{2}+\left(1-\zeta\right)^{2}\right]S_{\perp}\left(2\pi\right)^{2}\delta^{\left(2\right)}\left(\bm{p}+\bm{k}\right)\\
        &\times\int\dfrac{d^{2}p_{0}}{\left(2\pi\right)^{2}}\dfrac{\bm{p}\cdot\bm{p_{0}}}{\left(p^{2}+\epsilon^{2}\right)\left(p_{0}^{2}+\epsilon^{2}\right)}\int d^{2}\bm{\Delta} \ e^{-i\left(\bm{p}-\bm{p}_{0}\right)\cdot\bm{\Delta}} \ e^{-\frac{Q_{s}^{2}}{4}\Delta^{2}},
    \end{split}
\end{equation}
while the after-after~\eqref{eq.2.48} term is purely eikonal.
Taking the correlation limit in the above expressions we obtain
\begin{equation}
    \begin{split}
        \left\langle\overline{\sum}\left|M\right|^{2}\right\rangle^{\left(0\right)}=H_{\gamma^{\ast}_{T}g\to q\bar{q}}\left[\dfrac{16\pi N_{c}}{\alpha_{s}}S_{\perp}\int d^{2}\Delta_{b} \ \dfrac{e^{-i\bm{q}\cdot\bm{\Delta_{b}}}}{\Delta_{b}^{2}}\left(1-e^{-\frac{Q_{s}^{2}\Delta_{b}^{2}}{2}}\right)+\mathcal{O}\left(\dfrac{Q_{s}^{2}}{P^{2}}\right)\right],
    \end{split}
\end{equation}
where the hard partonic cross section por the transverse photon 
reads
\begin{equation}
    H_{\gamma^{\ast}_{T}g\to q\bar{q}}=\alpha_{s}\alpha_{em}e^{2}_{q}\zeta\left(1-\zeta\right)\left[\zeta^{2}+\left(1-\zeta\right)^{2}\right]\dfrac{P^{4}+\epsilon^{4}}{\left(P^{2}+\epsilon^{2}\right)^{4}}\ .
\end{equation}

\section{$\displaystyle 2^{nd}$-order term in the shockwave expansion for a transverse photon}
At second order in the shockwave expansion $\displaystyle \left(L/q^{+}\right)^{2}$ (next-to-next-to-eikonal) for a transverse photon we obtain the following contributions:
\begin{equation}
    \begin{split}
        &\left\langle\overline{\sum}\left|\mathcal{M}_{bef}\right|^{2}\right\rangle^{\left(2\right)}=N_{c}e_{q}^{2}e^{2}\left[\zeta^{2}+\left(1-\zeta\right)^{2}\right]S_{\perp}\dfrac{Q_{s}^{4}}{180\zeta\left(1-\zeta\right)}\left(\dfrac{L^{+}}{q^{+}}\right)^{2}\int d^{2}\Delta_{b} \\
        &\times e^{-i\bm{q}\cdot\bm{\Delta_{b}}}\int d^{2}r \ d^{2}\bar{r} \ e^{-i\bm{P}\cdot\left(\bm{r}-\bm{\bar{r}}\right)}\int \dfrac{d^{2}p_{0}}{\left(2\pi\right)^{2}}\dfrac{d^{2}\bar{p}_{0}}{\left(2\pi\right)^{2}}e^{i\bm{p_{0}}\cdot\bm{r}}e^{-i\bm{\bar{p}_{0}}\cdot\bm{\bar{r}}}\dfrac{\bm{p_{0}}\cdot\bm{\bar{p}_{0}}}{\left(p^{2}_{0}+\epsilon^{2}\right)\left(\bar{p}_{0}^{2}+\epsilon^{2}\right)}\\
        &\times\dfrac{Q_{s}^{2}}{2}\left[\Delta_{b}^{2}-\dfrac{\left(\bm{r}-\bm{\bar{r}}\right)^{2}}{4}\right]\int^{1}_{0}dt \ e^{-\frac{Q_{s}^{2}\left(1-t\right)}{2}\Delta_{b}^{2}}e^{-\frac{Q_{s}^{2}\left(1-t\right)}{8}\left(\bm{r}-\bm{\bar{r}}\right)^{2}}e^{-\frac{Q_{s}^{2}t}{4}r^{2}}e^{-\frac{Q_{s}^{2}t}{4}\bar{r}^{2}} t^{2}\\
        &\times \Biggl\lbrace t\dfrac{\bar{p}_{0}^{2}}{Q_{s}^{2}}\left(120-45i\bm{\bar{p}_{0}}\cdot\bm{\bar{r}}-45i\bm{\bar{p}_{0}}\cdot\bm{r}\right)-t^{2}\dfrac{15}{2}\Bigl[-8+3\left(\bm{\bar{p}_{0}}\cdot\bm{\bar{r}}\right)^{2}+2\bar{p}_{0}^{2}\bar{r}^{2}+6i\bm{\bar{p}_{0}}\cdot\bm{r}\Bigr.\Biggr.\\
        &\Bigl.+\bm{\bar{p}_{0}}\cdot\bm{\bar{r}}\left(14i+3\bm{\bar{p}_{0}}\cdot\bm{r}\right)+3\bar{p}_{0}^{2}\left(\bm{r}\cdot\bm{\bar{r}}\right)+\bar{p}_{0}^{2}r^{2}\Bigr]+\dfrac{3}{4}it^{3}Q_{s}^{2}\Bigl[\bar{r}^{2}\left(32i+15\bm{\bar{p}_{0}}\cdot\bm{\bar{r}}+5\bm{\bar{p}_{0}}\cdot\bm{r}\right)\Bigr.\\
        &\Biggl.\Bigl.+\left(25i+15\bm{\bar{p}_{0}}\cdot\bm{\bar{r}}\right)\left(\bm{r}\cdot\bm{\bar{r}}\right)+5r^{2}\left(i+\bm{\bar{p}_{0}}\cdot\bm{\bar{r}}\right)\Bigr]+\dfrac{5}{8}t^{4}Q_{s}^{4}\bar{r}^{2}\left(2\bar{r}^{2}+3\bm{r}\cdot\bm{\bar{r}}+r^{2}\right)\Biggr\rbrace
    \end{split}
\end{equation}
for the before-before contribution,
\begin{equation}
    \begin{split}
        &\left\langle\overline{\sum}\mathcal{M}_{bef}\mathcal{M}_{in\left(fact\right)}^{\ast}\right\rangle^{\left(2\right)}=-\dfrac{N_{c}e_{q}^{2}e^{2}}{Q^{2}}\left[\zeta^{2}+\left(1-\zeta\right)^{2}\right]S_{\perp}\int d^{2}\Delta_{b} \ e^{-i\bm{q}\cdot\bm{\Delta_{b}}}\\
        &\times\int d^{2}r \ d^{2}\bar{r} \ e^{-i\bm{P}\cdot\left(\bm{r}-\bm{\bar{r}}\right)}\int\dfrac{d^{2}p_{0}}{\left(2\pi\right)^{2}}\dfrac{d^{2}\bar{p}_{0}}{\left(2\pi\right)^{2}} \ e^{i\bm{p_{0}}\cdot\bm{r}} \ e^{-i\bm{\bar{p}_{0}}\cdot\bm{\bar{r}}}\left(\dfrac{\bm{p_{0}}\cdot\bm{\bar{p}_{0}}}{p_{0}^{2}+\epsilon^{2}}\right)\int^{1}_{0}dt \ e^{-\frac{Q_{s}^{2}\left(1-t\right)}{2}\Delta_{b}^{2}} \ \\
        &\times\ e^{-\frac{Q_{s}^{2}\left(1-t\right)}{8}\left(\bm{r}-\bm{\bar{r}}\right)^{2}} \ e^{-\frac{Q_{s}^{2}t}{4}r^{2}}t\Biggl\lbrace\dfrac{\lambda^{4}}{2}+\dfrac{\lambda^{2}\kappa^{2}}{2\zeta\left(1-\zeta\right)}\biggl[\dfrac{t}{2}+\dfrac{\bar{p}_{0}^{2}}{Q_{s}^{2}}-i\dfrac{t}{2}\bm{\bar{p}_{0}}\cdot\bm{r}-\dfrac{t^{2}Q_{s}^{2}}{12}r^{2}\biggr]\Biggr\rbrace
    \end{split}
\end{equation}
for the factorisable part of the before-inside contribution,
\begin{equation}
    \begin{split}
        &\left\langle\overline{\sum}\mathcal{M}_{bef}\mathcal{M}_{in\left(Nfact\right)}^{\ast}\right\rangle^{\left(2\right)}=-\dfrac{N_{c}e_{q}^{2}e^{2}}{Q^{2}}\left[\zeta^{2}+\left(1-\zeta\right)^{2}\right]S_{\perp}\int d^{2}\Delta_{b} \ e^{-i\bm{q}\cdot\bm{\Delta_{b}}}\\
        &\times\int d^{2}r \ d^{2}\bar{r} \ e^{-i\bm{P}\cdot\left(\bm{r}-\bm{\bar{r}}\right)}\int\dfrac{d^{2}p_{0}}{\left(2\pi\right)^{2}}\dfrac{d^{2}\bar{p}_{0}}{\left(2\pi\right)^{2}} \ e^{i\bm{p_{0}}\cdot\bm{r}} \ e^{-i\bm{\bar{p}_{0}}\cdot\bm{\bar{r}}}\left(\dfrac{\bm{p_{0}}\cdot\bm{\bar{p}_{0}}}{p_{0}^{2}+\epsilon^{2}}\right)\dfrac{Q_{s}^{2}}{2}\\
        &\times\left[\Delta^{2}_{b}-\dfrac{\left(\bm{r}-\bm{\bar{r}}\right)^{2}}{4}\right]\int^{1}_{0}dt\int^{1}_{t}dt^{\prime} \ e^{-\frac{Q_{s}^{2}\left(1-t^{\prime}\right)}{2}\Delta_{b}^{2}} \ e^{-\frac{Q_{s}^{2}\left(1-t^{\prime}\right)}{8}\left(\bm{r}-\bm{\bar{r}}\right)^{2}} \ e^{-\frac{Q_{s}^{2}t^{\prime}}{4}r^{2}} \ e^{-\frac{Q_{s}^{2}\left(t^{\prime}-t\right)}{4}\bar{r}^{2}}\\
        &\times\Biggl\lbrace\dfrac{\lambda^{4}}{2}t+\dfrac{\lambda^{2}\kappa^{2}}{2\zeta\left(1-\zeta\right)}\biggl[t^{\prime2}-\dfrac{t^{2}}{2}+t\dfrac{\bar{p}_{0}^{2}}{Q_{s}^{2}}+\dfrac{i}{2}\left(t^{2}-t^{\prime2}\right)\bm{\bar{p}_{0}}\cdot\bm{\bar{r}}\biggr.\Biggr.\\
        &\Biggl.\biggl.\hskip 1cm +\dfrac{1}{24}\left(-2t^{3}+9tt^{\prime2}-7t^{\prime3}\right)Q_{s}^{2}\bar{r}^{2}-i\dfrac{t^{\prime2}}{2}\bm{\bar{p}_{0}}\cdot\bm{r}\biggr.\Biggr.\\
        &\Biggl.\biggl.\hskip 1cm +\dfrac{1}{24}\left(3t-5t^{\prime}\right)t^{\prime2}Q_{s}^{2}r^{2}+\dfrac{1}{8}t^{\prime2}\left(t^{\prime}-t\right)Q_{s}^{2}\left(\bm{r}-\bm{\bar{r}}\right)^{2}\biggr]\Biggr\rbrace
    \end{split}
\end{equation}
for the non-factorisable part of the before-inside contribution,
\begin{equation}
    \begin{split}
        &\left\langle\overline{\sum}\left|\mathcal{M}_{in}\right|^{2}\right\rangle^{\left(2\right)}=\dfrac{N_{c}e_{q}^{2}e^{2}}{Q^{4}\zeta\left(1-\zeta\right)}\left[\zeta^{2}+\left(1-\zeta\right)^{2}\right]S_{\perp}\lambda^{4}\int d^{2}\Delta_{b} \ e^{-i\bm{q}\cdot\bm{\Delta_{b}}}\\
        &\times\int d^{2}r \ d^{2}\bar{r} \ e^{-i\bm{P}\cdot\left(\bm{r}-\bm{\bar{r}}\right)}\int\dfrac{d^{2}p_{0}}{\left(2\pi\right)^{2}}\dfrac{d^{2}\bar{p}_{0}}{\left(2\pi\right)^{2}} \ e^{i\bm{p_{0}}\cdot\bm{r}} \ e^{-i\bm{\bar{p}_{0}}\cdot\bm{\bar{r}}}\left(\bm{p_{0}}\cdot\bm{\bar{p}_{0}}\right)\\
        &\times\int^{1}_{0}dt\int^{1}_{t}dt^{\prime}\Biggl\lbrace e^{-\frac{Q_{s}^{2}\left(1-t^{\prime}\right)}{2}\Delta_{b}^{2}} \ e^{-\frac{Q_{s}^{2}\left(1-t^{\prime}\right)}{8}\left(\bm{r}-\bm{\bar{r}}\right)^{2}} \ e^{-\frac{Q_{s}^{2}\left(t^{\prime}-t\right)}{4}r^{2}}+\dfrac{Q_{s}^{2}}{4}\left[\Delta_{b}^{2}-\dfrac{\left(\bm{r}-\bm{\bar{r}}\right)^{2}}{4}\right]\Biggr.\\
        &\Biggl.\hskip 1cm \times \int^{1}_{t^{\prime}}dt^{\prime\prime} \ e^{-\frac{Q_{s}^{2}\left(1-t^{\prime\prime}\right)}{2}\Delta_{b}^{2}} \ e^{-\frac{Q_{s}^{2}\left(1-t^{\prime\prime}\right)}{8}\left(\bm{r}-\bm{\bar{r}}\right)^{2}} \ e^{-\frac{Q_{s}^{2}\left(t^{\prime\prime}-t\right)}{4}r^{2}} \ e^{-\frac{Q_{s}^{2}\left(t^{\prime\prime}-t^{\prime}\right)}{4}\bar{r}^{2}}\Biggr\rbrace
    \end{split}
\end{equation}
for the inside-inside contribution,
\begin{equation}
    \begin{split}
        &\left\langle\overline{\sum}\mathcal{M}_{bef}\mathcal{M}_{aft}^{\ast}\right\rangle^{\left(2\right)}=2N_{c}e_{q}^{2}e^{2}\zeta\left(1-\zeta\right)\left[\zeta^{2}+\left(1-\zeta\right)^{2}\right]S_{\perp}\int d^{2}\Delta_{b} \ e^{-i\bm{q}\cdot\bm{\Delta_{b}}}\\
        &\times\int d^{2}r \ d^{2}\bar{r} \ e^{-i\bm{P}\cdot\left(\bm{r}-\bm{\bar{r}}\right)}\int\dfrac{d^{2}p_{0}}{\left(2\pi\right)}\dfrac{d^{2}\bar{p}_{0}}{\left(2\pi\right)^{2}} \ e^{i\bm{p_{0}}\cdot\bm{r}} \ e^{-i\bm{\bar{p}_{0}}\cdot\bm{\bar{r}}}\dfrac{\bm{p_{0}}\cdot\bm{\bar{p}_{0}}}{\left(p^{2}_{0}+\epsilon^{2}\right)\left(\bar{p}^{2}_{0}+\epsilon^{2}\right)} \ e^{-\frac{Q_{s}^{2}}{4}\bar{r}^{2}}\\
        &\times\Biggl\lbrace\dfrac{\lambda^{2}}{8}+\dfrac{\lambda^{2}\kappa^{2}}{4\zeta\left(1-\zeta\right)}\biggl[\dfrac{1}{2}+\dfrac{P^{2}}{Q_{s}^{2}}+\dfrac{i}{2}\bm{P}\cdot\bm{\bar{r}}-\dfrac{Q_{s}^{2}}{12}\bar{r}^{2}\biggr]\Biggr.\\
        &+\dfrac{\kappa^{4}}{360\zeta^{2}\left(1-\zeta\right)^{2}}\biggl[\dfrac{75}{4}+45\dfrac{P^{4}}{Q_{s}^{4}}+75\dfrac{P^{2}}{Q_{s}^{2}}+i45\dfrac{P^{2}}{Q_{s}^{2}}\bm{P}\cdot\bm{\bar{r}}-\dfrac{45}{4}\left(\bm{P}\cdot\bm{\bar{r}}\right)^{2}\biggr.\\
        &\Biggl.\biggl.-\dfrac{15}{2}P^{2}\bar{r}^{2}+i\dfrac{165}{4}\bm{P}\cdot\bm{\bar{r}}-i\dfrac{15}{4}Q_{s}^{2}\bar{r}^{2}\bm{P}\cdot\bm{\bar{r}}-\dfrac{27}{4}Q_{s}^{2}\bar{r}^{2}+\dfrac{5}{16}Q_{s}^{2}\bar{r}^{4}\biggr]\Biggr\rbrace
    \end{split}
\end{equation}
for the before-after contribution and, finally,
\begin{equation}
    \begin{split}
        &\left\langle\overline{\sum}\mathcal{M}_{in}\mathcal{M}_{aft}^{\ast}\right\rangle^{\left(2\right)}=-\dfrac{N_{c}e_{q}^{2}e^{2}}{Q^{2}}\left[\zeta^{2}+\left(1-\zeta\right)^{2}\right]S_{\perp}\int d^{2}\Delta_{b} \ e^{-i\bm{q}\cdot\bm{\Delta_{b}}}\int d^{2}r \ d^{2}\bar{r} \\
        &\times e^{-i\bm{P}\cdot\left(\bm{r}-\bm{\bar{r}}\right)}\int\dfrac{d^{2}p_{0}}{\left(2\pi\right)}\dfrac{d^{2}\bar{p}_{0}}{\left(2\pi\right)^{2}} \ e^{i\bm{p_{0}}\cdot\bm{r}} \ e^{-i\bm{\bar{p}_{0}}\cdot\bm{\bar{r}}} \ \dfrac{\bm{p_{0}}\cdot\bm{\bar{p}_{0}}}{p_{0}^{2}+\epsilon^{2}}\int^{1}_{0}dt\left(1-t\right)e^{-\frac{Q_{s}^{2}\left(1-t\right)}{4}\bar{r}^{2}}\\
        &\times\Biggl\lbrace\dfrac{\lambda^{4}}{2}+\dfrac{\lambda^{2}\kappa^{2}}{2\zeta\left(1-\zeta\right)}\biggl[\dfrac{P^{2}}{Q_{s}^{2}}+\dfrac{\left(1-t\right)}{2}+\dfrac{i}{2}\left(1-t\right)\bm{P}\cdot\bm{\bar{r}}-\dfrac{1}{12}\left(1-t\right)^{2}Q_{s}^{2}\bar{r}^{2}\biggr]\Biggr\rbrace
    \end{split}
\end{equation}
for the inside-after contribution.
Taking the correlation limit in the above expressions we obtain
\begin{equation}
    \begin{split}
        \left\langle\overline{\sum}\left|M\right|^{2}\right\rangle^{\left(2\right)}=&H_{\gamma^{\ast}_{T}g\to q\bar{q}}\dfrac{P^{4}}{Q_{s}^{4}}\left(\dfrac{L^{+}}{q^{+}}\right)^{2}\dfrac{1}{\zeta^{2}\left(1-\zeta\right)^{2}}\left[\dfrac{4\pi N_{c}S_{\perp}}{\alpha_{s}}\int d^{2}\Delta_{b} \ \dfrac{e^{-i\bm{q}\cdot\bm{\Delta_{b}}}}{\Delta_{b}^{6}}\right.\\
        &\times\left.\left(-16+16e^{-\frac{Q_{s}^{2}\Delta_{b}^{2}}{2}}+8\Delta_{b}^{2}Q_{s}^{2}-2\Delta^{4}_{b}Q_{s}^{4}+\dfrac{1}{3}\Delta^{6}_{b}Q_{s}^{6}\right)+\mathcal{O}\left(\dfrac{Q_{s}^{2}}{P^{2}}\right)\right]
    \end{split}
\end{equation}
which, as for the longitudinal photon, does not diverge due to the cancellation of the first terms with those coming from the expansion of the exponential.

\bibliographystyle{JHEP}
\bibliography{mybib}

\end{document}